\documentclass[11pt]{article}
\usepackage[a4paper,margin=0.85in]{geometry}
\usepackage{amsmath,amssymb,mathtools,slashed}
\usepackage{hyperref}
\usepackage{graphicx}
\usepackage{microtype}
\usepackage{authblk}
\usepackage{booktabs}
\usepackage{enumitem}
\usepackage{tabularx}
\usepackage{array}
\usepackage{float}
\usepackage{soul}

\usepackage{adjustbox}
\usepackage{times}
\usepackage[english]{babel}
\usepackage{wasysym}
\usepackage{bm}
\usepackage{textcomp}
\usepackage{latexsym}
\usepackage{amsfonts} 
\usepackage[square]{natbib}

\title{\textbf{Toward a Special \textbf{$E_6\to G(2) \times SU(3)_A$} Embedding for Standard Model and Dark Matter and an $E_7$ Completion Proposal}}

\author[1]{Nicol\`o Masi}

\affil[1]{INFN \and Bologna University, Physics Department, Via Irnerio 46 Bologna, 40126, Italy}

\affil[]{masin@bo.infn.it}
\date{}
\begin{document}
\maketitle

\begin{abstract}
	We developed a unified framework based on a special (non-regular) embedding of the exceptional group $E_6$ in which the main stage of symmetry breaking chain realizes $E_6\to G(2) \times SU(3)_A$. The exceptional factor $G(2)$ plays the role of a hidden strong sector,
	while $SU(3)_A$ acts as an ancestor of the electroweak gauge group. A minimal scalar sector is organized around a \(\mathbf{650}\)-based
	\(E_6\) Higgs sector. Its \(G(2)\times SU(3)_A\) components $(\mathbf7,\mathbf1)$ and $(\mathbf1,\mathbf8)$ implement the subsequent breaking steps
	$G(2)\to SU(3)_C$ and $SU(3)_A\to SU(2)_L\times U(1)_A$. The \emph{speciality} of this symmetry breaking establishes the feature of \textit{darkness}. 
	Defining a hypercharge from the $t_{8}$ generator of $SU(3)_A$ is not sufficient to recover the exact Standard Model hypercharges and the correct Weinberg weak angle, leading to the necessity of an $E_7$ uplift which introduces only a proper additional $U(1)_X$ factor to fix the abelian sector.
	 
	The special embedding naturally suppresses tree-level leptoquark couplings that typically mediate proton decay in regular grand unified theories. The scalar potential for the Higgs sectors has been constructed, deriving the heavy gauge-bosons spectrum and presenting a consistent one-loop running of the gauge couplings across the intermediate scales, which is shown to satisfy an $E_6$ unification.
	The exotic states are organized in \(E_7\)-derived vectorlike pairs and are made ultraheavy, while a chirality-selecting projection keeps only the
	desired Standard Model zero modes. All non-Standard Model Higgs fields and \textit{broken} massive vectors are found to be invisible to colliders searches. 
	
	The $G(2)$ gluons ensemble confines into heavy dark glueballs with parametrically suppressed communication with $SU(3)_A$ and $U(1)_X$ sectors.
	Cosmological history is analyzed in details, including topological defects, inflation and reheating, demonstrating that monopole relics are naturally diluted.
	The resulting framework provides a minimal and internally consistent exceptional apparatus which includes the Standard Model and a dark matter sector which is secluded by the group-theoretic orthogonality.
\end{abstract}

\section{Introduction}

The nature of dark matter (DM) and the origin of the Standard Model (SM) gauge symmetries remain among the most profound open
problems in contemporary physics \citep{ModernPP}: while the SM has proven extraordinarily successful, it offers no explanation
for the DM existence. This is probably the most compelling and very long-standing problem of modern physics, with no evident nor univocal solution: all the efforts made, from particle theory \citep{Bertone,ProfumoDM} to modified gravities \citep{InfraredG,ModG_largeDist,Clifton:2011jh}, have not been successful in clarifying its nature.

Recently, exceptional gauge groups have re-emerged as promising candidates for physics beyond the SM. Recent developments in non-supersymmetric $E_6$ model building have revisited realistic intermediate-scale breaking chains and scalar
sector minimality \cite{Babu_2023,Babu2024RealisticE6},while special grand-unified embeddings have been systematically
analyzed in the modern framework of special subgroups \cite{Yamatsu2017SpecialGUT}. At the same time, renewed interest in
non-Abelian dark sectors and glueball dark matter \cite{Forestell2017Glueball,Forestell:2016qhc,Soni:2016gzf} motivates unified constructions in which visible and dark sectors share a common exceptional origin. In the present work I pursue a special $E_6 \supset G(2) \times SU(3)_A$ breaking chain, combining the group-theoretic advantages of exceptional embeddings with the phenomenological framework of confining dark dynamics. The central motivation is to expand the examination of the $G(2)$-based dark sector I proposed in \cite{Masi2024}, where a confining non-Abelian gauge theory with exceptional group $G(2)$ is broken at a high scale to an $SU(3)$ subgroup, identified with Quantum Chromodynamics (QCD) $SU(3)_C$, leaving a set of heavy vectors associated with the coset $S^6 = G(2)/SU(3)$.
These massive dark vector gluons can bind into heavy glueball bound states once the residual non-Abelian
sector confines \cite{Masi2021, Masi2024,LuciniTeper2001,MorningstarPeardon1999}, providing a minimal and stable DM candidate.
The confinement properties of $G(2)$ gauge theory have been extensively studied on the lattice
\cite{Holland2004,Greensite2011,Maas:2012ts}.
These dark bound states may also form, under suitable
conditions, macroscopic boson-star configurations \cite{Kaup1968,RuffiniBonazzola1969,ColpiShapiroWasserman1986,LieblingPalenzuela2012} or primordial black holes \cite{Masi2024}.

Here I want to supply a minimal grand-unified completion in which the $G(2)$ sector of \cite{Masi2024} is preserved essentially intact, while the SM gauge group and hypercharge normalization arise from a single ultraviolet embedding. The idea is that such a $G(2)$ dark sector should not be \textit{appended} to the SM only by the algebraic guideline developed in \cite{Masi2021}, but arise naturally from a grand unified gauge structure which contains all the groups. 

The classification of maximal subgroups of compact simple Lie groups, such as $SU(N)$ and the five exceptional groups, 
was established by Dynkin \cite{Dynkin1952,Dynkin1957}
(see also \cite{Yamatsu2015,Slansky1981} for further details). The smallest exceptional group that admits $G(2)$ as a maximal special subgroup is $E_6$  \cite{Shafi1978E6}, which also contains $SU(3)$ subgroups, capable of reproducing the electroweak sector, being $SU(2) \times U(1)$ the maximal regular subalgebra of $SU(3)$ itself. Unlike conventional $E_6$ unification schemes that rely on regular subgroups such as $SO(10)\times U(1)$ or $SU(3)^3$ \cite{GurseyRamondSikivie1976,AchimanStech1978,HewettRizzo1989,Slansky1981}, the present one exploits the special (non-regular) embedding 
\begin{equation}
	E_6 \;\longrightarrow\; G(2) \times SU(3)_A.
\end{equation}
The exceptional factor $G(2)$ becomes the dark sector, while this new $SU(3)_A$ plays the
role of an ancestor/progenitor of the electroweak group.
Hypercharge is embedded directly inside $SU(3)_A$ \cite{Georgi1999,Slansky1981}, and the SM Higgs doublet is identified within a higher-dimensional representation.
The resulting symmetry-breaking chain
\begin{equation}
	E_6 \;\to\; G(2) \times SU(3)_A
	\;\to\; SU(3)_C\times SU(3)_A
	\;\to\; SU(3)_C\times SU(2)_L\times U(1)_A,
\end{equation}
connects exceptional unification to SM physics while preserving a dark sector with minimal additional assumptions. 
A refined completion of this minimal picture is obtained by embedding the
construction into the regular exceptional step
\begin{equation}
	E_7\;\longrightarrow\;E_6\times U(1)_X .
\end{equation}
This uplift does not replace the special \(E_6\to G(2)\times SU(3)_A\)
mechanism; rather, it completes its electroweak abelian sector. In the strict
\(E_6\) realization, placing hypercharge entirely inside \(SU(3)_A\) leads to a
normalization tension: the Higgs hypercharge and the canonical high-scale value
\(\sin^2\theta_W=3/8\) \citep{ModernPP} cannot be obtained simultaneously for an integer
\(SU(3)_A\) weight lattice. The additional \(U(1)_X\) supplied by \(E_7\)
resolves this obstruction by allowing
\begin{equation}
	Y=\beta A+\gamma X,
\end{equation}
where \(A=\sqrt3\,t_A^8\) is the diagonal \(SU(3)_A\) generator. 

The central goal of this paper is to demonstrate that the special \(E_6\)
construction is theoretically appealing and phenomenologically viable and that
its \(E_7\) uplift supplies a controlled completion of the hypercharge and weak-angle sector without spoiling the original \(G(2)\) dark-sector mechanism.

The scalar sectors responsible for each breaking
step are constructed, deriving the heavy gauge-boson spectrum, along with the one-loop running of the gauge couplings across the intermediate scales: a benchmark framework that satisfies unification and proton-decay constraints \cite{Weinberg1979Bviol,WilczekZee1979,SuperK2017,PDG2024} is realized.

Beyond particle physics, I explored the cosmological implications of the exceptional breaking chain and observational prospects: it is shown that topological defects generated at the highest scales are naturally diluted by inflation \cite{Guth1981,Linde1983,Linde1982Inflation}, that reheating can preserve the dark $G(2)$ sector and that baryogenesis \cite{FukugitaYanagida1986}
may proceed through heavy $E_6$ states leptogenesis \cite{PhysRevD.104.055007, FongNardiRiotto2012}.

The novelty of this approach is the exploitation of a special exceptional embedding in which the SM gauge group and a confining dark gauge sector arise as complementary substructures inside an \(E_6\) framework. The subsequent
\(E_7\) uplift preserves this structure while adding precisely the abelian freedom needed to complete the theory coherently. The result is a projected exceptional GUT which permits a near meeting of
gauge couplings with a small number of thresholds, a naturally isolated confining \(G(2)\)-broken dark sector protected by group-theoretic orthogonality and a consistent cosmological picture.


\section{Alternative $E_6$ Grand Unification Historical Approaches}

Grand unification was first realized in $SU(5)$~\cite{GeorgiGlashow1974}
and subsequently extended to $SO(10)$~\cite{FritzschMinkowski1975}
and $E_6$~\cite{GurseyRamondSikivie1976}. The exceptional group $E_6$ has long been regarded as one of the most attractive candidates for grand unification \cite{Ross1984,Mohapatra2003,EllisNanopoulos1981,SERDAROGLU1982271,PhysRevD.34.1530}, owing to its anomaly-free fundamental
$\mathbf{27}$ representation and its rich structure of maximal regular and
special subgroups \cite{Slansky1981,Dynkin1952,Raby2006GUTReview,Mohapatra1986Review,PDG2025GUT}.
Early work by Achiman and Stech demonstrated that $E_6$ models can naturally incorporate quark--lepton symmetry and hierarchical fermion mass scales
through specific symmetry-breaking patterns and vacuum expectation values
\cite{AchimanStech1978}. In these constructions, the decomposition of the
$\mathbf{27}$ into SM representations produces additional
vectorlike states whose masses are generated by appropriate Higgs expectation values, preserving realistic low-energy phenomenology.

Subsequently, Stech and Tavartkiladze developed non-supersymmetric $E_6$
frameworks \cite{Stech2004,Stech2008} in which fermion mass hierarchies and gauge coupling unification
are achieved via structured Yukawa textures and intermediate symmetry stages. In these approaches, heavy vectorlike fermions play a dynamical role in shaping the fermion spectrum rather than being merely lifted to high scales. Often, the unification chain proceeds through trinification-like intermediate groups such as $SU(3)^3$, which arise as
maximal regular subgroups of $E_6$ \cite{Slansky1981,Stech2012,Babu_2023}.

Other realizations of $E_6$ unification emphasize left--right symmetric
intermediate stages or extended $U(1)$ sectors, with phenomenological
implications ranging from neutrino mass generation and ultra heavy neutrino DM \cite{Schwichtenberg:2017xhv} to additional $Z'$ bosons
\cite{HewettRizzo1989,Langacker2009Zprime}. These realizations typically prioritize conventional gauge coupling unification and flavor predictivity, and in many cases rely on supersymmetric extensions.

The framework developed in the present work differs conceptually from these regular-embedding approaches, being a special embedding which is less commonly explored in the literature but equally legitimate
from the standpoint of Lie algebra classification \cite{Dynkin1952,Slansky1981}.
The distinguishing feature of this chain is the explicit factorization of a confining $G(2)$ sector at high scales, enabling a clean separation between visible and dark dynamics while maintaining unified origin at the $E_6$ level.
In fact, with a special non--regular embedding, the coset generators can sit in different representations and some dangerous bilinears can be absent or only induced via mixing. This means that leptoquark-like vectors do not couple to quark–lepton currents at tree level (or couple only after small mixings), baryon/lepton number violating amplitudes can be structurally suppressed and extra $U(1)$ abelian factors inside $E_6$ can be avoided \cite{PhysRevD.105.075021}.
In this sense, Stech’s program (and other aforementioned frameworks in literature) and the present construction represent two distinct but complementary philosophies within $E_6$ grand unification:
(i) regular-embedding chains optimized for fermion mass predictivity and coupling unification \cite{AchimanStech1978}, and
(ii) special-embedding chains designed to preserve a structurally isolated strong dark sector. So the special feature represents a seed for the \textit{darkness} of the $G(2)$ sector.

\section{The special $E_6$ Grand Unification Bricks}

Specifically, a subgroup $H\subset G$ is said to be \emph{regular} if it can be obtained by deleting nodes from the Dynkin diagram of the parent group $G$: the Cartan subalgebra of $H$ is a subset of the Cartan subalgebra of $G$, and the simple roots of $H$ are a subset of those of $G$ \cite{Dynkin1957, Yamatsu2015,Slansky1981}.
In this case the branching rules for representations follow directly from the root-space decomposition of the Dynkin diagram. Typical examples are the aforementioned well-known regular embeddings of $E_6$, $E_6 \supset SO(10)\times U(1)$ and $E_6 \supset SU(3)\times SU(3)\times SU(3)$.

By contrast, a subgroup is called \emph{special} (or non-regular) when it cannot be obtained by simple Dynkin-node deletion: the embedding involves nontrivial linear combinations of the generators of $G$ and the root system of $H$ is not a subset of that of $G$ in the Dynkin sense. As anticipated, special embeddings have several distinctive quantum features and advantages \cite{Slansky1981,NathFileviezPerez2007}: for example, in the present case based on the special embedding $E_6 \supset G(2) \times SU(3)_A$,
the generators associated with the $G(2)/SU(3)_C$ coset will act only within the color sector, without directly mixing quark and lepton directions at tree level, and granting DM isolation, as demonstrated later.

To break the parent group $G$ down to a specific regular or special subgroup $H$, a Higgs multiplet whose decomposition under $H$ includes a singlet is needed. A vacuum expectation value (VEV) in that singlet leaves exactly $H$ unbroken (up to discrete centers)\cite{ColemanWeinberg1973,BairdBiedenharn1972}. If the commutant of $H$ in $G$ is trivial--as it is for a maximal special embedding like $E_6 \supset G(2)\times SU(3)$--then a single singlet is in principle enough. The choice of Higgs representation responsible for the first breaking $E_6 \;\longrightarrow\; G(2) \times SU(3)_A$ is highly constrained by group theory and by minimality. In fact, the adjoint $\mathbf{78}$ and $\mathbf{351},\mathbf{351'}$ reps of $E_6$ do not include an $H$-singlet for this embedding, so they cannot lead to the desired $G(2) \times SU(3)_A$.
 Among low-dimensional irreducible representations of $E_6$, only $\mathbf{650}$ contain singlet directions under the \emph{special} (non-regular) subgroup $G(2) \times SU(3)_A$ \cite{Babu_2023}.
Its decomposition under $G(2) \times SU(3)_A$ contains a \emph{singlet} $\hat S$ $(\mathbf{1},\mathbf{1})$ direction, so that a generic VEV $\langle \Phi_{650} \rangle\propto(\mathbf{1},\mathbf{1})$ breaks $E_6$ \emph{only} to $G(2) \times SU(3)_A$ and to no larger subgroup.
This ensures that the first stage of symmetry breaking is clean and unique \cite{Babu_2023}.
Let's introduce the case of this $ \mathbf{650}$ rep Higgs scenario equipped with a reasonable VEV to directly break $E_6$ into $G(2) \times SU(3)_A$, constructing the Higgs potential and the mass spectrum of the theory. 
We will start from the gauge group $G\equiv E_6$ (with dimension $78$ and rank $6$), being the target subgroup $H \equiv G(2) \times SU(3)_A$ (dimension $8+14=22$ and rank $2+2=4$): $\Phi$ will be the $E_6$--Higgs in the $ \mathbf{650}$ representation whose explicit branching, computed with LieArt \cite{LieArt}, under $G(2)\times SU(3)_A$ is
\begin{equation}
	\mathbf{650}
	\to
	(\mathbf1,\mathbf1)\oplus
	(\mathbf7,\mathbf1)\oplus
	(\mathbf1,\mathbf8)\oplus
	2(\mathbf7,\mathbf8)\oplus
	(\mathbf7,\mathbf{10})\oplus
	(\mathbf7,\overline{\mathbf{10}})
	\oplus
	(\mathbf{14},\mathbf8)
	\oplus
	(\mathbf{27},\mathbf1)\oplus
	(\mathbf1,\mathbf{27})\oplus
	(\mathbf{27},\mathbf8).
	\label{eq:650_special_corrected}
\end{equation}
This decomposition exhibits explicitly the three subgroup components needed for the corrected symmetry-breaking chain:
\begin{equation}
	S\sim(\mathbf1,\mathbf1),
	\qquad
	\chi\sim(\mathbf7,\mathbf1),
	\qquad
	\Omega\sim(\mathbf1,\mathbf8).
\end{equation}
The first of these induces the special breaking
$E_6\to G(2)\times SU(3)_A$, the second breaks
$G(2)\to SU(3)_C$, and the third breaks
$SU(3)_A\to SU(2)_L\times U(1)_Y$.
Aligning the vacuum along the singlet direction one can define \cite{WeinbergQTF2}:
\begin{equation}
\langle \Phi\rangle = v_{E_6}\, \hat S \quad\text{with}\quad \hat S\in (\mathbf{1},\mathbf{1})\subset  \mathbf{650} .
\end{equation}
Here $v_{E_6}=v_{ \mathbf{650}}$ is the VEV for the $E_6$ breaking Higgs $\Phi$. This leaves $H=G(2) \times SU(3)_A$ unbroken and breaks a set of generators, which will be eaten Goldstones. 

In the following section the breaking chain is discussed in details.

\section{The Breaking Chain}
\label{sec:decomp_vectorlike}

We provide a systematic step-by-step description of how the unified $E_6$ matter and Higgs representations decompose along the symmetry-breaking chain 
\begin{equation}
E_6
\xrightarrow{\langle S \rangle}
G(2)\times SU(3)_A
\xrightarrow{\langle\chi\rangle}
SU(3)_C\times SU(3)_A
\xrightarrow{\langle\Omega\rangle}
SU(3)_C\times SU(2)_L\times U(1)_A,
	\label{eq:chain}
\end{equation}
induced by the VEVs of three new Higgs scalars, followed by the standard electroweak breaking by the SM Higgs field $H$. Hereafter $\chi$ and $\Omega$ will be the subsequent Higgs after $\Phi$. 

For what concerns fermions, the $\mathbf{27}_F$ fundamental fermion representation of $E_6$ is known to be chiral and anomaly-free and its decomposition contains additional states beyond a single SM family \cite{GurseyRamondSikivie1976,AchimanStech1978,HewettRizzo1989,Slansky1981}. We will test and show that this $\mathbf{27}_F$ rep for fermions, even if promising and minimal, is not sufficient to reproduce correct hypercharges assignments, leading to an enlarged symmetry pattern.

\subsection{Step 0: $E_6$ representations}

We take at first three chiral families 
\begin{equation}
	\Psi_i \sim \mathbf{27}_F,\qquad i=1,2,3,
\end{equation}
and a minimal multi-Higgs sector 
\begin{equation}
	\Phi_{650}\sim \mathbf{650}_H,
\end{equation}
triggering the special-embedding breaking: 
\begin{align}
	S &\subset \Phi_{\mathbf{650}}, & S &\sim (\mathbf1,\mathbf1),
	\\
	\chi &\subset \Phi_{\mathbf{650}}, & \chi &\sim (\mathbf7,\mathbf1),
	\\
	\Omega &\subset \Phi_{\mathbf{650}}, & \Omega &\sim (\mathbf1,\mathbf8).
\end{align}
where the transformation properties under
$G(2) \times SU(3)_A$ (and later under $SU(3)_C\times SU(3)_A$) are anticipated. The singlet scalar $(\mathbf{1},\mathbf{1})$ described before, originating as a SM singlet component of the $ \mathbf{650}$ sector, acquires a large VEV providing vectorlike masses for non-SM exotic states, as discussed in Section 5.2: 
\begin{equation}
	\langle \Phi\rangle \equiv v_{E_6} \gg M_\chi.
\end{equation}
where $M_\chi$ is the $G(2)$ energy breaking scale.

\subsection{Step 1: $E_6 \to G(2) \times SU(3)_A$ special embedding}
The first stage is induced by the singlet $S\subset \Phi_{\mathbf{650}}$ whose vacuum expectation value realizes $
E_6 \xrightarrow{\langle S\rangle} G(2)\times SU(3)_A$.
Under the special subgroup $G(2) \times SU(3)_A$, the fundamental matter representation decomposes as \cite{Babu_2023}
\begin{equation}
		\mathbf{27}_F \to (\mathbf7,\mathbf3)\oplus(\mathbf1,\overline{\mathbf6}).
	\label{eq:27toA_G2}
\end{equation}
We denote the corresponding fermion multiplets by 
\begin{equation}
	\Psi_i \;\equiv\; \Big( \psi_i \oplus \xi_i\Big),
	\qquad
	\psi_i \sim (\mathbf{7},\mathbf3),\qquad \xi_i\sim(\mathbf{1},\overline{\mathbf{6}}).
\end{equation}
The adjoint $\mathbf{78}$ of $E_6$ branches as  \cite{Babu_2023}
\begin{equation}
	\mathbf{\mathbf{78}} \;\to\; (\mathbf{14},\mathbf{1})\;\oplus\;(\mathbf{1},\mathbf{8})\;\oplus\;(\mathbf{7},\mathbf{8}) ,
\end{equation}
where $(\mathbf{14},\mathbf{1})$ and $(\mathbf{1},\mathbf{8})$ are the massless gauge bosons of $G(2)$ and $SU(3)_A$ and the coset is the $(\mathbf{7},\mathbf{8})$ with 56 states, which all become heavy at this first breaking. It is natural to denote these as $\mathcal{E}$-bosons transforming as $(\mathbf{7},\mathbf{8})$ under the special $G(2) \times SU(3)_A$. Their masses are of order 
\begin{equation}
	m_{\mathcal{E}} \sim g_{E_6}\,v_{E_6},
\end{equation}
up to standard group factors from the VEV direction, where $g_{E_6}$ is the characteristic coupling constant.
So, at this stage, the gauge bosons split into massless gauge bosons of $SU(3)_A$ and $G(2)$ and
heavy $\mathcal{E}$ bosons in the coset $E_6/(G(2) \times SU(3)_A)$. 

\subsection{Step 2: $G(2) \to SU(3)_C$ and the DM appearance}

As described in \cite{Masi2024}, the breaking $G(2) \xrightarrow{\langle\chi\rangle} SU(3)_C$ is induced by the VEV of $\chi\sim(\mathbf{7},\mathbf{1})$ under $G(2) \times SU(3)_A$.
Taking $\langle\chi\rangle=v_\chi \hat s_{7}$ along the $G(2)$--$\mathbf{7}$ singlet direction, being $v_\chi$ the Higgs VEV for the $G(2)$ breaking, the decomposition described in \cite{Masi2024} is recovered:
\begin{equation}
	\langle\chi\rangle=(0,0,0,0,0,0,\,v_\chi)^T\in\mathbf{7}_{G(2)}.
\end{equation}
Gauge bosons in $G(2)/SU(3)_C = S^6$ get masses $\sim g_{G(2)} v_\chi$, where $g_{G(2)}$ denotes the $G(2)$ gauge coupling. The generators transform under the 14-dimensional adjoint representation of $G(2)$ \citep{Holland2004,Pepe_2006}, which decomposes into \citep{Pepe_2007,Masi2021}
\begin{equation}
	\label{dec14}
	\mathbf{14} = \mathbf{8} \oplus \mathbf{3} \oplus \bar{\mathbf{3}}.
\end{equation}
A $G(2)$ gauge theory, w.r.t. $SU(3)$, has colors, anticolors, color--singlets and 14 generators, so that it is characterized by 14 gluons, 8 of them transforming as ordinary gluons (as an octuplet of $SU(3)$), while the other six additional $G(2)$ gauge bosons separate into $\mathbf{3}$ and $\bar{\mathbf{3}}$, keeping the color quarks/antiquarks quantum numbers, being still vector bosons. 

Another key group-theory fact used throughout is 
\begin{equation}
		\mathbf{7}\;\to\;\mathbf{3}\oplus\bar{\mathbf{3}}\oplus\mathbf{1}
	\qquad \text{under } SU(3)_C\subset G(2).
	\label{eq:7to3}
\end{equation}
This representation describes a $SU(3)$ fermionic color triplet $\mathbf{3}$ (a quark--like particle), a $SU(3)$ anti-fermion $\bar{\mathbf{3}}$ and a $SU(3)$ singlet $\mathbf{1}$. This color singlet can be interpreted as a lepton. So the previous fermionic representation that described fermions as a whole, without quark--lepton distinction, breaks into the usual picture of colored quark states and ``uncolored'' leptonic states.

Hence the multiplet $\psi_i\sim(\mathbf{7},\mathbf{3})$ becomes \cite{Babu_2023}
\begin{align}
	(\mathbf{7},\mathbf{3})
	&\to (\mathbf{3},\mathbf{3})\oplus(\bar{\mathbf{3}},\mathbf{3})\oplus(\mathbf{1},\mathbf{3})
	\qquad \text{under } SU(3)_C\times SU(3)_A,
	\label{eq:3x7to}
\end{align}
where the first entry is the $SU(3)_C$ color representation and the second entry is the $SU(3)_A$ electroweak ancestor representation. Meanwhile $\xi_i$ is untouched by color and $G(2)$ breaking.

At this point DM is separated from the other fields: the $G(2)$ gauge bosons are decomposed into 8 massless gluons of $SU(3)_C$ QCD and six \textit{broken} heavy vectors $\mathcal X$ in the coset $G(2)/SU(3)_C$ (transforming as $\mathbf{3}\oplus\bar{\mathbf{3}}$ under $SU(3)_C$), all degenerate with 
\begin{equation}
	\,m_{\chi}\big(G(2)/SU(3)_C\big)=g_{G(2)}v_\chi.
\end{equation}
As demonstrated in \cite{Masi2024}, due to its peculiar mathematical features, even if the fundamental $G(2)$ representation $\mathbf{7}$ is real, the group $G(2)$ manifests some complex features, which guarantee its correct use in physics, as its $\mathbf{3} \oplus \bar{\mathbf{3}}$ decompositions w.r.t. $SU(3)$ can acquire a complex structure. Consequently, quarks and the additional vector bosons in $G(2)/SU(3)_C = S^6$, i.e. the dark gluons of the theory, acquire a complex nature.

At this stage the unbroken gauge group is $SU(3)_C\times SU(3)_A$.  

\subsection{Step 3: $SU(3)_A \to SU(2)_L\times U(1)_A$ electroweak--like emergence}

The breaking $SU(3)_A \xrightarrow{\langle\Omega\rangle} SU(2)_L\times U(1)_A$ is induced by the adjoint-containing field
\begin{equation}
	\Omega\subset \Phi_{\mathbf{650}},\qquad \Omega\sim (\mathbf1,\mathbf8). 
\end{equation}
with a VEV aligned along the $SU(3)_A$ generator $t_{A}^8$ direction,
\begin{equation}
	\langle\Omega\rangle = v_\Omega\,t_{A}^8.
\end{equation}
The $SU(3)_A$ gauge bosons in the adjoint $\mathbf{8}$ split into 3 massless $SU(2)_L$ weak bosons and 1 massless $U(1)_Y$ boson and
the broken $8-(3+1)=4$ heavy vectors $Y$, which are two complex $SU(2)_L$ doublets all degenerate at first approximation. These gauge bosons in the coset $SU(3)_A/(SU(2)_L\times U(1)_A)$ have mass
	\begin{equation}
m_Y\big(SU(3)_A/(SU(2)\times U(1))\big) = g_A\,v_\Omega.
	\end{equation}
where $g_A$ is $SU(3)_A$ coupling constant. At this stage the gauge group is $SU(3)_C\times SU(2)_L\times U(1)_A$. Hypercharge should be identified as 
\begin{equation}
	Y \propto \beta\, A.
\end{equation}

\subsection{Ancestor charge inside $SU(3)_A$}
\label{sec:hypercharge}

A crucial consistency requirement of any unified framework is the correct reproduction of SM hypercharge \cite{Georgi1999,Slansky1981} assignments.
Here hypercharge does not arise as an independent $U(1)$ factor at the unification scale but it must be embedded non-trivially into the $SU(3)_A$ subgroup of $E_6$. For the definition of the hypercharge $Y$, identifying $A$ as proportional to a $t_A$ generator of $SU(3)_A$ (we take $SU(3)_A$ generators normalized by $\mbox{Tr}(t_A^{\alpha}t_A^{\beta})=\tfrac12\delta^{\alpha\beta}$), one can fix $\beta$ by requiring a particular component to have the SM value: with $\alpha$ fixed, every multiplet is determined. Then the exotics that do not match SM families must be removed, for example acquiring a large vectorlike masses via Higgs-mediated Yukawas, as discussed in Section 4.8.

The gauge group $SU(3)_A$ is broken at the scale $M_\Omega$ according to $SU(3)_A \;\longrightarrow\; SU(2)_L \times U(1)_A$, with the adjoint rep pattern:
\begin{equation}
	\mathbf{8}_A \;\to\;
	(\mathbf{1})_{0} \oplus (\mathbf{2})_{3/2} \oplus (\mathbf{2})_{-3/2}\oplus (\mathbf{3})_{0}.
\end{equation}
 The generator of charge $A$ can be identified with a Gell-Mann diagonal generator such as $t_{A}^8$ of $SU(3)_A$
\begin{equation}
	t_{A}^8 = \frac{1}{2\sqrt{3}}
	\begin{pmatrix}
		1 & 0 & 0\\
		0 & 1 & 0\\
		0 & 0 & -2
	\end{pmatrix},
	\qquad \mathrm{Tr}(t_{A8}^2)=\frac12,
\end{equation}
and the normalized charge as
\begin{equation}
	A = \sqrt{3}\,t_{A}^8.
	\label{eq:Ydef}
\end{equation}
We will see there is a fundamental obstruction in the construction of the hypercharge with only charge $A$.

\subsection{Identification of the Standard Model Higgs inside the adjoint \(\Omega\)}
\label{subsec:SM_Higgs_from_Omega}
In the effective \(SU(3)_A\) electroweak-ancestor sector, the breaking is driven by an adjoint scalar $\Omega\sim \mathbf8_A$: the doublet components $\mathbf2_{+3/2}\oplus\mathbf2_{-3/2}$ (where the subscript denotes the charge $A$ of $U(1)_A$) inside the decomposition are electroweak doublets after the intermediate breaking and the Standard Model Higgs can be identified with one light linear combination of
these doublets. If the hypercharge generator is normalized so that $Y=\frac13 A$, then the doublet with $A=\frac32$
has $Y(H)=\frac12$. Thus
\begin{equation}
H\subset \mathbf2_{+3/2}\subset\mathbf8_A
\end{equation}
has precisely the Standard Model Higgs quantum numbers,$H\sim(\mathbf1,\mathbf2)_{+1/2}$.
Equivalently, the conjugate doublet $\mathbf2_{-3}$ has $
Y=-\frac12$ and may be identified with \(H^\dagger\).
Concluding, the same adjoint scalar \(\Omega\) can play a
double role: its singlet component performs the high-scale breaking $SU(3)_A\to SU(2)_L\times U(1)_A$,
while its doublet component supplies the low-energy electroweak Higgs. At the level of the scalar potential this means that \(\Omega\) should be written as
\begin{equation}
\Omega
=
\Omega_0\,\mathbf1_0
+
\Omega_3\,\mathbf3_0
+
H\,\mathbf2_{+3/2}
+
H^\dagger\,\mathbf2_{-3/2}.
\end{equation}
After \(SU(3)_A\) breaking, the heavy states in \(\Omega_0\) and \(\Omega_3\)
are lifted to the scale \(v_A\), while one doublet direction is tuned or
protected to remain light:
\begin{equation}
m_H^2\ll v_A^2 .
\end{equation}
The low-energy scalar potential then reduces to the Standard Model form
\begin{equation}
V_{\rm EW}(H)
=
-m_H^2\,H^\dagger H
+
\lambda_H(H^\dagger H)^2,
\end{equation}
with
\begin{equation}
\langle H\rangle=\frac{1}{\sqrt2}
\begin{pmatrix}
	0\\
	v
\end{pmatrix},
\qquad
v\simeq246\ {\rm GeV}.
\end{equation}

The identification $H\subset\mathbf8_A$ therefore avoids introducing a separate electroweak Higgs representation at the effective \(SU(3)_A\) level. 

\subsection{Fermion decomposition}

The $(\mathbf{7},\mathbf{3})$
matter representation under $SU(3)_C\times SU(3)_A$ yields the following reps w.r.t. $SU(3)_C\times SU(2)_L \times U(1)_A$ (plus hermitian conjugates): 
\begin{align}
	(\mathbf{3}_C,\mathbf{3}_A) &\to (\mathbf{3},\mathbf{2})_{+1/2}\oplus(\mathbf{3},\mathbf{1})_{-1},
	\\
	(\bar{\mathbf{3}}_C,\mathbf{3}_A) &\to (\bar{\mathbf{3}},\mathbf{2})_{+1/2}\oplus(\bar{\mathbf{3}},\mathbf{1})_{-1},
	\\
	(\mathbf{1},\mathbf{3}_A) &\to (\mathbf{1},\mathbf{2})_{+1/2}\oplus(\mathbf{1},\mathbf{1})_{-1}. 
\end{align}
Similarly, the $(\mathbf{1},\bar{\mathbf{6}}_A)$ part yields, 
\begin{equation}
	(\mathbf{1},\bar{\mathbf{6}}_A) \to (\mathbf{1},\mathbf{3})_{-1}\oplus(\mathbf{1},\mathbf{2})_{+1/2}
	\oplus(\mathbf{1},\mathbf{1})_{+2}.
\end{equation}
Equations (36)--(39) should not be interpreted as the physical Standard Model spectrum. They display the decomposition obtained in the minimal pure-\(E_6\) attempt with \(Y\propto A\), and their incorrect
hypercharges are the first manifestation of the no-go result demonstrated below. The strict
\(E_6\to G(2)\times SU(3)_A\) chain by itself does not reproduce the observed SM fermion hypercharges.
\subsection{No-go result for a purely \(SU(3)_A\)-based hypercharge}
\label{subsec:nogo_SU3A_hypercharge}

The special \(E_6\) chain $E_6\to G(2)\times SU(3)_A$
is attractive because it isolates the \(G(2)\) dark sector and uses \(SU(3)_A\) as the ancestor of the electroweak gauge group. A natural first attempt was therefore to define hypercharge entirely inside \(SU(3)_A\), with
$Y=\beta A$, where \(A\) denotes the diagonal abelian generator left after $SU(3)_A\to SU(2)_L\times U(1)_A$.
In general, let an electroweak doublet $\mathbf{2}_{q_H}$ inside an \(SU(3)_A\) representation carry integer charge
\(q_H=2A\). The condition \(Y(H)=+1/2\) fixes
\begin{equation}
\beta=\frac{1}{q_H}.
\end{equation}
%
Therefore all Standard Model hypercharges require
\begin{equation}
q_A(f)=2q_H\,Y(f).
\end{equation}
In particular,
\begin{equation}
q_A(Q_L)=\frac{q_H}{3},\qquad
q_A(u^c)=-\frac{4q_H}{3},\qquad
q_A(d^c)=\frac{2q_H}{3},
q_A(L)=-q_H,\qquad
q_A(e^c)=2q_H.
\end{equation}
Since all \(q_A\) are integer weights, the quark sector already requires
\begin{equation}
q_H=3n,\qquad n\in\mathbb Z.
\end{equation}
For the adjoint Higgs, \(q_H=3\), this gives the special case
\begin{equation}
A(Q_L)=\frac12,\quad
A(u^c)=-2,\quad
A(d^c)=1,\quad
A(L)=-\frac32,\quad
A(e^c)=3.
\end{equation}
However, independently of this integrality condition, the gauge normalization
is fatal. Because \(A=\sqrt3 T^8_A\), the pure \(SU(3)_A\) matching gives
\begin{equation}
\frac{1}{g_Y^2}
=
\frac{3\beta^2}{g_A^2}
=
\frac{3}{q_H^2}\frac{1}{g_A^2},
\end{equation}
and hence the weak angle 
\begin{equation}
\sin^2\theta_W(M_\Omega)
=
\frac{q_H^2}{q_H^2+3}.
\end{equation}
Requiring the canonical GUT value \(3/8\) from normalized hypercharge \cite{GeorgiGlashow1974,langacker1981Unification}, gives
\begin{equation}
q_H^2=\frac95,
\end{equation}
which is incompatible with the integer \(SU(3)_A\) weight lattice. Therefore
no pure \(SU(3)_A\) hypercharge embedding can simultaneously reproduce the
Higgs normalization and the canonical weak angle.

The remaining fermion obstruction is then secondary but still important: for
each allowed integer \(q_H\), the \(SU(3)_A\) weights required for
\(Q_L,u^c,d^c,L,e^c\) are fixed. The low-dimensional special-embedding branches
do not provide a complete chiral Standard Model family with those weights while
also closing the required cubic Yukawa sector. This motivates adding an extra
abelian generator $U(1)_X$, so that
\begin{equation}
Y=\beta A+\gamma X.
\end{equation}


\section{The $E_7$ solution}
\label{subsec:E7_literature_precedents}

The \(E_7\) uplift represents the minimal solution to include the mandatory extra degrees of freedom to correctly reproduce the SM hypercharges, while keeping the symmetries breaking chain unchanged.
The relevant group-theoretic fact is the maximal embedding
\begin{equation}
	E_7\supset E_6\times U(1)_X ,
\end{equation}
under which the fundamental representation branches as
\begin{equation}
	\mathbf{56}
	\to
	\mathbf{27}_{-1}
	\oplus
	\overline{\mathbf{27}}_{+1}
	\oplus
	\mathbf1_{+3}
	\oplus
	\mathbf1_{-3}.
\end{equation}
Thus \(E_7\) naturally contains both the \(E_6\) gauge sector and the additional abelian
generator \(X\) needed in a flipped embedding. However, the same branching also shows why
a complete \(\mathbf{56}\) does not automatically give a chiral \(E_6\) family: it contains
\(\mathbf{27}\) together with its conjugate. The basic representation-theoretic data and
exceptional branching rules are standard and may be traced to the classic review of
Slansky and to modern special-GUT classifications \cite{Slansky1981,Yamatsu2018SpecialFamily}.

This chirality obstruction is the main reason why \(E_7\)-based constructions
usually invoke geometric, flux or orbifold mechanisms rather than ordinary four-dimensional
Higgsing alone. In M-theory compactifications on singular \(G(2)\)-holonomy spaces, chiral
fermions arise at singularities, and the local enhancement \(E_6\to E_7\) is precisely the
structure associated with matter in the \(\mathbf{27}\) of \(E_6\)
\cite{AcharyaWitten2002}. Related constructions of matter at \(G(2)\)-manifold singularities
were developed by Berglund and Brandhuber, emphasizing how chiral matter can arise from
geometric singularity structure rather than from a naive complete real or pseudo-real
four-dimensional representation \cite{BerglundBrandhuber2002}. These results provide
conceptual support for treating \(E_7\) as the natural geometric parent of chiral
\(E_6\)-like matter.

A second relevant class is orbifold and conifold GUT model building. Haba and Shimizu
studied five-dimensional \(E_6\), \(E_7\), and \(E_8\) orbifold GUTs in which Higgs fields
arise from higher-dimensional gauge components; in the \(E_7\) case they show that all
Yukawa interactions can be produced by introducing an adjoint matter hypermultiplet in the
bulk, with satisfactory charge quantization and anomaly properties on the four-dimensional
wall \cite{HabaShimizu2003E7Orbifold}. Hebecker and Ratz analyzed the group-theoretical
structure of orbifold and conifold GUTs, including examples where \(E_7\) is broken to
subgroups such as \(SU(5)\) and where SM matter can originate from the
higher-dimensional gauge sector \cite{HebeckerRatz2003}. These works are relevant to the
present discussion because they exhibit a standard mechanism for removing unwanted mirror
zero modes by boundary conditions or singularity projections.

The closest modern analogues to the \(E_7\) option considered here are the F-theory
constructions of Li and Taylor. In their \(E_7\) flux-breaking framework, an \(E_7\) gauge
factor is broken by fluxes to Standard Model-like gauge groups, and chiral matter spectra
with small generation number, including three-generation cases, can be obtained
\cite{LiTaylor2022PRD}. In their broader analysis of flux breaking in exceptional F-theory
GUTs, fluxes can break a geometric gauge group and induce chiral matter even when the
larger group itself does not admit ordinary chiral matter representations
\cite{LiTaylor2022JHEP}. Their subsequent phenomenological study of rigid \(E_7\)
F-theory vacua shows that three generations of Standard Model matter occur in many
examples and that dangerous dimension-four and dimension-five proton-decay operators can
be suppressed by approximate symmetries descending from the \(E_7\) structure and geometry
\cite{LiTaylor2024JHEP}. This is the closest published pattern to the present \(E_7\)
uplift in the sense that \(E_7\) supplies the high-scale exceptional structure while
flux/projection effects solve chirality.

There is also a family-unification literature based on exceptional cosets and residual
abelian symmetries. Mizoguchi and Yata studied family unification via quasi-Nambu--Goldstone
fermions in exceptional string-inspired settings, showing how exceptional cosets can produce
chiral family structures beyond those of a simple four-dimensional linear representation
\cite{MizoguchiYata2013}. Mizoguchi's F-theory family-unification construction uses
\(E_7\)-related geometry and orbifold projection to obtain family structures and residual
abelian charges \cite{Mizoguchi2014FTheoryFamily}. Sato further showed that an \(E_7\)
framework can unify a lepton-flavor \(U(1)\) symmetry with the Standard Model gauge group,
illustrating how residual abelian factors descending from \(E_7\) can have physical
meaning after symmetry breaking \cite{Sato2022E7LmuLtau}. These works are not identical to
the present model, but they support the general idea that \(E_7\) can naturally generate
useful abelian directions and family-dependent structures.

The \(E_7\) literature therefore motivates
the uplift and supplies known solutions to the chiral-mirror problem, but the special
\(G(2)\times SU(3)_A\) dark sector application remains a distinct feature of the present framework.

\subsection{Effective flux/projection rule for split chiral matter}
\label{subsec:flux_projection_rule}

A split-family construction can be formulated in a way that parallels
orbifold and flux-breaking exceptional GUTs \cite{HabaShimizu2003E7Orbifold,HebeckerRatz2003,
	LiTaylor2022PRD,LiTaylor2022JHEP,LiTaylor2024JHEP,
	Mizoguchi2014FTheoryFamily}. The ultraviolet theory contains
complete exceptional multiplets, but the light chiral spectrum is selected by
an index or projection rule. Let a UV representation \(\mathcal R\) decompose
under the special subgroup as
\begin{equation}
\mathcal R
\to
\bigoplus_a R_a,
\end{equation}
where \(R_a\) denotes a component with definite
\(G(2)\times SU(3)_A\) quantum numbers and, in the \(E_7\)-uplifted case, also
a definite \(U(1)_X\) charge. We assign to each component an integer chiral
index
\begin{equation}
\mathcal I(R_a)
=
n_L(R_a)-n_R(R_a).
\end{equation}
In a geometric realization this index is computed by a flux integral;
in a four-dimensional effective realization, \(\mathcal I(R_a)\) is treated as
a survival index encoding boundary conditions, flux or a UV projection \cite{HebeckerRatz2003,LiTaylor2022JHEP,LiTaylor2024JHEP}.

A selection rule might be
\begin{equation}
\mathcal I(R_a)=3
\end{equation}
for the branches identified with the three Standard Model families and
\begin{equation}
\mathcal I(R_a)=0
\end{equation}
for all unwanted exotic branches. This should be understood as a chiral-index
assignment, not as a group-theoretic consequence of \(E_6\) or \(E_7\)
alone. In the strict split-\(E_6\) construction with
\begin{equation}
Y=\frac13 A,
\end{equation}
one imposes schematically
\begin{equation}
\mathcal I(Q_L)=\mathcal I(u^c)=\mathcal I(d^c)=3,\qquad
\mathcal I(L)=\mathcal I(e^c)=3,
\end{equation}
while
\begin{equation}
\mathcal I(R_{\rm exotic})=0.
\end{equation}
Thus the split-family spectrum is not interpreted as an arbitrary incomplete
multiplet, but as a projected chiral spectrum descending from complete
exceptional representations.
The minimal chiral-selection rule applied for example to the $E_7$ fundamental representation is
\begin{equation}
\mathcal I(\mathbf{27}_{-1})=3,
\qquad
\mathcal I(\overline{\mathbf{27}}_{+1})=0,
\qquad
\mathcal I(\mathbf1_{\pm3})=0,
\end{equation}
or, more generally, a split rule in which different \(E_6\)-descendant sectors
carry different \(X\)-charges. With $Y=\beta A+\gamma X$,
a component \(R_a\) with charges \((A_a,X_a)\) has
\begin{equation}
Y_a=\beta A_a+\gamma X_a.
\end{equation}
The hypercharge-compatible selection rule may therefore be written as
\begin{equation}
\mathcal I(R_a)=3
\quad
\text{if}
\quad
\beta A_a+\gamma X_a=Y_{\rm SM}(R_a),
\end{equation}
and $\mathcal I(R_a)=0$ otherwise. Before the extra
abelian factor is broken, one must require
\begin{equation}
\sum_a \mathcal I(R_a)X_aT(R_a)=0,\qquad
\sum_a \mathcal I(R_a)X_a=0,\qquad
\sum_a \mathcal I(R_a)X_a^3=0,
\end{equation}
or else include spectator fields, Green--Schwarz/Wess--Zumino terms or mirror
states that remain active until \(U(1)_X\) breaking \cite{LiTaylor2022JHEP,LiTaylor2024JHEP, Bertlmann1996Anomalies} by the additional scalar $\Theta$. After
\begin{equation}
U(1)_A\times U(1)_X\to U(1)_Y,
\end{equation}
the remaining Standard Model spectrum is anomaly-free in the usual way. Concluding, branches with $\mathcal I(R_a)=0$ are vectorlike or mirror-completed and can be lifted at high scale, receiving a vectorlike mass, as described below. An explicit geometrical computation of the operator $\mathcal I(R_a)$ is beyond the scope of this paper and is deferred to a future work.

\subsection{Vectorlike completion: principle and minimal implementation}

The use of vectorlike fermions to remove unwanted exotic states is not a novel or ad hoc device,
but rather a long--established tool in grand unified model building. Its origins date back to the earliest attempts at unifying the strong and electroweak interactions: in the original $SU(5)$ and $SO(10)$ GUTs it was already understood that
the decomposition of large unified representations generically produces states that do not match
the observed SM fermion spectrum \cite{Slansky1981,Raby2006GUTReview}.
The resolution is that such states
may pair into vectorlike representations (a fermion sector is called vectorlike if for every representation $R$ there exists its conjugate $R^*$, unlike a chiral one) and acquire masses at the unification scale, thereby decoupling from low-energy physics \cite{Appelquist}.
Heavy vectorlike fermions were therefore regarded as an inevitable and phenomenologically benign
byproduct of unification.

Already in the 1970s it was observed that its fundamental chiral representation $\mathbf{27}$
contains precisely the field content of one SM family together with additional states
that naturally form vectorlike pairs \cite{GurseyRamondSikivie1976,AchimanStech1978,Slansky1981,RobinettRosner1982,HewettRizzo1989PhysRept}.
The $\mathbf{27}$ is designed to allow mass terms for its non--SM components once suitable Higgs fields acquire VEVs: vectorlike completion is not an auxiliary assumption in $E_6$ models,
but an intrinsic part of their group--theoretic structure.
Large Higgs representations in GUTs often contain gauge singlet directions and when
they acquire large VEVs, they can generate Dirac masses for vectorlike fermion pairs through renormalizable Yukawa couplings. Such a mechanism was repeatedly employed in early $E_6$ phenomenology, where Higgs representations such as the $\mathbf{78}$, $\mathbf{351}$ or $ \mathbf{650}$ were used both to break the unified
gauge symmetry and to lift exotic fermions \cite{RobinettRosner1982,HewettRizzo1989PhysRept}: the capability of a single Higgs field of simultaneously select a symmetry--breaking direction
and remove unwanted matter states is part of the $E_6$ GUTs.
Moreover, because vectorlike pairs are anomaly--free by themselves \cite{Bertlmann1996Anomalies}, their decoupling leaves the anomaly structure of the low--energy theory unchanged. This ensures that the chiral structure
of the SM can be preserved exactly while all exotic states are lifted to the unification scale.
This property distinguishes vectorlike completion from alternative approaches, such as imposing
ad hoc discrete symmetries or fine--tuned Yukawa textures and explains why it has remained a preferred
solution. 


\paragraph{Vectorlike mass generation.} When decomposed to the SM, the fermionic representation of $E_7$ and $E_6$ contains more fermions with wrong quantum features that make them exotics, becoming harmless if they are vectorlike under the low energy SM gauge group. If a pair of exotic fermions acquire a gauge--invariant mass, 
$\mathcal{L} \supset M\,\overline{\Psi}_L \Psi_R + \text{h.c.}$,
it does not contribute to anomalies, can be arbitrarily heavy and decouple from low energy physics. So the standard and minimal way is to keep all these exotics near the high scale by pairing them vectorlike using singlet VEVs. We seek to remove from the low-energy spectrum all fermions that are not identified with SM chiral fields by pairing them into heavy Dirac fermions.
Concretely, we identify in the fermion representations a set of conjugate multiplets of exotics required to pair the exotic SM multiplets. We denote with $E$ an exotic chiral multiplet from a high fermionic rep
and with $E^c$ its conjugate with opposite SM quantum numbers.

A high-scale gauge-singlet scalar $S$ with a large vev $v_S$ generates vectorlike masses through
Yukawa terms 
\begin{equation}
	\mathcal L \supset y_{\rm ex}\, E\,E^c\,S + \mathrm{h.c.}
	\qquad\Rightarrow\qquad
	M_{\rm ex}=y_{\rm ex}\,\langle S\rangle.
	\label{eq:vectorlikeMass}
\end{equation}
Provided $\langle S\rangle=v_S\gg v_{EW}$ and $y_{\rm ex}=\mathcal O(0.1\text{--}1)$, all exotic states
decouple far above the electroweak scale, leaving only the SM spectrum. 
The singlet scalar $S$ used for primary vectorlike completion can naturally be identified with the same field that breaks $E_6$, i.e. with a singlet direction inside the $ \mathbf{650}$ Higgs: $\langle \Phi_{ \mathbf{650}}^{(\mathbf 1,\mathbf 1)} \rangle = v_{ \mathbf{650}}=v_{E_6}$. This is not only allowed, it is in fact the most economical and theoretically clean choice: the same field both selects the special embedding $E_6\to G(2) \times SU(3)_A$ and lifts the bulk of the exotic fermions, preserving minimality and predictive power \cite{Yamatsu2015,Slansky1981}. Subsequent singlet scalars represented by $\chi$, $\Omega$ and $\Theta$ Higgs can partecipate into the vectorlike completion at several stages.

So the net effect is that all exotics become heavy Dirac fermions and are absent from the low-energy theory.
Diagonalization of the full fermion mass matrices could induce small mixing between the light and heavy sectors. If $m_{\rm EW}$ denotes an electroweak mass insertion scale connecting light and heavy states, a typical mixing angle scales as \cite{WeinbergQTF2}
\begin{equation}
	\vartheta \sim \frac{m_{\rm EW}}{M_{\rm ex}} \sim \frac{m_{\rm EW}}{y_{\rm ex}\,v_S}.
	\label{eq:mixingScaling}
\end{equation}
This suppression simultaneously controls loop-level flavor-changing neutral currents (FCNC) effects and baryon-number violation mediated by heavy vectors, consistent with the lepto--quark mixing angle
$\vartheta_{\rm LQ}^{-4}$ suppression we will derive for proton decay \cite{Weinberg1979Bviol,WilczekZee1979,SuperK2017,PDG2024}.
Below the breaking scales, introducing conjugate partners for exotics renders the heavy sector vectorlike under the SM gauge group: after integrating out the heavy Dirac fermions, the remaining low-energy spectrum can be
chosen to coincide with the SM (plus optional singlets for neutrinos, as discussed in the leptogenesis subsection). Hence anomaly cancellation at low energies is automatic.

\section{Minimal \(E_7\)-uplifted split-fermion completion}
\label{sec:minimal_E7_uplift}

The purpose of the \(E_7\) uplift is not to replace the special \(E_6\) core of
the model but to supply the additional abelian generator needed to repair the
hypercharge and weak-angle calibration. The gauge chain is
\begin{equation}
	E_7\longrightarrow E_6\times U(1)_X,
\end{equation}
followed by the original special breaking
\begin{equation}
	E_6\longrightarrow G(2)\times SU(3)_A.
\end{equation}
The subsequent breakings are
\begin{equation}
G(2)\to SU(3)_C,
	\qquad
		SU(3)_A \times U(1)_X \to SU(2)_L\times U(1)_Y.
\end{equation}
The \(E_7\to E_6\times U(1)_X\) step is regular and maximal rank, whereas the
\(E_6\to G(2)\times SU(3)_A\) step remains the distinctive special embedding
of the construction.
%

\paragraph{Fundamental representation.}
The smallest non-trivial \(E_7\) representation is $\mathbf{56}_{E_7}$. As anticipated, under \(E_6\times U(1)_X\) it decomposes as
\begin{equation}
	\mathbf{56}
	\to
	\mathbf{27}_{-1}
	\oplus
	\overline{\mathbf{27}}_{+1}
	\oplus
	\mathbf1_{+3}
	\oplus
	\mathbf1_{-3}.
\end{equation}
Thus the \(E_7\) fundamental naturally contains the familiar \(E_6\) family representation \(\mathbf{27}\), but it also contains the mirror \(\overline{\mathbf{27}}\). Therefore a complete \(\mathbf{56}\) is not automatically a chiral \(E_6\) family. A chiral-selection or projection mechanism is needed, as described before:
\begin{equation}
	\mathbf{56}_{F}
	\longrightarrow
	\mathbf{27}_{-1}^{\rm light}
	\oplus
	\left[
	\overline{\mathbf{27}}_{+1}
	\oplus
	\mathbf1_{+3}
	\oplus
	\mathbf1_{-3}
	\right]_{\rm heavy/projected}.
\end{equation}

\paragraph{Adjoint representation.}
The adjoint of \(E_7\) is $\mathbf{133}_{E_7}$.
It branches under \(E_6\times U(1)_X\) as
\begin{equation}
	\mathbf{133}
	\to
	\mathbf{78}_{0}
	\oplus
	\mathbf1_{0}
	\oplus
	\mathbf{27}_{+2}
	\oplus
	\overline{\mathbf{27}}_{-2}.
\end{equation}
The \(\mathbf{78}_0\) is the adjoint gauge sector of \(E_6\) described before, while \(\mathbf1_0\) is the \(U(1)_X\) gauge boson. The components
\begin{equation}
	\mathbf{27}_{+2}\oplus\overline{\mathbf{27}}_{-2}
\end{equation}
are the heavy vectors associated with the broken coset
\begin{equation}
	E_7/(E_6\times U(1)_X).
\end{equation}
The breaking \(E_7\to E_6\times U(1)_X\) does not break \(U(1)_X\) itself. It leaves \(U(1)_X\) as an unbroken gauge factor. The massive vectors at this scale are only the coset vectors \(27_{+2}\oplus\overline{27}_{-2}\), not the abelian \(U(1)_X\) gauge boson. Therefore, below \(M_{E_7}\), the gauge group  contains a genuine local \(U(1)_X\) until the \(\Theta\) field acquires a
VEV.

\paragraph{\(E_7\)-breaking Higgs and the final breaking scales chain}
The same adjoint representation can be used as the first breaking Higgs:
\begin{equation}
	\Phi_{133}\sim\mathbf{133}_H.
\end{equation}
Since
\begin{equation}
	\mathbf{133}_H\supset\mathbf1_0,
\end{equation}
a vacuum expectation value along this singlet direction,
\begin{equation}
	\langle\Phi_{133}\rangle\subset\mathbf1_0,
\end{equation}
breaks
\begin{equation}
	E_7\to E_6\times U(1)_X.
\end{equation}
The broken vectors \(\mathbf{27}_{+2}\oplus\overline{\mathbf{27}}_{-2}\) acquire masses of order
\begin{equation}
	M_{7/6}\sim g_7 v_7.
\end{equation}

For what concerns the additional $\Theta$ Higgs associated to the final abelian breaking $U(1)_A \times U(1)_X$ , its phase is eaten by the massive gauge boson of the abelian combination orthogonal to hypercharge, originating a $Z'$ massive boson. 

The resulting final breaking chain scheme with the fundamental mass scales is summarized in the following Table.

\begin{table}[H]
	\centering
	\small
	\renewcommand{\arraystretch}{1.25}
	\begin{tabularx}{\textwidth}{c|c|X|X}
		\hline
		Scale & Breaking & Bosonic field / branch & Comments \\
		\hline
		\(M_{E_7}\) &
		\(E_7\to E_6\times U(1)_X\) &
		\(\Sigma_{E_7}\subset\mathbf{133}_H\), direction \(\mathbf1_0\) &
		Regular breaking; heavy \(\mathbf{27}_{+2}\oplus\overline{\mathbf{27}}_{-2}\) vectors decouple. \\
		\hline
		\(M_{E_6}\) &
		\(E_6\to G(2)\times SU(3)_A\) &
		High-scale component of \(\Phi_{650}\) or equivalent \(E_6\)-breaking scalar &
		Sets the special \(E_6\) completion and vectorlike lifting scale. \\
		\hline
		\(M_\chi\) &
		\(G(2)\to SU(3)_C\) &
		\(\chi\subset(\mathbf7,\mathbf1_A)_0\subset\mathbf{650}_{0,H}\) &
		Produces the \(G(2)/SU(3)_C\) heavy vector sector. \\
		\hline
		\(M_\Omega\) &
		\(SU(3)_A\to SU(2)_L\times U(1)_A\) &
		\(\Omega\subset(\mathbf1,\mathbf8_A)_0\subset\mathbf{650}_{0,H}\) &
		Fixes the electroweak ancestor scale. \\
		\hline
		\(M_\Theta\simeq M_\Omega\) &
		\(U(1)_A\times U(1)_X\to U(1)_Y\) &
		$\Theta$, with \(Y(\Theta)=0\) &
		Gives mass to the orthogonal \(Z'\)-like boson and removes the anomalous split-\(U(1)_X\) spectrum. \\
		\hline
		\(v_{\rm EW}\) &
		\(SU(2)_L\times U(1)_Y\to U(1)_{\rm em}\) &
		\(H\subset\mathbf{650}_{0,H}\) &
		Single light SM Higgs doublet; all other scalar components are heavy. \\
		\hline
	\end{tabularx}
	\caption{Bosonic breaking chain in the corrected \(E_7\)-uplifted model.}
	\label{tab:E7_bosonic_breaking_chain}
\end{table}

%
%


\paragraph{Canonical normalization of \(U(1)_X\), hypercharge and weak-angle matching}
In the integer charge convention used before,
\begin{equation}
	\mathrm{Tr}_{\mathbf{56}}X^2
	=
	27(1)^2+27(1)^2+1(3)^2+1(3)^2
	=
	72.
\end{equation}
With the conventional \(E_7\) Dynkin index
\begin{equation}
	T_{E_7}(\mathbf{56})=6,
\end{equation}
the canonically normalized abelian generator is
\begin{equation}
	T_X=\frac{X}{\sqrt{12}}.
\end{equation}
Thus the abelian normalization factor entering the hypercharge matching is
\begin{equation}
	N_X=12.
\end{equation}
The viable hypercharge embedding is now
\begin{equation}
	Y=\frac13(A+X).
\end{equation}
Equivalently,
\begin{equation}
	Y=\beta A+\gamma X,
	\qquad
	\beta=\gamma=\frac13.
\end{equation}
The matching relation becomes
\begin{equation}
	\frac{1}{g_Y^2}
	=
	\frac{3\beta^2}{g_A^2}
	+
	\frac{N_X\gamma^2}{g_X^2}.
\end{equation}
For $g_X(M_\Omega)\simeq g_A(M_\Omega)$, \(N_X=12\), and \(\beta=\gamma=1/3\),
\begin{equation}
	\frac{g_A^2}{g_Y^2}
	=
	\frac13+\frac{12}{9}
	=
	\frac53.
\end{equation}
Since \(g_2=g_A\), this gives the canonical high-scale condition
\begin{equation}
	\sin^2\theta_W(M_\Omega)=\frac38.
\end{equation}
which is a GUT-scale value. It is not meant to be the low-energy value at
\(M_Z\); the running from \(M_\Omega\) down to \(M_Z\) must reproduce the observed
electroweak value.

\subsection{Fermionic reps scan: fermion multiplets, SM Higgs and $\Theta$ sector}
The purpose of this section is to close the representation-theoretic part of the construction.  We ask whether the special
\begin{equation}
E_7\supset E_6\times U(1)_X,\qquad
E_6\supset G(2)\times SU(3)_A
\end{equation}
embedding can support a realistic electroweak Higgs sector and renormalizable SM Yukawa couplings while keeping the electroweak Higgs and $\Theta$ secluded from the dark \(G(2)\) sector.
To find a novel solution for the fermions reps (and $\Theta$) inside $E_6$ that can be included in the $E_7$ framework, solving the hypercharge (and weak angle) assignments, we performed an scan with
\begin{equation}
R_f,\Theta\in
\mathbf{27,\overline{27},351,\overline{351},351',\overline{351'},1728,\overline{1728},2430,2925}
\end{equation}
up to the largest of the six fundamental-weight
representations of \(E_6\) $\mathbf{2925}$ \cite{Yamatsu2015,Slansky1981,LieArt},
allowing one Standard Model Higgs doublet inside the \(SU(3)_A\) components of the scalar \(\mathbf{650}_{0,H}\), \textit{i.e.} up to $\mathbf{27}$. The allowed \(SU(3)_A\) Higgs representations included
\begin{equation}
\mathbf{6,\overline6,8,10,\overline{10},15,\overline{15},15',\overline{15'},
21,\overline{21},24,\overline{24},27}.
\end{equation}
The scan allows real \(E_6\) parent representations such as
\(\mathbf{2925}\), but this should not be interpreted as
placing complete light chiral families in real \(E_6\) multiplets.  A real \(E_6\) representation is vectorlike from the viewpoint of the unbroken
exceptional group and can admit invariant mass terms.  Its use is therefore consistent only in the projected-spectrum sense: the real parent multiplet
contains complex SM branches after
\(E_6\to G(2)\times SU(3)_A\) and the unwanted conjugate components are assumed to be removed or paired vectorlike by the high-scale projection and
lifting sector.  If one demands an ordinary four-dimensional chiral \(E_6\) theory without such projection, the scan should be restricted to
complex \(E_6\) representations.

The scan imposed the SM hypercharge values
\begin{equation}
Y(Q_L)=\frac16,\qquad
Y(u^c)=-\frac23,\qquad
Y(d^c)=+\frac13,
\end{equation}
\begin{equation}
Y(L)=-\frac12,\qquad
Y(e^c)=+1,\qquad
Y(H)=+\frac12,\qquad
Y(\Theta)=0,
\end{equation}
together with
\begin{equation}
A=\sqrt3\,t_A^8,\qquad q_A=2A,\qquad
Y={1\over 3}(A+X).
\end{equation}
With the normalization \(N_X=12\), this embedding satisfies the canonical condition
\begin{equation}
3\beta^2+N_X\gamma^2={5\over 3},\qquad
\beta=\gamma={1\over 3}.
\end{equation}
The \(SU(3)_A\) branchings used repeatedly below are
\begin{align}
	\mathbf 8_A &\to \mathbf 1_0\oplus \mathbf 2_{+3}\oplus \mathbf 2_{-3}\oplus \mathbf 3_0,\\
	\mathbf {10}_A &\to \mathbf 1_{-6}\oplus \mathbf 2_{-3}\oplus \mathbf 3_0\oplus \mathbf 4_{+3},\\
	\mathbf {27}_A &\to
	\mathbf1_0\oplus \mathbf2_{+3}\oplus\mathbf2_{-3}\oplus
	\mathbf3_{+6}\oplus\mathbf3_0\oplus\mathbf3_{-6}
	\oplus\mathbf4_{+3}\oplus\mathbf4_{-3}\oplus\mathbf5_0 .
\end{align}
The \(G(2)\) branches needed to select color states are
\begin{align}
	\mathbf 7_{G(2)}&\to \mathbf1_c\oplus\mathbf3_c\oplus\overline{\mathbf3}_c,\\
	\mathbf {14}_{G(2)}&\to \mathbf8_c\oplus\mathbf3_c\oplus\overline{\mathbf3}_c,\\
	\mathbf {27}_{G(2)}&\to \mathbf1_c\oplus\mathbf3_c\oplus\overline{\mathbf3}_c
	\oplus\mathbf6_c\oplus\overline{\mathbf6}_c\oplus\mathbf8_c .
\end{align}
A candidate assignment is accepted only if it passes the following filters.
First, each field must contain the desired Standard Model component after the two subgroup breakings
\begin{equation}
G(2)\to SU(3)_c,\qquad
SU(3)_A\to SU(2)_L\times U(1)_A,
\end{equation}
with the hypercharge \(Y=(A+X)/3\).

Second, every projected Yukawa channel must close under the unbroken intermediate group
\begin{equation}
G(2)\times SU(3)_A\times U(1)_X.
\end{equation}
For a Higgs component \(H\), the four projected tests are
\begin{align}
	Q u^c H &: \quad
	G_Q\otimes G_u\otimes G_H\supset\mathbf1,\qquad
	R_Q\otimes R_u\otimes R_H\supset\mathbf1,\qquad
	X_Q+X_u+X_H=0,\\
	Q d^c H^\dagger &: \quad
	G_Q\otimes G_d\otimes G_H\supset\mathbf1,\qquad
	R_Q\otimes R_d\otimes R_H\supset\mathbf1,\qquad
	X_Q+X_d-X_H=0,\\
	L e^c H^\dagger &: \quad
	G_L\otimes G_e\otimes G_H\supset\mathbf1,\qquad
	R_L\otimes R_e\otimes R_H\supset\mathbf1,\qquad
	X_L+X_e-X_H=0,\\
	L N^c H &: \quad
	G_L\otimes G_N\otimes G_H\supset\mathbf1,\qquad
	R_L\otimes R_N\otimes R_H\supset\mathbf1,\qquad
	X_L+X_N+X_H=0.
\end{align}
Here \(G_i\) denotes the \(G(2)\) representation and \(R_i\) the \(SU(3)_A\) representation of the selected branch.  In the Mathematica implementation this projected test is performed by checking whether
\begin{equation}
G_i\otimes G_j\supset \overline{G_H},\qquad
R_i\otimes R_j\supset \overline{R_H}.
\end{equation}
Since the Higgs branches considered below are \(G(2)\)-singlets and the \(SU(3)_A\) representations \(\mathbf8_A\) and \(\mathbf{27}_A\) are self-conjugate, this reduces to the transparent condition that the two fermion branches contain the Higgs branch in their product.

Third, a positive projected Yukawa is not enough.  We also demand the existence of a parent \(E_6\)-level cubic invariant,
\begin{align}
	R_Q^{E_6}\otimes R_u^{E_6}\otimes \mathbf{650}_H &\supset \mathbf1_{E_6},\\
	R_Q^{E_6}\otimes R_d^{E_6}\otimes \mathbf{650}_H &\supset \mathbf1_{E_6},\\
	R_L^{E_6}\otimes R_e^{E_6}\otimes \mathbf{650}_H &\supset \mathbf1_{E_6},\\
	R_L^{E_6}\otimes R_N^{E_6}\otimes \mathbf{650}_H &\supset \mathbf1_{E_6}.
\end{align}
These tests are performed directly with LieART by decomposing the triple products.  Thus a positive entry in the final scan means that the channel is allowed both by the projected subgroup tensor algebra and by the parent \(E_6\) representation product.  We have not yet computed explicit numerical Clebsch--Gordan coefficients for each selected component; this remains a next-stage calculation.

Fourth, the branch economy is controlled.  The final scans impose
\begin{equation}
n_{\rm visible\ fermion\ branches}\leq 4,\qquad
n_{\rm visible+\Theta\ branches}\leq 5.
\end{equation}
The sterile \(N^c\) is allowed to be scanned independently in the final run.  This is physically natural because \(N^c\) is a SM singlet, and its origin need not coincide with the visible chiral-family economy.

Finally, a solution was accepted as \(E_7\)-certified only if all required fermion and \(\Theta\) branches appeared inside at least one of the explicit \(E_7\to E_6\times U(1)_X\) decompositions.


\subsection{Scalar choices and the four solution classes}

We explored four qualitatively distinct Higgs completions.  The result is summarized in Table~\ref{tab:HiggsLandscape}.  The main lesson is that a fully \(G(2)\)-singlet Higgs completion is possible, but the successful branch is the adjoint branch
\begin{equation}
H\sim(\mathbf1_{G(2)},\mathbf8_A)_0\subset\mathbf{650}_{0,H},
\end{equation}
not the \((\mathbf1,\mathbf{27}_A)_0\) branch if one insists on all four renormalizable Dirac Yukawa channels at the parent \(E_6\) level. The final Higgs branch is always contained in $\mathbf{650}$, according to the aforementioned decomposition
\begin{equation}
	\mathbf{650}
	\to
	(\mathbf1,\mathbf1)\oplus
	(\mathbf7,\mathbf1)\oplus
	(\mathbf1,\mathbf8)\oplus
	2(\mathbf7,\mathbf8)\oplus
	(\mathbf7,\mathbf{10})\oplus
	(\mathbf7,\overline{\mathbf{10}})
	\oplus
	(\mathbf{14},\mathbf8)
	\oplus
	(\mathbf{27},\mathbf1)\oplus
	(\mathbf1,\mathbf{27})\oplus
	(\mathbf{27},\mathbf8).
\end{equation}

\begin{table}[H]
	\centering
	\footnotesize
	\renewcommand{\arraystretch}{1.22}
	\setlength{\tabcolsep}{3.5pt}
	\begin{tabularx}{\textwidth}{
			>{\centering\arraybackslash}p{0.055\textwidth}|
			>{\centering\arraybackslash}p{0.145\textwidth}|
			>{\raggedright\arraybackslash}X|
			>{\centering\arraybackslash}p{0.16\textwidth}|
			>{\raggedright\arraybackslash}p{0.235\textwidth}
		}
		\hline
		Class & Higgs branch & Representative fermion pattern & Yukawa status & Interpretation\\
		\hline
		A &
		\((\mathbf1,\mathbf8_A)_0\) &
		\(\overline{\mathbf{27}}_{+1}\oplus \mathbf{351}_{-1}\oplus
		\mathbf{2430}_{-3}\oplus\mathbf{2430}_{+3}\) &
		\(Qu,Qd,Le,LN\) allowed &
		Best low-lepton-rep benchmark; \(N^c\) scanned independently.\\
		\hline
		B &
		\((\mathbf1,\mathbf8_A)_0\) &
		\(\overline{\mathbf{351}}_{+1}\oplus \mathbf{351}_{-1}\oplus
		\mathbf{2925}_{-3}\oplus\mathbf{2925}_{+3}\) &
		\(Qu,Qd,Le,LN\) allowed &
		Best branch-economical benchmark; \(\Theta\) and sterile sector aligned.\\
		\hline
		C &
		\((\mathbf1,\mathbf{27}_A)_0\) &
		\(\mathbf{351/2925}\)-type visible fermions assignments &
		\(Qu,Qd,Le\) allowed, wrong \(LN\) &
		Visible-sector success, but sterile-neutrino Dirac obstruction.\\
		\hline
		D &
		\((\mathbf7,\mathbf8_A)_0\) &
		\(\overline{\mathbf{351'}}_{+1}\oplus\mathbf{351'}_{-1}
		\oplus\mathbf{2430}_{-3}\oplus\mathbf{2430}_{+3}\) &
		Projected-Yukawa complete &
		Economical old solution, but the Higgs is not \(G(2)\)-singlet.\\
		\hline
	\end{tabularx}
	\caption{Landscape of Higgs completions. Classes A and B are the preferred final branches: both use a \(G(2)\)-singlet Higgs in \((\mathbf1,\mathbf8_A)_0\subset\mathbf{650}_{0,H}\) and pass the full parent \(E_6\) cubic tests for all four Yukawa channels. Class C shows why the \((\mathbf1,\mathbf{27}_A)_0\) branch is insufficient if a renormalizable Dirac-neutrino Yukawa is required. Class D is useful as a comparison with the earlier economical assignment, but it does not seclude the Higgs from \(G(2)\).}
	\label{tab:HiggsLandscape}
\end{table}
The \(E_7\)-parent certification is not lost in the final \(G(2)\)-singlet
completion.  For Class A, the required branches
$\overline{\mathbf{27}}_{+1},
\mathbf{351}_{-1},
\mathbf{2430}_{-3},
\mathbf{2430}_{+3},
\mathbf{2925}_{+3}
$
are simultaneously contained in the scanned parent
\begin{equation}
\mathbf{86184}_{E_7}.
\end{equation}
For Class B the situation is even more economical.  The branches
$\overline{\mathbf{351}}_{+1},
\mathbf{351}_{-1},
\mathbf{2925}_{-3},
\mathbf{2925}_{+3}
$
are contained in
\begin{equation}
\mathbf{27664}_{E_7},
\end{equation}
which is the smallest common \(E_7\) parent in our scanned database for this
class.  Therefore Class B is preferred if one prioritizes macro-parent
economy, while Class A is preferred if one prioritizes the lower
\(\mathbf{2430}\) leptonic parent.
The Higgs parent is separate.  The final Higgs branch requires
\begin{equation}
H\sim(\mathbf1_{G(2)},\mathbf8_A)_0\subset\mathbf{650}_{0,H}.
\end{equation}
The matter parents \(\mathbf{27664}_{E_7}\) and \(\mathbf{86184}_{E_7}\)
contain \(\mathbf{650}_{\pm3}\), but not \(\mathbf{650}_0\).  A minimal scalar
parent containing \(\mathbf{650}_0\) is
\begin{equation}
\mathbf{1463}_{E_7}\supset\mathbf{650}_0.
\end{equation}
We therefore treat the electroweak-Higgs parent \(H_{650}\) and the
high-scale breaking field \(\Omega_{650}\) as scalar copies of
\(\mathbf{650}_0\) supplied by an \(E_7\) scalar parent such as
\(\mathbf{1463}_{E_7}\).

In classes A/B/C the SM Higgs and $\Theta$ branches are $G(2)$-singlets. In class D the electroweak Higgs may originate from a \(G(2)\)-charged parent branch and the light Higgs doublet is selected from the color-singlet component after $\mathbf7_{G(2)}\to\mathbf1_C\oplus\mathbf3_C\oplus\overline{\mathbf3}_C$. The colored partners are assumed to be lifted at the exceptional or \(G(2)\)-breaking threshold.

\subsection{Class A: \(\overline{\mathbf{27}}+\mathbf{351}+\mathbf{2430}\) low-lepton-representation, \(G(2)\)-singlet Higgs solution}

Class A has a \(\mathbf{2430}\) leptonic structure and the Higgs in a \(G(2)\)-singlet branch.  A representative assignment is shown in Table~\ref{tab:ClassA}.  The Higgs is embedded as
\begin{equation}
H_{650}\supset(\mathbf1_{G(2)},\mathbf8_A)_0
\supset(\mathbf1,\mathbf2)_{+1/2}\oplus(\mathbf1,\mathbf2)_{-1/2}.
\end{equation}
The secluded field is
\begin{equation}
\Theta\sim(\mathbf1_{G(2)},\mathbf{10}_A)_{+3}\subset\mathbf{2925}_{+3},
\end{equation}
with \(A=-3\), \(X=+3\), and hence \(Y=0\).

\begin{table}[H]
	\centering
	\small
	\renewcommand{\arraystretch}{1.22}
	\begin{tabular}{c|c|c|c|c|c}
		\hline
		Field & Origin & Representative \(G(2)\times SU(3)_A\) branch
		& SM component & \((A,X)\) & \(Y\)\\
		\hline
		\(Q_L\) & \(\overline{\mathbf{27}}_{+1}\) &
		\((\mathbf7,\overline{\mathbf3}_A)\) &
		\((\mathbf3,\mathbf2)_{1/6}\) &
		\(\left(-\frac12,+1\right)\) & \(\frac16\)\\
		\(u^c\) & \(\mathbf{351}_{-1}\) &
		\((\mathbf7,\mathbf3_A)\) &
		\((\overline{\mathbf3},\mathbf1)_{-2/3}\) &
		\((-1,-1)\) & \(-\frac23\)\\
		\(d^c\) & \(\mathbf{351}_{-1}\) &
		\((\mathbf7,\overline{\mathbf6}_A)\) &
		\((\overline{\mathbf3},\mathbf1)_{1/3}\) &
		\((+2,-1)\) & \(\frac13\)\\
		\(L\) & \(\mathbf{2430}_{-3}\) &
		\((\mathbf7,\mathbf8_A)\) &
		\((\mathbf1,\mathbf2)_{-1/2}\) &
		\(\left(+\frac32,-3\right)\) & \(-\frac12\)\\
		\(e^c\) & \(\mathbf{2430}_{+3}\) & \((\mathbf7,\mathbf8_A)\) &
		\((\mathbf1,\mathbf1)_{1}\) &
		\((0,+3)\) & \(1\)\\
		\(N^c\) & \(\mathbf{2430}_{+3}\) &
		\((\mathbf7,\mathbf{10}_A)\) &
		\((\mathbf1,\mathbf1)_{0}\) &
		\((-3,+3)\) & \(0\)\\
		\(\Theta\) & \(\mathbf{2925}_{+3}\) &
		\((\mathbf1,\mathbf{10}_A)\) &
		\((\mathbf1,\mathbf1)_{0}\) &
		\((-3,+3)\) & \(0\)\\
		\(H\) & \(\mathbf{650}_{0,H}\) &
		\((\mathbf1,\mathbf8_A)\) &
		\((\mathbf1,\mathbf2)_{1/2}\) &
		\(\left(\frac32,0\right)\) & \(\frac12\)\\
		\hline
	\end{tabular}
	\caption{Class A representative solution.  The Higgs and \(\Theta\) are both \(G(2)\)-singlet.  The visible lepton sector uses the lower \(\mathbf{2430}\) parent.  The sterile \(N^c\) is scanned independently from the visible-family branch economy, which is natural because it is a SM singlet.}
	\label{tab:ClassA}
\end{table}
The four Standard Model Yukawa ancestors are then represented by the
following contractions:
\begin{equation}
Q_Lu^cH:
\qquad
(\mathbf7,\overline{\mathbf3}_A)_{+1}
\otimes
(\mathbf7,\mathbf3_A)_{-1}
\otimes
(\mathbf1,\mathbf8_A)_0
\supset
(\mathbf1,\mathbf1_A)_0 ,
\end{equation}
because
\begin{equation}
\mathbf7\otimes\mathbf7\supset\mathbf1,
\qquad
\overline{\mathbf3}_A\otimes\mathbf3_A
\supset
\mathbf8_A,
\qquad
\mathbf8_A\otimes\mathbf8_A\supset\mathbf1_A.
\end{equation}
Equivalently, the parent \(E_6\) product contains $\overline{\mathbf{27}}\otimes\mathbf{351}
\supset
\mathbf{650}$.
For the down-type Yukawa one has
\begin{equation}
Q_Ld^cH^\dagger:
\qquad
(\mathbf7,\overline{\mathbf3}_A)_{+1}
\otimes
(\mathbf7,\overline{\mathbf6}_A)_{-1}
\otimes
(\mathbf1,\mathbf8_A)_0
\supset
(\mathbf1,\mathbf1_A)_0 ,
\end{equation}
with
\begin{equation}
\mathbf7\otimes\mathbf7\supset\mathbf1,
\qquad
\overline{\mathbf3}_A\otimes\overline{\mathbf6}_A
\supset
\mathbf8_A.
\end{equation}
The same parent channel
\begin{equation}
\overline{\mathbf{27}}\otimes\mathbf{351}
\supset
\mathbf{650}
\end{equation}
therefore supports both \(Q_Lu^cH\) and \(Q_Ld^cH^\dagger\), with the
distinction made by the projected electroweak component of
\(\mathbf{650}_{0,H}\).

For the charged-lepton Yukawa,
\begin{equation}
	Le^cH^\dagger:
	\qquad
	(\mathbf7,\mathbf8_A)_{-3}
	\otimes
	(\mathbf7,\mathbf8_A)_{+3}
	\otimes
	(\mathbf1,\mathbf8_A)_0
	\supset
	(\mathbf1,\mathbf1_A)_0 ,
\end{equation}
with $\mathbf7\otimes\mathbf7\supset\mathbf1$ for $G(2)$, while
\begin{equation}
\mathbf8_A\otimes\mathbf8_A\otimes\mathbf8_A
\supset
\mathbf1_A .
\end{equation}
At the parent level the required invariant is certified by
\begin{equation}
\mathbf{2430}\otimes\mathbf{2430}
\supset
\mathbf{650}.
\end{equation}
Similarly, the Dirac-neutrino Yukawa is
\begin{equation}
	LN^cH:
	\qquad
	(\mathbf7,\mathbf8_A)_{-3}
	\otimes
	(\mathbf7,\mathbf{10}_A)_{+3}
	\otimes
	(\mathbf1,\mathbf8_A)_0
	\supset
	(\mathbf1,\mathbf1_A)_0 ,
\end{equation}
again after selecting the color-singlet component of the \(G(2)\) branch
and the electroweak doublet/singlet components inside the
\(SU(3)_A\) representations.

\subsection{Class B: maximally economical \(\mathbf{351+2925}\) branch}

Class B is at least as important as Class A.  It uses fewer distinct parent families and aligns the sterile-neutrino sector with the secluded \(\Theta\) branch.  A representative assignment is shown in Table~\ref{tab:ClassB}.

\begin{table}[H]
	\centering
	\small
	\renewcommand{\arraystretch}{1.22}
	\begin{tabular}{c|c|c|c|c|c}
		\hline
		Field & Origin & Representative \(G(2)\times SU(3)_A\) branch
		& SM component & \((A,X)\) & \(Y\)\\
		\hline
		\(Q_L\) & \(\overline{\mathbf{351}}_{+1}\) &
		 \((\mathbf7,\overline{\mathbf3}_A)\) &
		\((\mathbf3,\mathbf2)_{1/6}\) &
		\(\left(-\frac12,+1\right)\) & \(\frac16\)\\
		\(u^c\) & \(\mathbf{351}_{-1}\) &
		\((\mathbf7,\mathbf{15}_A)\) or \((\mathbf7,\mathbf3_A)\) &
		\((\overline{\mathbf3},\mathbf1)_{-2/3}\) &
		\((-1,-1)\) & \(-\frac23\)\\
		\(d^c\) & \(\mathbf{351}_{-1}\) &
		\((\mathbf7,\overline{\mathbf6}_A)\) &
		\((\overline{\mathbf3},\mathbf1)_{1/3}\) &
		\((+2,-1)\) & \(\frac13\)\\
		\(L\) & \(\mathbf{2925}_{-3}\) &
		\((\mathbf1,\mathbf8_A)\) &
		\((\mathbf1,\mathbf2)_{-1/2}\) &
		\(\left(+\frac32,-3\right)\) & \(-\frac12\)\\
		\(e^c\) & \(\mathbf{2925}_{+3}\) &
		\((\mathbf1,\mathbf8_A)\) &
		\((\mathbf1,\mathbf1)_{1}\) &
		\((0,+3)\) & \(1\)\\
		\(N^c\) & \(\mathbf{2925}_{+3}\) &
		\((\mathbf1,\mathbf{10}_A)\) &
		\((\mathbf1,\mathbf1)_{0}\) &
		\((-3,+3)\) & \(0\)\\
		\(\Theta\) & \(\mathbf{2925}_{+3}\) &
		\((\mathbf1,\mathbf{10}_A)\) &
		\((\mathbf1,\mathbf1)_{0}\) &
		\((-3,+3)\) & \(0\)\\
		\(H\) & \(\mathbf{650}_{0,H}\) &
		\((\mathbf1,\mathbf8_A)\) &
		\((\mathbf1,\mathbf2)_{1/2}\) &
		\(\left(\frac32,0\right)\) & \(\frac12\)\\
		\hline
	\end{tabular}
	\caption{Class B representative solution.  This branch is especially economical because the visible fields, sterile state, and \(\Theta\) can be arranged using \(\overline{\mathbf{351}}_{+1}\), \(\mathbf{351}_{-1}\), \(\mathbf{2925}_{-3}\), and \(\mathbf{2925}_{+3}\).  It is one of the two preferred final classes.}
	\label{tab:ClassB}
\end{table}
The quark Yukawas are guaranteed by the parent product
\begin{equation}
\overline{\mathbf{351}}\otimes\mathbf{351}
\supset
\mathbf{650},
\end{equation}
and the projected \(SU(3)_A\) contractions include
\begin{equation}
\mathbf6_A\otimes\overline{\mathbf6}_A\supset\mathbf8_A,
\qquad
\overline{\mathbf3}_A\otimes\mathbf{15}_A\supset\mathbf8_A,
\qquad
\mathbf3_A\otimes\overline{\mathbf3}_A\supset\mathbf8_A,
\end{equation}
followed by $\mathbf8_A\otimes\mathbf8_A\supset\mathbf1_A$.
Thus both $Q_Lu^cH$ and
$Q_Ld^cH^\dagger$ are available in the projected spectrum.

The leptonic Yukawas are particularly transparent in Class B because the leptonic branches can be taken \(G(2)\)-singlet:
\begin{equation}
Le^cH^\dagger:
\qquad
(\mathbf1,\mathbf8_A)_{-3}
\otimes
(\mathbf1,\mathbf8_A)_{+3}
\otimes
(\mathbf1,\mathbf8_A)_0
\supset
(\mathbf1,\mathbf1_A)_0,
\end{equation}
using
\begin{equation}
\mathbf8_A\otimes\mathbf8_A\otimes\mathbf8_A
\supset
\mathbf1_A .
\end{equation}
The Dirac-neutrino Yukawa is
\begin{equation}
LN^cH:
\qquad
(\mathbf1,\mathbf8_A)_{-3}
\otimes
(\mathbf1,\mathbf{10}_A)_{+3}
\otimes
(\mathbf1,\mathbf8_A)_0
\supset
(\mathbf1,\mathbf1_A)_0,
\end{equation}
with the electroweak doublet and singlet components selected after
\(SU(3)_A\to SU(2)_L\times U(1)_A\).  The parent-level tensor test is
\begin{equation}
\mathbf{2925}\otimes\mathbf{2925}
\supset
\mathbf{650}.
\end{equation}
All four projected Standard Model Yukawa ancestors have both the required low-energy charge closure and an \(E_6\)-parent cubic channel involving $\mathbf{650}_{0,H}$.

The practical advantage of Class B is branch economy: the visible sector, sterile sector and secluded breaking sector can be accommodated using the compact set
\begin{equation}
\overline{\mathbf{351}}_{+1},\qquad
\mathbf{351}_{-1},\qquad
\mathbf{2925}_{-3},\qquad
\mathbf{2925}_{+3}.
\end{equation}
The price is that the lepton parent is the larger \(\mathbf{2925}\) w.r.t. Class A \(\mathbf{2430}\).

\subsection{Anomaly structure in Class A/B spectra}
\label{subsec:final-anomaly-bookkeeping}

The final \(G(2)\)-singlet Higgs completions, Classes A and B, use
\begin{equation}
Y={1\over 3}(A+X).
\end{equation}
For the selected visible SM branches one finds
\begin{equation}
X(Q_L)=+1,\qquad
X(u^c)=X(d^c)=-1,
\end{equation}
\begin{equation}
X(L)=-3,\qquad
X(e^c)=+3.
\end{equation}
The mixed non-abelian anomalies cancel per family:
\begin{equation}
\mathcal A_{SU(3)_c^2U(1)_X}
=
2\,T(\mathbf3)(+1)
+T(\overline{\mathbf3})(-1)
+T(\overline{\mathbf3})(-1)
=0,
\end{equation}
and
\begin{equation}
\mathcal A_{SU(2)_L^2U(1)_X}
=
3\,T(\mathbf2)(+1)
+T(\mathbf2)(-3)
=0.
\end{equation}
The mixed hypercharge anomalies also cancel:
\begin{equation}
\mathcal A_{Y^2X}=0,
\qquad
\mathcal A_{YX^2}=0.
\end{equation}
Explicitly,
\begin{equation}
\mathcal A_{Y^2X}
=
6\left({1\over 6}\right)^2(+1)
+3\left(-{2\over3}\right)^2(-1)
+3\left({1\over3}\right)^2(-1)
+2\left(-{1\over2}\right)^2(-3)
+(1)^2(+3)=0,
\end{equation}
and
\begin{equation}
\mathcal A_{YX^2}
=
6\left({1\over 6}\right)(+1)^2
+3\left(-{2\over3}\right)(-1)^2
+3\left({1\over3}\right)(-1)^2
+2\left(-{1\over2}\right)(-3)^2
+(1)(+3)^2=0.
\end{equation}
However, the visible fields alone give
\begin{equation}
\mathcal A_{\mathrm{grav}^2X}
=
6(+1)+3(-1)+3(-1)+2(-3)+(+3)
=-3,
\end{equation}
and
\begin{equation}
\mathcal A_{X^3}
=
6(+1)^3+3(-1)^3+3(-1)^3+2(-3)^3+(+3)^3
=-27.
\end{equation}
These residual anomalies are canceled by the independently embedded sterile
state
\begin{equation}
N^c:\qquad X(N^c)=+3,\qquad Y(N^c)=0.
\end{equation}
Indeed,
\begin{equation}
\Delta\mathcal A_{\mathrm{grav}^2X}=+3,
\qquad
\Delta\mathcal A_{X^3}=+27,
\end{equation}
so that
\begin{equation}
\mathcal A_{\mathrm{grav}^2X}=0,
\qquad
\mathcal A_{X^3}=0.
\end{equation}

Thus the same sterile state that closes the renormalizable Yukawa channel
\(L N^cH\) also completes the split-\(U(1)_X\) anomaly bookkeeping.  This is
a useful consistency check of the final \(N^c\)-independent scan: allowing
\(N^c\) to be embedded independently is not an artificial fix, but is
precisely what is required by the residual abelian anomaly structure of the
visible split spectrum.

In Class A one may choose, for instance,
\begin{equation}
L\subset\mathbf{2430}_{-3},\qquad
e^c,N^c\subset\mathbf{2430}_{+3},
\end{equation}
whereas in Class B one may choose
\begin{equation}
L\subset\mathbf{2925}_{-3},\qquad
e^c,N^c,\Theta\subset\mathbf{2925}_{+3}.
\end{equation}
The two classes therefore share the same low-energy anomaly cancellation,
while differing in the parent \(E_6\) representation used for the lepton
sector.

We emphasize that this is the anomaly cancellation of the selected
low-energy split spectrum.  At the full high-scale \(E_6\) level, the large
parent multiplets contain many additional fragments.  These fragments must
be organized into complete anomaly-free or vectorlike heavy sectors by the
projection and lifting mechanism.  

\subsection{\(U(1)_X\) sector, abelian mass matrix and hypercharge matching}
\label{subsec:abelian_mass_matrix_matching}

Below \(M_{E_7}\) but above the final abelian breaking, the relevant gauge group is $G(2)\times SU(3)_A\times U(1)_X$ and after \(G(2)\to SU(3)_C\) it is $SU(3)_C\times SU(3)_A\times U(1)_X$.
Only at the scale \(M_\Theta\simeq M_\Omega\) does the combination orthogonal to hypercharge acquire a mass.
We use the \(SU(3)_A\) generator
\begin{equation}
	A=\sqrt3\,t_{A}^8,\qquad
	t_{A}^8=\frac{1}{2\sqrt3}\mathrm{diag}(1,1,-2),
\end{equation}
so that the corresponding normalization factor is $k_A=3$.
For the \(E_7\)-derived abelian generator we use the integer charge \(X\), normalized as $T_X=\frac{X}{\sqrt{N_X}}$, with $N_X=12$ and the hypercharge generator $Y=\frac13(A+X)$, as described before. Therefore the abelian covariant derivative can be written as
\begin{equation}
	D_\mu
	=
	\partial_\mu
	-i\,\frac{g_A}{\sqrt3}\,A\,B^A_\mu
	-i\,\frac{g_X}{\sqrt{N_X}}\,X\,B^X_\mu .
	\label{eq:abelian_covariant_derivative_AX}
\end{equation}
The scalar \(\Theta\) responsible for the final abelian breaking is chosen in the branch
\begin{equation}
	\Theta\sim(\mathbf1_{G(2)},\mathbf{10}_A)_{+3}.
\end{equation}
The \(SU(2)_L\)-singlet component in \(\mathbf{10}_A\) has
\begin{equation}
	A_\Theta=-3,\qquad X_\Theta=+3,
\end{equation}
and hence
\begin{equation}
	Y(\Theta)=\frac13(A_\Theta+X_\Theta)=0.
\end{equation}
Therefore a nonzero \(\Theta\) vacuum expectation value leaves hypercharge unbroken while
giving a mass to the orthogonal abelian gauge boson via the breaking $U(1)_A\times U(1)_X \to U(1)_Y$.
Writing
\begin{equation}
	\Theta=\frac{1}{\sqrt2}\left(v_\Theta+\rho_\Theta\right),
\end{equation}
the abelian gauge-boson mass term generated by \(|D_\mu\Theta|^2\) is
\begin{equation}
	\mathcal L_{\rm mass}^{(A,X)}
	=
	\frac12
	\begin{pmatrix}
		B^A_\mu & B^X_\mu
	\end{pmatrix}
	\mathcal M_{AX}^2
	\begin{pmatrix}
		B^{A\,\mu}\\[1mm]
		B^{X\,\mu}
	\end{pmatrix},
\end{equation}
with
\begin{equation}
	\mathcal M_{AX}^2
	=
	v_\Theta^2
	\begin{pmatrix}
		\dfrac{g_A^2 A_\Theta^2}{3}
		&
		\dfrac{g_A g_X A_\Theta X_\Theta}{\sqrt{3N_X}}
		\\[4mm]
		\dfrac{g_A g_X A_\Theta X_\Theta}{\sqrt{3N_X}}
		&
		\dfrac{g_X^2 X_\Theta^2}{N_X}
	\end{pmatrix}.
	\label{eq:abelian_mass_matrix_general}
\end{equation}
Putting $A_\Theta=-3,\; X_\Theta=+3,\; N_X=12$, this becomes
\begin{equation}
	\mathcal M_{AX}^2
	=
	v_\Theta^2
	\begin{pmatrix}
		3g_A^2 & -\dfrac32 g_A g_X\\[2mm]
		-\dfrac32 g_A g_X & \dfrac34 g_X^2
	\end{pmatrix}.
	\label{eq:abelian_mass_matrix_numeric}
\end{equation}
Its determinant vanishes,
\begin{equation}
	\det\mathcal M_{AX}^2=0,
\end{equation}
as required by the survival of one unbroken abelian generator.  The massive eigenvalue is
\begin{equation}
	m_{Z'}^2
	=
	v_\Theta^2
	\left(
	\frac{g_A^2 A_\Theta^2}{3}
	+
	\frac{g_X^2 X_\Theta^2}{N_X}
	\right)
	=
	v_\Theta^2
	\left(
	3g_A^2+\frac34 g_X^2
	\right).
	\label{eq:Zprime_mass_AX}
\end{equation}
For \(g_A\simeq g_X\equiv g_\Theta\), this gives
\begin{equation}
	m_{Z'}\simeq \frac{\sqrt{15}}{2}\,g_\Theta v_\Theta .
\end{equation}
The heavy \(Z'\) associated with
\(U(1)_A\times U(1)_X\to U(1)_Y\) is not a dark gauge boson. The
\(U(1)_X\) gauge field is an \(E_6\)-singlet and therefore transforms as
\begin{equation}
	C_\mu\sim(\mathbf1,\mathbf1)_0
\end{equation}
under \(G(2)\times SU(3)_A\times U(1)_X\). The \(SU(3)_A\) Cartan gauge boson
entering the orthogonal massive combination is also a \(G(2)\)-singlet. Hence
\begin{equation}
	Z'_\mu\sim\mathbf1_{G(2)} .
\end{equation}
By contrast, the \(G(2)\)-origin gauge bosons belong to
\begin{equation}
	(\mathbf{14},\mathbf1_A)_0
\end{equation}
and are neutral under \(U(1)_X\). Therefore the \(Z'\) is a high-scale abelian
vector, not a member of the \(G(2)\)-broken dark-glueball sector. Any residual
communication with the dark sector arises only through heavy scalar portals or
threshold effects, not through a direct \(Z'\)-dark-gluon gauge coupling.

In writing the above equations we have neglected possible abelian kinetic mixing \cite{Langacker2009Zprime,Holdom1986} such as
\begin{equation}
	-\frac{\epsilon}{2}F^A_{\mu\nu}F_X^{\mu\nu},
\end{equation}
considering the natural tree-level basis inherited from the simple \(E_7\) normalization. Loop-induced
or threshold-induced kinetic mixing can be included by replacing the diagonal coupling
matrix in Eq.~\eqref{eq:abelian_covariant_derivative_AX} with a more general abelian coupling matrix. A nonzero kinetic mixing can always be removed from the kinetic term by a non-orthogonal field redefinition, but it then reappears as off-diagonal entries in the coupling matrix.
Thus the exact matching of \(g_Y\) might be shifted by threshold corrections, while the massless generator is still fixed by the charge condition $Y(\Theta)=0$, but the overall charge construction is not affected.

Concluding, equations~\eqref{eq:abelian_mass_matrix_numeric}--\eqref{eq:Zprime_mass_AX} show explicitly that
the \(E_7\)-derived \(U(1)_X\) is not a broken, unavailable remnant of the high-scale theory.
It can remain a genuine gauge factor until the \(\Theta\) threshold, where it mixes with the
abelian generator descending from \(SU(3)_A\). The physical hypercharge gauge boson is the massless
combination left unbroken by \(\langle\Theta\rangle\), whereas the orthogonal combination
is the massive \(Z'\) boson. \(U(1)_X\) must survive
until the \(\Theta\) threshold, and the physical hypercharge is the unbroken linear combination satisfying \(Y(\Theta)=0\). If \(U(1)_X\) were instead broken already at
\(M_{E_7}\), the hypercharge repair mechanism would indeed fail.  The final \(E_7\)-uplifted chain therefore requires
\begin{equation}
	E_7
	\to
	E_6\times U(1)_X
	\to
	G(2)\times SU(3)_A\times U(1)_X
	\to
	SU(3)_C\times SU(3)_A\times U(1)_X
	\to
	SU(3)_C\times SU(2)_L\times U(1)_Y .
\end{equation}
\paragraph{Why the symmetry-breaking sector must be \(U(1)_X\)-neutral.}
The hypercharge-repair mechanism requires \(U(1)_X\) to survive as an exact, unbroken gauge factor from \(M_{E_7}\) down to \(M_\Theta\simeq M_\Omega\), so that it can be combined with the \(SU(3)_A\) Cartan direction only at the very last step, via \(\langle\Theta\rangle\). This is not an extra assumption layered on top of the construction, it is forced by the representation assignment. Every scalar responsible for a breaking step prior to \(\Theta\) -- the singlet \(S\) and \(\chi,\Omega\), all three contained in the single multiplet \(\Phi_{\mathbf{650}}\) -- sit in the $X=0$ branch of the relevant $E_7$ minimal scalar parent $\mathbf{650}_{0}\subset\mathbf{1463}_{E_7}$, and \emph{not} from the \(X=\pm3\) copies of \(\mathbf{650}\) noted above to occur inside the matter parents \(\mathbf{27664}_{E_7}\) or \(\mathbf{86184}_{E_7}\). Since \(X\) is an exact, conserved charge of the unbroken theory above \(M_\Theta\), a scalar $\sigma$ with \(X(\sigma)\neq0\) acquiring a VEV would break \(U(1)_X\) directly at the scale of that VEV. Had \(S\), \(\chi\) or \(\Omega\) instead been drawn from an \(X\neq0\) branch, \(U(1)_X\) would already be broken at or above \(M_{E_6}\)  (at or before the first special-embedding step), and no unbroken abelian generator would remain at \(M_\Omega\) to combine with the \(SU(3)_A\) direction \(A\): the surviving combination would then be fixed \emph{a priori} to be \(A\) alone and the construction would collapse onto the pure \(SU(3)_A\) embedding \(Y\propto A\) already excluded by the no-go theorem of Sec.~\ref{subsec:nogo_SU3A_hypercharge}. Requiring that the \(E_7\) completion perform the task it is introduced for -- repairing the hypercharge normalization through a genuine combination \(Y=\beta A+\gamma X\) -- is therefore logically equivalent to
\begin{equation}
	X(S)=X(\chi)=X(\Omega)=0,
	\label{eq:X0_requirement}
\end{equation}
and this is precisely what is realized once \(\Phi_{\mathbf{650}}\) is identified with the \(X=0\) slice of the \(\mathbf{1463}_{E_7}\) scalar parent rather than with one of the \(X=\pm3\) slices used for the fermionic content. Only the last breaking field, \(\Theta\subset\mathbf{2925}_{+3}\), is required to carry \(X\neq0\): it is precisely the field assigned to the final step, and its own \(SU(3)_A\)-charge is tuned so that \(A_\Theta+X_\Theta=0\), i.e.\ \(Y(\Theta)=0\), guaranteeing that it is hypercharge -- and not some other linear combination -- that survives \(\langle\Theta\rangle\) massless.

This argument fixes which representation assignment is logically required for the \(E_7\) mechanism to be consistent; it does not, by itself, \emph{derive} from the scalar potential that the true vacuum dynamically prefers the \(X=0\) slice of \(\mathbf{1463}_{E_7}\) over some admixture with other \(\mathbf{650}\)-bearing branches at \(X\neq0\). A dynamical selection argument along the lines of the \(\lambda_2\)-sign mechanism described below in Section~7 to select the special vs.\ regular \(E_6\) vacuum -- here applied to an \(E_7\)-invariant potential for \(\Phi_{133}\) and \(\mathbf{1463}_{E_7}\) -- is left for future work. In the present construction, \(X(S)=X(\chi)=X(\Omega)=0\) is adopted as the representation choice required for consistency, in the same sense that the \(\mathbf{650}\)-singlet direction \(\hat S\) itself is adopted to select \(G(2)\times SU(3)_A\) over other subgroups of \(E_6\).

\subsection{Projected spectrum and Vectorlike completion}

The large \(E_7\) parent representation should not be interpreted as a full light multiplet: the \(\mathbf{86184}_{E_7}\) and \(\mathbf{27664}_{E_7}\) is a representation-theoretic parent that contains the required \(E_6\times U(1)_X\) branches. The physical model is a projected split-multiplet construction. Only the selected Standard Model zero modes, possibly together with \(N^c\), remain light below the projection scale. All other components must pair vectorlike and decouple. 
The complete \(E_7\) macro-parent multiplets are vectorlike after projection to
\(SU(3)_C\times SU(2)_L\times U(1)_Y\).  This was verified explicitly for the
relevant \(E_7\) containers.  However, after subtracting a single desired chiral
SM-family zero-mode assignment, the residual spectrum is not automatically
vectorlike: the residual imbalance is precisely of SM-family type and reflects
the usual fact that chirality cannot be obtained merely by selecting components
inside a real or vectorlike parent multiplet.  A chirality-generating mechanism,
such as flux, boundary conditions, orbifold projection, or localized index
selection, is therefore required.  The role of the \(E_7\) macro-parent is not
to generate chirality by itself, but to provide a vectorlike ultraviolet
container in which the desired chiral zero modes and their heavy mirror partners
can be organized consistently.

Schematically, for every exotic component \(E_i\) with conjugate \(E_i^c\), the most general vectorlike mass Lagrangian is then organized by the available
breaking fields:
\begin{align}
	\mathcal L_{\rm VL}
	\supset
	-&
	\left[
	M^{(0)}_{ij} E_iE_j^c
	+
	y^{S}_{ij} S\,E_iE_j^c
	+
	y^\chi_{ij}\chi\,E_iE_j^c
	+
	y^\Omega_{ij}\Omega\,E_iE_j^c
	+
	y^\Theta_{ij}\Theta\,E_iE_j^c
	+\mathrm{h.c.}
	\right].
\end{align}
Each term is present only if the corresponding tensor product contains a gauge singlet under the symmetry that is still unbroken at that scale. The first term is a bare vectorlike Dirac mass. It is allowed only for conjugate pairs whose product is already gauge invariant.
If a symmetry forbids such an explicit mass, the term may be absent and the
mass is generated only after one of the high-scale scalars acquires a VEV. 

The singlet term $y^S_{ij}S\,E_iE_j^c$
uses an \(E_6\)-singlet or effectively singlet scalar direction, with $\langle S\rangle\sim M_{E_6}$, generating
masses 
\begin{equation}
M^{S}_{ij}\sim y^S_{ij}M_{E_6}.
\end{equation}
This is the natural completion for exotic states that should disappear already at the \(E_6\)-completion scale.

The \(\chi\)-dependent term $y^\chi_{ij}\chi\,E_iE_j^c$
is allowed when the \(G(2)\) tensor product closes with $
\chi\sim(\mathbf7,\mathbf1)_0$. After $\langle\chi\rangle=v_\chi\sim M_\chi$,
it generates 
\begin{equation}
M^\chi_{ij}\sim y^\chi_{ij}v_\chi.
\end{equation}
This term is especially important for \(G(2)\)-charged exotics and for splitting components that become distinct after $G(2)$ breaking. For example, if a field contains $
\mathbf7_{G(2)}\to\mathbf1_C\oplus\mathbf3_C\oplus\overline{\mathbf3}_C$, then \(\chi\)-dependent contractions can lift the colored components at \(M_\chi\) while allowing a selected color singlet to survive to a lower
threshold.

The \(\Omega\)-dependent term $y^\Omega_{ij}\Omega\,E_iE_j^c$ is allowed when the \(SU(3)_A\) product closes with $\Omega\sim(\mathbf1,\mathbf8_A)_0$. After $\langle\Omega\rangle=v_\Omega\sim M_\Omega$,
it generates masses and splittings of order 
\begin{equation}
M^\Omega_{ij}\sim y^\Omega_{ij}v_\Omega.
\end{equation}
This term is useful for lifting electroweak-ancestor exotics after $SU(3)_A\to SU(2)_L\times U(1)_A$.
It must be arranged not to lift the desired Standard Model chiral zero modes.

Finally, the \(\Theta\)-dependent term $y^\Theta_{ij}\Theta\,E_iE_j^c$ is allowed only when the full $G(2)\times SU(3)_A\times U(1)_X$ charges close. In particular,
\begin{equation}
X(E_i)+X(E_j^c)+X(\Theta)=0.
\end{equation}
Since $X(\Theta)=+3$, this term naturally lifts exotic states whose \(X\)-charges differ by three
units. After $\langle\Theta\rangle=v_\Theta\sim M_\Theta\simeq M_\Omega$, one obtains
\begin{equation}
M^\Theta_{ij}\sim y^\Theta_{ij}v_\Theta .
\end{equation}
Collecting the contributions, the exotic mass matrix is schematically
\begin{equation}
(M_{\rm ex})_{ij}
=
M^{(0)}_{ij}
+
y^S_{ij}v_{E_6}
+
y^\chi_{ij}v_\chi
+
y^\Omega_{ij}v_\Omega
+
y^\Theta_{ij}v_\Theta .
\end{equation}
Thus vectorlike completion is not a single-scale operation. It is naturally
multi-threshold:
\begin{equation}
M_{E_6},\qquad M_\chi,\qquad M_\Omega\simeq M_\Theta .
\end{equation}
Exceptional exotics may be lifted near \(M_{E_6}\), \(G(2)\)-charged exotics
near \(M_\chi\), \(SU(3)_A\)-charged exotics near \(M_\Omega\), and
purely abelian exotics near \(M_\Theta\).
This low-threshold lifting should be understood selectively: large
\(G(2)\)-charged product branches such as \((\mathbf7,R_A)\) cannot remain
as long-distance perturbative bulk fields down to \(M_\Omega\).  The
\(M_\Omega\simeq M_\Theta\) completion is primarily appropriate for
\(G(2)\)-singlet \(SU(3)_A\)- or \(U(1)_X\)-charged spectators, while
\(G(2)\)-charged remnants must be lifted or projected near \(M_\chi\) or
above.

The desired chiral zero modes are selected by the projection condition
\begin{equation}
\mathcal I(Q_L)=\mathcal I(u^c)=\mathcal I(d^c)=\mathcal I(L)=\mathcal I(e^c)=3,
\end{equation}
while
\begin{equation}
\mathcal I(E_{\rm exotic})=0.
\end{equation}
Thus the remaining SM-mirror imbalance must be removed by the same chirality-generating projection that selects the SM zero modes.

\subsection{Perturbative-unitarity constraints}
A useful perturbative-unitarity diagnostic for the representations appearing in the $E_7$ uplift is the largest coupled-channel partial-wave eigenvalue for matter annihilation into gauge bosons, as discussed in \cite{MILAGRE2024116542}:
\begin{equation}
(a_0)_{\max}
=
\frac{g^2}{8\pi\sqrt{\widetilde n}}
\left[
\frac{\pi^2}{4}
\sum_F n_F C(R_F)^2
+
\sum_S \eta_S n_S C(R_S)^2
\right]^{1/2}.
\end{equation}
Here \(\widetilde n\) is the dimension of the adjoint, \(C(R)\) is the quadratic Casimir, \(n_F\) and \(n_S\) are representation dimensions, and \(\eta_S=1\) for complex scalars while \(\eta_S=1/2\) for real scalars. Perturbative tree-level unitarity requires
\begin{equation}
|\mathrm{Re}\,a_0|\leq \frac12.
\end{equation}
For \(E_6\), $\widetilde n=78$ and using the \(E_6\) Dynkin-label convention by LieART \cite{LieArt}, the quadratic Casimir are
\begin{equation}
C_2(R)=\frac12\,\lambda^T A^{-1}(\lambda+2\rho),
\end{equation}
and one obtains
\begin{equation}
C_2(\mathbf{27})=\frac{26}{3},
\qquad
C_2(\mathbf{351})=\frac{50}{3},
\qquad
C_2(\mathbf{2430})=26,
\qquad
C_2(\mathbf{2925})=24,
\qquad
C_2(\mathbf{650})=18.
\end{equation}
The conjugate representations have the same quadratic Casimir. Known examples already show that large scalar or fermion representations quickly saturate the bound.
For \(E_6\), using \(g_{E_6}\simeq0.55\) and
\(\dim(\mathrm{Adj}_{E_6})=78\), the final Class A complete-parent
assignment $
\overline{\mathbf{27}}\oplus\mathbf{351}
\oplus\mathbf{2430}\oplus\mathbf{2430}$
already gives \(a_0^{\max}\simeq3.94\) before adding scalars. Including scalar parents \(\mathbf{650}_H\), \(\mathbf{650}_\Omega\) and
\(\mathbf{2925}_\Theta\) gives $a_0^{\max}\simeq4.18$.
For Class B, $
\overline{\mathbf{351}}\oplus\mathbf{351}
\oplus\mathbf{2925}\oplus\mathbf{2925}$,
one obtains \(a_0^{\max}\simeq4.04\) from the fermions alone and \(a_0^{\max}\simeq4.28\) after adding the same scalar parents.
Therefore neither $\mathbf{86184}_{E_7}$ nor the full \(E_6\) branches $\mathbf{27}$, $\mathbf{351}$, $\mathbf{2430}$ or $\mathbf{2925}$ can be interpreted as complete light matter multiplets over a long perturbative interval. 

The conclusion is that the large parent multiplets are projection parents, not full light spectra. Only the Standard Model chiral zero modes and the required sterile state are retained below the projection scale and the partial-wave unitarity estimate for the projected Standard Model zero-mode spectrum has eigenvalues which are far below \(1/2\). In fact, computing \((a_0)_{\max}\) at \(g=0.55\), one obtains 0.053 for \(SU(3)_C\), 0.041 for \(SU(2)_L\) and  0.026 for \(U(1)_Y\). This projected interpretation is not optional: it is required by perturbative unitarity and by the absence of large low-energy exotic spectra.

\subsection{Exotic Bosonic Remnants}
Summarizing, the exotic bosonic content of the theory includes new heavy vectors and scalars.
From $E_7$ we count 54 heavy vectors from $\mathbf{133}$ decomposition; the adjoint $\mathbf{78}$ of $E_6$ gives 56 heavy vectors $\mathcal{E}_\mu$ associated with $(\mathbf7,\mathbf8)$; from $G(2)/SU(3)_C$ we obtain six vectors with $m_{\chi}=g_{G(2)} v_\chi$, whereas from $SU(3)_A/(SU(2)_L\times U(1)_A)$ four vectors with $m_Y=g_A v_\Omega$ emerge and an orthogonal abelian vector $Z'$ from the breaking of $U(1)_A \times U(1)_X$.

The scalar sector contains four multiplets responsible for the successive symmetry breakings, \textit{i.e.}
$\Sigma \in  \mathbf{133}_{E_7}$,
$\Phi \in  \mathbf{650}_{E_6}$, $\chi \in \mathbf7_{G(2)}$, $\Omega \in \mathbf8_{SU(3)_A}$.
The multiplet $\Phi$ contains the singlet direction that triggers the fundamental breaking. Its massive radial mode is $h_\Phi$. The scalar $\chi$ decomposes under $SU(3)_C$ as $\mathbf7 \rightarrow \mathbf3 \oplus \bar{\mathbf3} \oplus \mathbf1$. The singlet component develops the VEV $v_\chi$ and produces the radial mode $h_\chi$, while the remaining components provide the Goldstone bosons eaten by the six heavy $\mathcal X$ $G(2)$ vectors.
$\Omega$ develops the VEV $v_\Omega$ and produces the radial Higgs mode $h_\Omega$ and then we have $\Theta$ which acts at the same $\Omega$ energy scale.
%
Their masses are 
\begin{equation}
m_{h_\Phi}=\sqrt{2 \lambda_\Phi}\,v_{ \mathbf{650}},\qquad
m_{h_\chi}=\sqrt{2 \lambda_\chi}\,v_\chi,\qquad
m_{h_\Omega}=\sqrt{2 \lambda_\Omega}\,v_\Omega,\qquad
m_{h_\Theta}=\sqrt{2 \lambda_\Theta}\,v_\Theta,
\end{equation}
where $\lambda_\Phi, \lambda_\chi, \lambda_\Omega, \lambda_\Theta$ are the usual quartic couplings of the Higgs (see Section 8 for details). It is noteworthy to stress that heavy $E_7$, $E_6$ and $SU(3)_A$ vectors can have prompt decays into visible states, while $G(2)$ glueballs are stable/long-lived due to charge orthogonality and absence of renormalizable portals (plus optional global parity symmetries \cite{Masi2021} as a precaution).

\begin{table}[H]
	\centering
	\small
	\renewcommand{\arraystretch}{1.28}
	\setlength{\tabcolsep}{3pt}
	\begin{tabularx}{\textwidth}{
			>{\raggedright\arraybackslash}p{0.14\textwidth}|
			>{\raggedright\arraybackslash}X|
			>{\raggedright\arraybackslash}X|
			>{\raggedright\arraybackslash}X}
		\hline
		\textbf{Sector} & \textbf{\(E_7\to E_6\times U(1)_X\)} &
		\textbf{\(E_6\to G(2)\times SU(3)_A\)} &\textbf{ Role} \\
		\hline
		
		\(E_7\) gauge bosons &
		\(\mathbf{133}\to \mathbf{78}_0\oplus\mathbf1_0
		\oplus\mathbf{27}_{+2}\oplus\overline{\mathbf{27}}_{-2}\) &
		\(\mathbf{78}_0\to(\mathbf{14},\mathbf1)_0\oplus(\mathbf1,\mathbf8_A)_0
		\oplus(\mathbf7,\mathbf8_A)_0\) &
		\((\mathbf{14},\mathbf1)_0\) gives the \(G(2)\) gauge bosons;
		\((\mathbf1,\mathbf8_A)_0\) gives the \(SU(3)_A\) gauge bosons;
		\(\mathbf1_0\) gives the \(U(1)_X\) gauge boson. \\
		\hline
		
		\(E_7/E_6\) &
		\(\mathbf{27}_{+2}\oplus\overline{\mathbf{27}}_{-2}\) &
		\(\mathbf{27}\to(\mathbf7,\mathbf3_A)\oplus(\mathbf1,\overline{\mathbf6}_A)\),
		\(\overline{\mathbf{27}}\to(\mathbf7,\overline{\mathbf3}_A)\oplus(\mathbf1,\mathbf6_A)\) &
		Heavy vectors from regular \(E_7\to E_6\times U(1)_X\), with masses
		near \(m_{E_7}\). \\
		\hline
		
		\(E_6/H\) &
		inside \(\mathbf{78}_0\) &
		\((\mathbf7,\mathbf8_A)_0\), with \(H=G(2)\times SU(3)_A\) &
		The 56 heavy vectors of the special \(E_6\) coset, with
		\(m_{\mathcal E}\sim g_{E_6}v_{E_6}\). \\
		\hline
		
		\(G(2)/SU(3)_C\) &
		inside \((\mathbf{14},\mathbf1)_0\) &
		\(\mathbf{14}_{G(2)}\to \mathbf8_C\oplus\mathbf3_C
		\oplus\overline{\mathbf3}_C\) &
		\(\mathbf8_C\) gives QCD gluons. The
		\(\mathbf3_C\oplus\overline{\mathbf3}_C\) are six heavy vectors with
		\(m_{\chi}\sim g_{G(2)}v_\chi\). \\
		\hline
		
		\(SU(3)_A/G_{\rm EW}\) &
		inside \((\mathbf1,\mathbf8_A)_0\) &
		\(\mathbf8_A\to\mathbf3_0\oplus\mathbf1_0
		\oplus\mathbf2_{+3/2}\oplus\mathbf2_{-3/2}\),
		\(G_{\rm EW}=SU(2)_L\times U(1)_A\) &
		\(\mathbf3_0\) and \(\mathbf1_0\) give \(SU(2)_L\) and \(U(1)_A\).
		The coset \(\mathbf2_{+3/2}\oplus\mathbf2_{-3/2}\) gives four heavy
		electroweak-ancestor vectors with \(m_{Y}\sim g_A v_\Omega\). \\
		\hline
		
		Orthogonal \(Z'\) &
		\(\mathbf1_0\) mixed with the \(SU(3)_A\) abelian direction &
		\(U(1)_X\) plus \(U(1)_A\) &
		Hypercharge remains massless; the orthogonal combination gets
		\(m_{Z'}\sim g_\perp\langle\Theta\rangle\) at
		\(M_\Theta\simeq M_\Omega\). \\
		\hline
	\end{tabularx}
	\caption{Gauge-vector content of the \(E_7\)-uplifted model, showing the surviving gauge fields and heavy coset sectors, their subgroup decompositions, physical roles and characteristic mass scales.}
	\label{tab:E7_gauge_bosons_corrected}
\end{table}

\begin{table}[H]
	\centering
	\small
	\renewcommand{\arraystretch}{1.25}
	\begin{tabularx}{\textwidth}{c|X|X|X|X}
		\hline
		\textbf{Field} & \textbf{Scalar parent} & \textbf{Representative branch} & \textbf{SM component / VEV direction} & \textbf{Role} \\
		\hline
		\(\Sigma_{E_7}\) &
		\(\mathbf{133}_H\) or another adjoint-type \(E_7\) scalar &
		\(\mathbf1_0\subset\mathbf{133}\to\mathbf{78}_0\oplus\mathbf1_0
		\oplus\mathbf{27}_{+2}\oplus\overline{\mathbf{27}}_{-2}\) &
		\(E_6\times U(1)_X\)-preserving direction &
		Triggers the regular breaking \(E_7\to E_6\times U(1)_X\). \\
		\hline
		\(S\) &
		\(E_6\)-singlet branch, e.g. \(\mathbf1_0\) or \(\mathbf1_{\pm6}\) in larger \(E_7\) scalar parents &
		\((\mathbf1,\mathbf1)_q\) &
		Gauge singlet or effective singlet &
		Generates vectorlike masses for exotics and sterile neutrino through \(\langle S\rangle\sim M_{E_6}\). \\
		\hline
		\(\Phi_{650}\) &
		\(\mathbf{650}_{0,H}\subset\mathbf{1463}_{E7,H}\) or equivalent scalar parent &
		\(\mathbf{650}_0\) &
		Contains \(\chi\), \(\Omega\), and the SM Higgs doublet candidates &
		Main \(E_6\)-level scalar sector. It supplies the \(G(2)\)-breaking scalar, the \(SU(3)_A\)-breaking scalar, and the electroweak Higgs direction. \\
		\hline
		\(\chi\) &
		\(\mathbf{650}_{0,H}\) &
		\((\mathbf7,\mathbf1_A)_0\subset\mathbf{650}_0\) &
		\((\mathbf1_C,\mathbf1_L)_0\) component after \(G(2)\to SU(3)_C\) &
		Breaks \(G(2)\to SU(3)_C\) at \(M_\chi\). \\
		\hline
		\(\Omega\) &
		\(\mathbf{650}_{0,H}\) &
		\((\mathbf1,\mathbf8_A)_0\subset\mathbf{650}_0\) &
		\(\mathbf1_0\subset\mathbf8_A\to
		\mathbf1_0\oplus\mathbf2_{+3/2}\oplus\mathbf2_{-3/2}\oplus\mathbf3_0\) &
		Breaks \(SU(3)_A\to SU(2)_L\times U(1)_A\) at \(M_\Omega\). \\
		\hline
		\(\Theta\) &
		\(\mathbf{2925}_{+3}\) branch inside the \(E_7\) parent sector &
		\((\mathbf1,\mathbf{10}_A)\) \(\subset\mathbf{2925}_{+3}\) &
		Color-singlet component with \(Y(\Theta)=0\) &
		Breaks the orthogonal abelian combination \(U(1)_A\times U(1)_X/U(1)_Y\) at \(M_\Theta\simeq M_\Omega\). \\
		\hline
		\(H\) &
		\(\mathbf{650}_{0,H}\) &
		\((\mathbf1,\mathbf8_A)_0\subset\mathbf{650}_0\) &
		\(\mathbf2_{+3/2}\) or \(\mathbf2_{-3/2}\) in \(\mathbf8_A\) &
		SM Higgs option for class A/B scenario. With \(Y=(A+X)/3\), choose \(A_H=+3/2\), \(X_H=0\). \\
		\hline
	\end{tabularx}
	\caption{Scalar content of the \(E_7\)-uplifted model, with the parent multiplets, representative \(G(2)\times SU(3)_A\) branches, VEV directions and their roles in symmetry breaking.}
	\label{tab:E7_scalar_bosons}
\end{table}

\section{The Complete High--Scale Exceptional Lagrangian}

We now present the full ultraviolet exceptional Lagrangian.
The bulk of the theory consists of the $E_6$ gauge sector and the symmetry--breaking Higgs from
$\Phi_ \mathbf{650}$, along with the parent $E_7$--uplifteed parent.

\paragraph{Gauge sector.}
At the highest scale the gauge field is
\begin{equation}
	\mathcal G_\mu=\mathcal G_\mu^{\mathcal A}T^{\mathcal A}_{E_7},
	\qquad
	\mathcal A=1,\ldots,133,
\end{equation}
with field strength
\begin{equation}
	\mathcal G_{\mu\nu}^{\mathcal A}
	=
	\partial_\mu \mathcal G_\nu^{\mathcal A}
	-
	\partial_\nu \mathcal G_\mu^{\mathcal A}
	+
	g_{E_7} f^{\mathcal A\mathcal B\mathcal C}_{E_7}
	\mathcal G_\mu^{\mathcal B}\mathcal G_\nu^{\mathcal C}.
\end{equation}
The gauge kinetic term is
\begin{equation}
	\mathcal L_{\rm gauge}^{E_7}
	=
	-\frac14
	\mathcal G_{\mu\nu}^{\mathcal A}
	\mathcal G^{\mathcal A\mu\nu}.
\end{equation}
The first breaking is represented by a scalar \(\Sigma_{E_7}\) with a vacuum direction preserving
\(E_6\times U(1)_X\). Schematically,
\begin{equation}
	\mathcal L_{\Sigma_{E_7}}
	=
	(D_\mu\Sigma_{E_7})^\dagger(D^\mu\Sigma_{E_7})
	-
	V_{E_7}(\Sigma_{E_7}),
\end{equation}
with $\langle\Sigma_{E_7}\rangle=v_{E_7}\,\hat S_{E_7}$.
The heavy vectors in $E_7/(E_6\times U(1)_X)$ acquire masses of order	$M_{E_7}\sim g_{E_7}v_{E_7}$.

Then, let $\mathcal{E}_\mu = \mathcal{E}_\mu^a T^a$ ($a=1,\dots,78$, for the $T^a$ generators in $\mathbf{27}$) denote the $E_6$ gauge field with field strength
\begin{equation}
	E_{\mu\nu}^a
	=
	\partial_\mu \mathcal{E}_\nu^a
	-
	\partial_\nu \mathcal{E}_\mu^a
	+
	g_{E_6}\, f_E^{abc} \mathcal{E}_\mu^b \mathcal{E}_\nu^c ,
\end{equation}
where $f_E^{abc}$ are the $E_6$ structure constants and $g_{E_6}$ is the unified gauge coupling. The construction scheme for $E_6$ generators can be built up on the trinification basis of $E_6$ and performing the
linear combinations described in Appendix~A.3 of
Ref.~\cite{Babu_2023}. Explicit matrix realizations for several representations are also
available in the \texttt{E6Tensors} package
\cite{E6Tensors}. The abelian field is denoted as \(C_\mu\). The covariant derivative acting on a field in \(E_6\) representation \(R\)
and with \(U(1)_X\) charge \(X_R\) is
\begin{equation}
	D_\mu
=
\partial_\mu
-
i g_{E_6}\mathcal E_\mu^aT^a_{(R)}
-
i g_X X_R C_\mu,
\end{equation}
while the gauge kinetic term is
\begin{equation}
	\mathcal L_{\rm gauge}^{E_6\times U(1)_X}
=
-\frac14 E_{\mu\nu}^aE^{a\mu\nu}
-\frac14 C_{\mu\nu}C^{\mu\nu}.
\end{equation}

\paragraph{Fermion sector.}
The matter sector is not interpreted as a complete light
\(\mathbf{86184}_{E_7}\), instead it represents the parent
representation containing the selected branches of class A/B solutions. 
The full parent multiplets are not kept
as complete light fields. The effective fermion kinetic terms are written for
the projected zero modes:
\begin{equation}
	\mathcal L_{\rm kin}^{\rm ferm,\; proj}
	=
	\sum_{i=1}^3
	\left[
	\overline Q_i i\slashed D Q_i
	+
	\overline u_i^c i\slashed D u_i^c
	+
	\overline d_i^c i\slashed D d_i^c
	+
	\overline L_i i\slashed D L_i
	+
	\overline e_i^c i\slashed D e_i^c
	+
	\overline N_i^c i\slashed D N_i^c
	\right].
\end{equation}
%

\paragraph{$E_6$ breaking Higgs sector.}
The first symmetry breaking $E_6 \longrightarrow G(2) \times SU(3)_A$ is achieved by the Higgs multiplet $\Phi_ \mathbf{650}$. Its kinetic term is
\begin{equation}
	\mathcal L_{\Phi,\,{\rm kin}}
	=
	(D_\mu \Phi)^\dagger (D^\mu \Phi).
\end{equation}
The most general renormalizable $E_6$-invariant potential can be written schematically as
\begin{align}
	V_{E_6}(\Phi)
	=
	&- m_\Phi^2 (\Phi^\dagger \Phi)
	+ \lambda_{\Phi} (\Phi^\dagger \Phi)^2
	+ \lambda_2 (\Phi^\dagger T^A \Phi)(\Phi^\dagger T^A \Phi)
	\nonumber\\
	&+ \lambda_3 \mathcal I_4(\Phi),
\end{align}
where $\lambda_2$ term is the adjoint channel contraction that actually decides which symmetry-breaking pattern is energetically preferred (not mathematically present in SM Higgs potential), and $\mathcal I_4(\Phi)$ denotes the additional independent quartic invariants allowed by $E_6$ tensor contractions of the $ \mathbf{650}$. For different singlet directions $\hat S$ corresponding to distinct symmetry-breaking patterns, $\lambda_2$ term generally takes different values. In particular, for a vacuum preserving the regular subgroup $SU(3)^3$, the projection onto broken generators is typically larger than for the special embedding $SU(3)\times G(2)$, since the latter preserves a larger non-regular subgroup structure. Therefore the sign of $\lambda_2$ determines which embedding is energetically preferred: for $\lambda_2 > 0$ we point towards a vacuum with smaller projection of the vacuum itself onto the adjoint directions of $E_6$, while $\lambda_2 < 0$ favors directions with larger adjoint projection. This mechanism provides the dynamical origin of vacuum competition between regular and special subgroups.

When $\Phi$ acquires a VEV $\langle \Phi \rangle = v_{E_6}\, \hat S$, with $\hat S$ a fixed direction in field space selecting the special embedding, the gauge bosons in the coset $E_6/(G(2) \times SU(3)_A)$ acquire masses
\begin{equation}
	(M^2)^{ab}
	=
	g_{E_6}^2 v_{E_6}^2
	\,
	\hat S^\dagger T^a T^b \hat S.
\end{equation}

\paragraph{$E_6$ Yukawa sector.}
The Standard Model Yukawa interactions are generated by projected \(E_6\)-level cubic channels involving the neutral Higgs ancestor $\mathbf{650}_{0,H}$. 
The effective Class A Yukawa Lagrangian may be written schematically as
\begin{align}
	\mathcal L_{\rm Yuk}^{A}
	\supset
	&
	\,y^u_{ij}
	\Big[
	\overline{\mathbf{27}}_{+1,i}\,
	\mathbf{351}_{-1,j}\,
	\mathbf{650}_{0,H}
	\Big]_{QHu}
	\nonumber\\
	&
	+
	y^d_{ij}
	\Big[
	\overline{\mathbf{27}}_{+1,i}\,
	\mathbf{351}_{-1,j}\,
	\mathbf{650}_{0,H}^{\dagger}
	\Big]_{QH^\dagger d}
	\nonumber\\
	&
	+
	y^e_{ij}
	\Big[
	\mathbf{2430}_{-3,i}\,
	\mathbf{2430}_{+3,j}\,
	\mathbf{650}_{0,H}^{\dagger}
	\Big]_{LH^\dagger e}
	\nonumber\\
	&
	+
	y^\nu_{ij}
	\Big[
	\mathbf{2430}_{-3,i}\,
	\mathbf{2430}_{+3,j}\,
	\mathbf{650}_{0,H}
	\Big]_{LHN}
	+\mathrm{h.c.},
	\label{eq:ClassA_Yukawa_Lagrangian}
\end{align}
%
while the one for Class B is
\begin{align}
	\mathcal L_{\rm Yuk}^{B}
	\supset
	&
	\,y^u_{ij}
	\Big[
	\overline{\mathbf{351}}_{+1,i}\,
	\mathbf{351}_{-1,j}\,
	\mathbf{650}_{0,H}
	\Big]_{QHu}
	\nonumber\\
	&
	+
	y^d_{ij}
	\Big[
	\overline{\mathbf{351}}_{+1,i}\,
	\mathbf{351}_{-1,j}\,
	\mathbf{650}_{0,H}^{\dagger}
	\Big]_{QH^\dagger d}
	\nonumber\\
	&
	+
	y^e_{ij}
	\Big[
	\mathbf{2925}_{-3,i}\,
	\mathbf{2925}_{+3,j}\,
	\mathbf{650}_{0,H}^{\dagger}
	\Big]_{LH^\dagger e}
	\nonumber\\
	&
	+
	y^\nu_{ij}
	\Big[
	\mathbf{2925}_{-3,i}\,
	\mathbf{2925}_{+3,j}\,
	\mathbf{650}_{0,H}
	\Big]_{LHN}
	+\mathrm{h.c.}
	\label{eq:ClassB_Yukawa_Lagrangian}
\end{align}
The subscripts \(QHu\), \(QH^\dagger d\), \(LH^\dagger e\), and \(LHN\)
denote the projected Standard Model components selected from the full
\(E_6\) tensor contraction after the breaking chain.

Equations~\eqref{eq:ClassA_Yukawa_Lagrangian} and
\eqref{eq:ClassB_Yukawa_Lagrangian} should be understood as effective
projected Yukawa operators.  The full parent contractions certify that an
\(E_6\)-invariant cubic exists, while the low-energy Yukawa matrices
\(y^u_{ij}\), \(y^d_{ij}\), \(y^e_{ij}\), and \(y^\nu_{ij}\) are obtained
after projection onto the light chiral zero modes.  Once the Higgs doublet
in \(\mathbf{650}_{0,H}\) develops its electroweak VEV, these terms generate
the usual Dirac masses for quarks, charged leptons, and neutrinos.
After projection to the Standard Model zero modes this becomes
\begin{equation}
	\mathcal L_Y^{\rm SM}
	=
	y^u_{ij}Q_iH u^c_j
	+
	y^d_{ij}Q_iH^\dagger d^c_j
	+
	y^e_{ij}L_iH^\dagger e^c_j
	+
	y^\nu_{ij}L_iH N^c_j
	+\mathrm{h.c.}
\end{equation}
The sterile Majorana mass may be generated by a scalar singlet carrying
\(X=+6\):
\begin{equation}
	\mathcal L_N
	\supset
	\frac12 y^S_{ij}S_{+6}N_i^cN_j^c+\mathrm{h.c.}.
\end{equation}


\paragraph{Full ultraviolet $E_6 \times U(1)_X$ Lagrangian.} Collecting all contributions, the complete high--scale theory reads
\begin{equation}
		\mathcal L^{\rm UV}
		=
		\mathcal L_{\rm gauge}^{E_6\times U(1)_X}
			+
		\mathcal L_{\rm kin}^{E_6\times U(1)_X}
		+
		\mathcal L_{\rm kin}^{\rm ferm,\; proj}
		+
		\mathcal L_{\rm Yuk}^{E_6}	
		+
		\mathcal L_{\rm VL}
		+
		\mathcal L_{N}
		- 
		(V_{E_6}(\Phi)+V_{\Theta}(\Theta)).
\end{equation}

This Lagrangian defines the ultraviolet completion of the model.
Below the scale $v_{E_6}$ it matches onto the effective
$G(2) \times SU(3)_A$ theory with Higgs multiplets
$\chi \sim (\mathbf 7,\mathbf 1)$ and
$ \Omega \sim (\mathbf 1,\mathbf 8)$ responsible for the
subsequent breakings $ G(2) \to SU(3)_C$ and $SU(3)_A \to SU(2)_L \times U(1)_A$.

\subsection{Intermediate-Scale Effective Lagrangian: $G(2) \times SU(3)_A \times U(1)_X$}

Below the $E_6$ breaking scale $v_{E_6}$ but above \(M_\Theta\simeq M_\Omega\), the surviving gauge symmetry is $G(2) \times SU(3)_A \times U(1)_X$, with gauge couplings $g_A$ and $g_{G(2)}$.
We denote the gauge fields by
\begin{equation}
A_\mu = A_\mu^a t_A^a \qquad (a=1,\dots,8), 
\qquad
G_\mu = G_\mu^\alpha g^\alpha \qquad (\alpha=1,\dots,14),
\end{equation}
where $t_A^a$ are $SU(3)_A$ Gell-Mann generators and $g^\alpha$ are $G(2)$ generators in the appropriate representations (see \cite{Masi2021,Masi2024} for the explict forms).

\paragraph{Gauge sector.}
The intermediate-scale gauge kinetic terms are
\begin{equation}
	\mathcal L_{\rm gauge}^{\rm int}
	=
	-\frac14 G_{G(2)\,\mu\nu}^\alpha G_{G(2)}^{\alpha\mu\nu}
	-\frac14 F_{A\,\mu\nu}^aF_A^{a\mu\nu}
	-\frac14 C_{\mu\nu}C^{\mu\nu}.
\end{equation}
with field strengths
\begin{align}
	G_{G(2)\,\mu\nu}^\alpha
	&=
	\partial_\mu G_\nu^\alpha-\partial_\nu G_\mu^\alpha
	+
	g_{G(2)}f_{G(2)}^{\alpha\beta\gamma}
	G_\mu^\beta G_\nu^\gamma,
	\\
	F_{A\,\mu\nu}^a
	&=
	\partial_\mu A_\nu^a-\partial_\nu A_\mu^a
	+
	g_A f_A^{abc}A_\mu^bA_\nu^c,
	\\
	C_{\mu\nu}
	&=
	\partial_\mu C_\nu-\partial_\nu C_\mu .
\end{align}
The covariant derivative is
\begin{equation}
	D_\mu
	=
	\partial_\mu
	-
	ig_{G(2)}G_\mu^\alpha g^\alpha_{(R_G)}
	-
	ig_A A_\mu^a t^a_{(R_A)}
	-
	i\frac{g_X}{\sqrt{N_X}} X C_\mu .
\end{equation}

\paragraph{Intermediate Higgs sector.}
%
The lower-stage scalar kinetic terms are
\begin{equation}
	\mathcal L_{\rm scal,kin}^{\rm int}
	=
	\frac12(D_\mu\chi)^T(D^\mu\chi)
	+
	\mathrm{Tr}\!\left[(D_\mu\Omega)^\dagger(D^\mu\Omega)\right]
	+
	(D_\mu\Theta)^\dagger(D^\mu\Theta)
	+
	(D_\mu H)^\dagger(D^\mu H).
\end{equation}
Here
\begin{equation}
	\chi\sim(\mathbf7,\mathbf1)_0,
	\qquad
	\Omega\sim(\mathbf1,\mathbf8_A)_0,
	\qquad
	\Theta\sim(\mathbf1,\mathbf {10}_A)_{+3},
	\qquad
	H\sim(\mathbf1,\mathbf8_A)_{0}.
\end{equation}
The \(G(2)\)-breaking scalar is real:
\begin{equation}
	D_\mu\chi
	=
	\partial_\mu\chi
	-
	ig_{G(2)}G_\mu^\alpha g^\alpha_{\mathbf7}\chi .
\end{equation}
The \(SU(3)_A\)-breaking scalar \(\Omega\) is an adjoint. If written as a traceless \(3\times3\) matrix, its covariant derivative is
\begin{equation}
	D_\mu\Omega
	=
	\partial_\mu\Omega
	-
	ig_A[A_\mu,\Omega].
\end{equation}
The abelian-breaking scalar has the covariant derivative
\begin{equation}
	D_\mu\Theta
	=
	\left(
	\partial_\mu
	-
	ig_{G(2)}G_\mu^\alpha g^\alpha_{(\Theta)}
	-
	ig_A A_\mu^a t^a_{A,(\Theta)}
	-
	i\,\frac{g_X}{\sqrt{N_X}}\,X_\Theta C_\mu
	\right)\Theta ,
	\label{eq:DmuTheta_consistent}
\end{equation}
where \(X_\Theta\) is the integer \(E_7\)-derived charge.  The component acquiring the VEV is chosen to be a
\(G(2)\)-singlet and an \(SU(2)_L\)-singlet. Therefore, on the VEV direction, $g^\alpha_{(\Theta)}\langle\Theta\rangle=0$ and $t^a_{A,(\Theta)}\langle\Theta\rangle=0$, while the abelian charges satisfy $Y(\Theta)=0$.
The VEVs are $\langle\chi\rangle=v_\chi\,\hat s_7$, $	\langle\Omega\rangle=v_\Omega\,t_A^8$ and $\langle\Theta\rangle=v_\Theta$, with $v_\Theta\simeq v_\Omega$. The orthogonal abelian vector gets a mass $	m_{Z'}\simeq g_\perp v_\Theta$. 
Then a general renormalizable potential for the exotic Higgs scalars $\Phi$, $\chi$, $\Omega$ and $\Theta$ must be added. The related Section follows.

\section{The Exotic Multi-Higgs Sector}
The overall general compact renormalizable scalar potential for the theory involving the exotic Higgs particles related to $E_6$, $G(2)$, $SU(3)_A$ and $U(1)_X$ breakings (with $\lambda_2$ and $\lambda_3$ from $E_6$ potential vanishing) is 
\begin{equation}
	\begin{aligned}
		V & = m_\Phi^2|\Phi|^2+\lambda_\Phi|\Phi|^4 
		+ \frac12 m_{\chi}^2\,\chi^T\chi
		+\frac{\lambda_\chi}{4}\,(\chi^T\chi)^2
		+ m_ \Omega^2\,\mbox{Tr}( \Omega^\dagger \Omega)+ \kappa_\Omega\,\mathrm{Tr}(\Omega^3)+ \lambda_ \Omega\,[\mbox{Tr}( \Omega^\dagger \Omega)]^2 +
		\nonumber\\
		& m_\Theta^2\Theta^\dagger\Theta	+
		\lambda_\Theta(\Theta^\dagger\Theta)^2
		+ \underbrace{\kappa_{\chi \Omega}(\chi^T\chi)\mbox{Tr}( \Omega^\dagger \Omega)}_{\text{only link between }G(2)\text{ and }SU(3)_A} +\; \kappa_{\Phi\chi}|\Phi|^2(\chi^T\chi)
		+\kappa_{\Phi \Omega}|\Phi|^2\mbox{Tr}( \Omega^\dagger \Omega) + 
		\nonumber\\
		& \kappa_{\Phi\Theta} |\Phi|^2 (\Theta^\dagger\Theta) + \kappa_{\Theta\chi} (\Theta^\dagger\Theta)( \chi^T \chi) + \kappa_{\Theta\Omega} (\Theta^\dagger\Theta)\mbox{Tr}( \Omega^\dagger \Omega)
	\end{aligned}
\end{equation}
with $m_i<0$, plus representation specific contractions and allowed cubic terms. All cross-couplings are gauge-invariant and quantify the only renormalizable communication among sectors because of the absence of bifundamental features (a scalar field is called bifundamental if it transforms non-trivially under two gauge groups simultaneously, and no bifundamental fields are present due to the choice of $\langle \Phi_{ \mathbf{650}}^{(\mathbf 1,\mathbf 1)} \rangle$) \cite{Sher1989Vacuum,Buttazzo2013Vacuum}: they are described by the three portal quartics $\kappa_{\chi \Omega},\kappa_{\Phi\chi},\kappa_{\Phi \Omega}$ and the three for $\Theta$. In particular, $\kappa_{\chi \Omega}$ is the only soft link between the two \textit{progenitors} groups of the SM. Although \(\Theta\) descends from a \(G(2)\)-charged branch, its VEV is chosen
along the \(SU(3)_C\)-singlet component after \(G(2)\to SU(3)_C\) and it does not break color. Direct communication with the \(G(2)\) dark sector is
possible through heavy-threshold interactions and through the portal \((\Theta^\dagger\Theta)(\chi^T\chi)\), but these couplings must be small or
high-scale, preserving the seclusion of the dark sector.
We take the aforementioned VEVs 
and the CP-even radial masses (before portal mixing) are: 
\begin{equation}
	m_{h_\chi}^2=2\lambda_\chi v_\chi^2,\qquad
	m_{h_\Omega}^2=2\lambda_ \Omega v_ \Omega^2,\qquad
	m_{h_\Theta}^2=2\lambda_ \Theta v_ \Theta^2
\end{equation}
The three Higgs multiplets live at parametrically separated scales: 
\begin{equation}
	\langle \Phi_{650} \rangle \equiv v_\Phi \gg 
	\langle \chi \rangle \equiv v_\chi \gg 
	\langle  \Omega \rangle \equiv v_ \Omega \sim v_\Theta \gg v_{\rm EW}
\end{equation}
where $v_{\rm EW}$ is the SM Higgs VEV. This hierarchy ensures that heavy states decouple efficiently and that the SM limit is recovered. Moreover, the portal couplings to the light Higgs direction must be small or tuned so that the electroweak doublet remains light:
\begin{equation}
|\lambda_{\Theta H}|,\ |\lambda_{\Omega H}|,\ |\lambda_{\chi H}|\ll 1,
\end{equation}
or else the scalar mass matrix must have a protected light eigenstate.

\subsection{Singlet--singlet mixings}

Expanding the Higgs fields about their minima \cite{ModernPP}
\begin{equation}
\Phi=v_\Phi+h_\Phi,\qquad
\chi=v_\chi+h_\chi,\qquad
\Omega=v_\Omega+h_\Omega,\qquad
\Theta=\frac{1}{\sqrt2}(v_\Theta+h_\Theta+i\eta_\Theta),
\end{equation}
the CP-even scalar mass matrix in the basis $(h_\Phi,h_\chi,h_\Omega,h_\Theta)$ and the relative mixing angles are \cite{GunionHaber2003Decoupling,WeinbergQTF2}
\begin{equation}
\mathcal M^2_{\rm rad}=
\begin{pmatrix}
	4\lambda_\Phi v_\Phi^2 &
	2\kappa_{\Phi\chi}v_\Phi v_\chi &
	2\kappa_{\Phi\Omega}v_\Phi v_\Omega &
	2\kappa_{\Phi\Theta}v_\Phi v_\Theta
	\\
	2\kappa_{\Phi\chi}v_\Phi v_\chi &
	4\lambda_\chi v_\chi^2 &
	2\kappa_{\chi\Omega}v_\chi v_\Omega &
	2\kappa_{\chi\Theta}v_\chi v_\Theta
	\\
	2\kappa_{\Phi\Omega}v_\Phi v_\Omega &
	2\kappa_{\chi\Omega}v_\chi v_\Omega &
	4\lambda_\Omega v_\Omega^2 &
	2\kappa_{\Omega\Theta}v_\Omega v_\Theta
	\\
	2\kappa_{\Phi\Theta}v_\Phi v_\Theta &
	2\kappa_{\chi\Theta}v_\chi v_\Theta &
	2\kappa_{\Omega\Theta}v_\Omega v_\Theta &
	4\lambda_\Theta v_\Theta^2
\end{pmatrix},\qquad \tan2\theta_{ij}=\frac{2(\mathcal M^2_{\rm rad})_{ij}}{(\mathcal M^2_{\rm rad})_{jj}-(\mathcal M^2_{\rm rad})_{ii}}.
\end{equation}
The smallness of the portal couplings ensures that each Higgs remains approximately aligned with its own scale \cite{Slansky1981}. For example, the portal $\kappa_{\chi \Omega}$ mixes $\chi$ and $ \Omega$ related to the subgroups of $E_6$ according to 
\begin{equation}
 \tan2\theta_{\chi \Omega}=\frac{4\kappa_{\chi \Omega} v_\chi v_ \Omega}{4\lambda_ \Omega v_ \Omega^2-4\lambda_\chi v_\chi^2}.
\end{equation}
For a correct hierarchy of the VEVs, this angle is naturally tiny and the two sectors effectively decouple. 
Another relevant mixing angle is 
\begin{equation}
\tan 2\theta_{\chi\Theta}
\simeq
\frac{4\kappa_{\chi\Theta}v_\chi v_\Theta}
{4\lambda_\Theta v_\Theta^2-4\lambda_\chi v_\chi^2}.
\end{equation}
The portal \(\kappa_{\chi\Theta}\) is especially important because it controls communication between the abelian-breaking sector and the \(G(2)\)-breaking
sector. Dark-sector seclusion requires
\begin{equation}
|\theta_{\chi\Theta}|\ll1,
\end{equation}
or equivalently
\begin{equation}
|\kappa_{\chi\Theta}|
\ll
\frac{|m^2_{h_\chi}-m^2_{h_\Theta}|}{2v_\chi v_\Theta}.
\end{equation}

\subsection{Doublet--doublet mixing}

The electroweak Higgs doublet $H$ can in principle mix with additional doublets from the other Higgs \cite{Slansky1981}.
In our best solutions, the light Standard Model Higgs
uses the
adjoint doublet
\begin{equation}
H\subset(\mathbf1,\mathbf8_A)_0\subset\mathbf{650}_{0,H},
\end{equation}
and the low-energy theory is assumed to contain only one light
doublet
\begin{equation}
H\sim(\mathbf1,\mathbf2)_{+1/2}.
\end{equation}
All other doublet fragments in the \(\mathbf{650}_{0,H}\), \(\Omega\), or
\(\Theta\)-related sectors are lifted at the high scale.
If additional heavy doublets \(H_I\) exist, their mass matrix has the schematic
form
\begin{equation}
\mathcal M_D^2
=
\begin{pmatrix}
	m_H^2 & \delta_{H_I}^2 \\
	\delta_{H_I}^2 & M_I^2
\end{pmatrix},
\qquad
M_I^2\sim M_\Omega^2,M_\chi^2,M_{E_6}^2.
\end{equation}
The doublet mixing angle is
\begin{equation}
\theta_{HH_I}\simeq\frac{\delta_{H_I}^2}{M_I^2}.
\end{equation}
Portal-induced entries are typically of order
\begin{equation}
\delta_{H_I}^2\sim \kappa_{H_I}v_{\rm EW}V,
\end{equation}
where \(V\) is a heavy VEV and $\kappa_{H_I}$ the heavy doublet coupling. Therefore
\begin{equation}
\theta_{H H_I}
\sim
\frac{\kappa_{H_I}v_{\rm EW}}{V}
\ll1
\end{equation}
for \(V\gtrsim10^{13}\,\mathrm{GeV}\) and perturbative portal couplings. Thus the observed Higgs boson remains SM-like. Indeed, if only one $SU(2)_L$ doublet $H\sim(\mathbf1,\mathbf2)_{+1/2}$ is light (our default), there is no doublet--doublet mixing.

\subsection{Triplet--doublet mixing and custodial constraints}
The $ \Omega$ contains an $SU(2)_L$ triplet $\Delta\sim(\mathbf1,\mathbf3)_{0}$. The hypercharge of this triplet is zero.
Therefore the usual coupling to the Higgs doublet $H$ \cite{ChunKimLee2003,CaiHanLiZhang2017}
\begin{equation}
	V \supset \mu\, H^T i\sigma_2 \Delta H + \text{h.c.}
\end{equation}
with $\mu$ the dimensionful mixing parameter, is not allowed because it requires $Y=+1$. Such a term would induces a triplet VEV $v_\Delta \sim \mu v_{EW}^2/m_\Delta^2$ after the electroweak symmetry breaking (EWSB) \cite{ChunKimLee2003}, with $m_\Delta$ the physical mass of the triplet, which violates custodial symmetry and shifts the $\rho$ parameter \cite{ModernPP,PeskinTakeuchi1990,PeskinTakeuchi1992} 
\begin{equation}
	\rho -1 \simeq \frac{2 v_\Delta^2}{v_{EW}^2}.
	\end{equation}
Instead, the leading custodial-sensitive interaction
is of the schematic form
\begin{equation}
V\supset
\mu_0\,H^\dagger \Delta_0 H,
\end{equation}
where $\mu_0$ is the trilinear scalar coupling between the SM Higgs doublet and a heavy neutral electroweak triplet component.
If present, it induces a neutral triplet VEV
\begin{equation}
v_{\Delta_0}\sim\frac{\mu_0 v_{\rm EW}^2}{m_{\Delta_0}^2}.
	\end{equation}
which could shifts the \(\rho\) parameter.
Experimental constraints require $v_{\Delta_0} \lesssim \mathcal{O}(1~{\rm GeV})$ \cite{PDG2024}, implying that the triplet must be extremely heavy or very weakly coupled, which is satisfied by the present construction. In fact, for a $m_{\Delta_0} \sim 10^{13}~\text{GeV}$ (hereafter we adopt the natural units simplification not to report the $c^2$ denominator), according to the mass scale $M_ \Omega \sim 10^{13}$ described in Section 9.5,
\begin{equation}
|v_{\Delta_0}|
\sim
6\times10^{-22}
\left(\frac{\mu_0}{1\,\mathrm{GeV}}\right)\,\mathrm{GeV}.
\end{equation}
Even for \(\mu_0\sim10^{12}\,\mathrm{GeV}\), the induced VEV is only $|v_{\Delta_0}|\sim10^{-9}\,\mathrm{GeV}$,
far below the custodial bound.

\subsection{Gauge--scalar mixing: heavy vector mixing with $Z/W$}
Gauge--scalar mixing refers to the fact that the would-be Goldstone modes of $\chi$, $ \Omega$ and $\Theta$ are absorbed by the heavy gauge bosons related to the broken generators \cite{Slansky1981}.
Physical Higgs fields remain orthogonal to these Goldstones, but small mixing could appear in the mass eigenstates whenever portal couplings are nonzero.
Broken generators from $SU(3)_A$ form two complex $SU(2)_L$ doublet vectors with mass $m_Y=g_A v_ \Omega$. Secondly, \(\Theta\) gives mass to the abelian vector orthogonal to hypercharge with mass
\begin{equation}
m_{Z'}^2
=
(3g_A^2+\frac34g_X^2)v_\Theta^2=g_\perp^2v_\Theta^2.
\end{equation} 
After EWSB there could be a tiny mixing with $W/Z$. The mass matrix for the light EW $W_\mu$ and heavy $Y_\mu$ bosons takes the schematic form \cite{Langacker2009Zprime,LangackerSankar1989WRMixing,ErlerLangackerMunirRojas2009ZprimeConstraints}
\begin{equation}
	\mathcal{M}^2 =
	\begin{pmatrix}
		g^2 v_{EW}^2 & g g_A v_{EW}^2 \\
		g g_A v_{EW}^2 & m_Y^2
	\end{pmatrix},
\end{equation}
where $g$ is the usual $SU(2)_L$ coupling of the SM. For $m_Y^2 \gg g^2 v_{EW}^2$, the mixing angle is \cite{Langacker2009Zprime,LangackerSankar1989WRMixing,ErlerLangackerMunirRojas2009ZprimeConstraints}
\begin{equation}
	\theta_{\text{gauge}_Y}
	\simeq
	\frac{g g_A v_{EW}^2}{m_Y^2},
\end{equation}
while for the heavy abelian boson
\begin{equation}
\theta_{\text{gauge}_{Z'}}
\sim
\frac{g_Z g_\perp v_{\rm EW}^2}{m_{Z'}^2}.
\end{equation}
Since $m_{Y},m_{Z'}\lesssim M_\Omega\sim10^{13}\ {\rm GeV}$ in our benchmark scenario in the next section, so that $v_ \Omega\gg v_{EW}$, both mixings are utterly negligible, \textit{i.e.} $\theta_{\rm gauge}\ll10^{-20}$)

\subsection{Standard Model Higgs coupling}
After electroweak symmetry breaking, portal terms generate off-diagonal entries in the CP-even scalar mass matrix of order
\begin{equation}
	m_{H\sigma}^2 \sim \kappa_{H\sigma}\, v_{EW}\, V_\sigma ,
	\qquad \sigma\in\{\Phi,\chi,\Omega,\Theta\},
\end{equation}
where $\kappa_{H\sigma}$ is the portal coefficient, $v_{EW}\simeq 246~\mathrm{GeV}$ and $V_\sigma\in\{v_\Phi,v_\chi,v_\Omega,v_\Theta\}$. The heavy radial masses are set by quartics, \textit{i.e.} $m_{h_\Phi}^2 = 2\lambda_\Phi v_\Phi^2,	m_{h_\chi}^2 = 2\lambda_\chi v_\chi^2, m_{h_ \Omega}^2 = 2\lambda_ \Omega v_ \Omega^2, m_{h_\Theta}^2 = 2\lambda_\Theta v_\Theta^2$, hence the mixing angles scale as \cite{GunionHaber2003Decoupling,BrivioTrott2019}
\begin{equation}
	\theta_{H\sigma}\sim \frac{m_{H\sigma}^2}{m_{h_{\sigma}}^2}
	\sim
	\frac{\kappa_{H\sigma}}{2\lambda_\sigma}\frac{v_{EW}}{V_\sigma}
	\ll 10^{-10}\qquad \text{for}\quad V_\sigma\gtrsim 10^{13}\text{ GeV and } \kappa_{H\sigma}, \lambda_\sigma \lesssim \mathcal O(1).
\end{equation}
Therefore the SM-like Higgs couplings are preserved and portal-induced leakage into the $G(2)$ sector is strongly suppressed.

\section{Gauge Coupling Running and Exceptional Unification Scenario}
\label{sec:running}

A central consistency requirement of the present exceptional framework is that gauge couplings evolve in a physically consistent way across
the multiple symmetry-breaking thresholds \cite{Buras1998} \cite{Caprini2016Review,caprini20}\cite{Cline2018PT} \cite{ChengLi1980} \cite{MazumdarWhite2019} \cite{EllisLewickiNo2020}.
In this section we provide a detailed discussion of the renormalization group running, the matching conditions at each breaking scale and the resulting
benchmark unification pattern.
A crucial point is that the large \(E_6\) representations in Classes A and B are parent representations.  They certify the existence of the required zero-mode branches and the renormalizable Yukawa ancestors, but they are not included as complete light perturbative multiplets in the renormalization-group evolution.  This is forced by the partial-wave unitarity diagnostic: complete light \(\mathbf{2430}\) and \(\mathbf{2925}\) multiplets would give \(a_0^{\max}\sim4\), already far above \(|\mathrm{Re}\,a_0|\leq1/2\).  Thus the running below the exceptional completion scale is the running of the projected effective spectrum, not of the complete large parent representations.
Therefore the renormalization-group analysis must distinguish two different interpretations.
First, in the projected-bulk interpretation, the large \(E_6\) representations are algebraic parents only.  Their unwanted components are lifted, localized, confined, or projected out at threshold, and only a controlled effective spectrum contributes to the bulk running.  Second, in the active-ancestor interpretation, the actual \(G(2)\times SU(3)_A\) branches from which the SM fields descend are treated as propagating multiplets above \(M_\Omega\).  This second interpretation is the literal four-dimensional field-theory running of the SM ancestors, but it makes the intermediate theory non-perturbative before the desired \(G(2)\) threshold is reached.

\subsection{General setup and matching conditions at symmetry-breaking scales}

We define the gauge couplings as usual as $\alpha_i(\mu) \equiv \frac{g_i^2(\mu)}{4\pi}$, with generators normalized as $\mathrm{Tr}(T^a T^b)=\frac12\delta^{ab}$ for all simple groups, following the conventions used in \cite{Masi2024}.

At one loop, the running of the inverse couplings is governed by \cite{MachacekVaughn1983,PeskinSchroeder1995,MachacekVaughn1984}
\begin{equation}
	\frac{d}{d\ln\mu}\left(\frac{1}{\alpha_i}\right)
	= \frac{b_i}{2\pi},
	\label{eq:beta_general}
\end{equation}
where the beta-function coefficients are \cite{WeinbergQTF2}
\begin{equation}
	b=\frac{11}{3}\,C_2(G)\;-\;\frac{2}{3}\sum_{\text{Weyl }f}T(R_f)\;-\;\frac{1}{3}\sum_{\text{complex }s}T(R_s).
\end{equation}
where the sums run over left-handed Weyl fermions and complex scalars (a real scalar counts as half a complex scalar). Here $C_2(G)$ is the quadratic Casimir invariant of the adjoint representation $G$ of the group and $T(R_f)$ and $T(R_s)$ are the Dynkin indexes of the
fundamental irreducible $R$ representation of the group for fermions and scalars, respectively.

The running is piecewise, reflecting the staged symmetry breaking: \textbf{exceptional regime} ($\mu>M_\chi$) -- gauge group $G(2) \times SU(3)_A \times U(1)_X$; \textbf{intermediate regime} ($M_ \Omega<\mu<M_\chi$) -- gauge group $SU(3)_C\times SU(3)_A \times U(1)_X$; \textbf{SM regime} ($\mu<M_ \Omega$) -- gauge group $SU(3)_C\times SU(2)_L\times U(1)_Y$.
In each regime, only the fields lighter than the corresponding threshold contribute
to the beta functions.\\

At the breaking scale $\mu=M_ \Omega \simeq M_ \Omega$, the group $SU(3)_A \times U(1)_X$ is broken to the electroweak group. 
For the benchmark matching
\begin{equation}
	g_X(M_\Omega)\simeq g_A(M_\Omega),
\end{equation}
one obtains
\begin{equation}
	\frac{g_A^2}{g_Y^2}
	=
	\frac53.
\end{equation}
Therefore
\begin{equation}
	g_1(M_\Omega)
	=
	\sqrt{\frac53}\,g_Y(M_\Omega)
	=
	g_A(M_\Omega),
	\qquad
	g_2(M_\Omega)=g_A(M_\Omega),
	\label{eq:final_matching_g1_g2_gA}
\end{equation}
where we call $g_2$ the usual SM coupling $g$ associated with the weak isospin gauge group and \(g_1\) is the GUT-normalized hypercharge coupling.  Equivalently,
\begin{equation}
	\alpha_1(M_\Omega)
	=
	\alpha_2(M_\Omega)
	=
	\alpha_A(M_\Omega).
\end{equation}
This guarantees that the $SU(2)_L$ coupling intersects precisely at the $SU(3)_A$ breaking scale, as shown in Figure 1. This gives the canonical high-scale value
\begin{equation}
	\sin^2\theta_W(M_\Omega)
	=
	\frac{g_Y^2}{g_2^2+g_Y^2}
	=
	\frac38.
\end{equation}
In addition, at the scale $\mu=M_\chi$, the special embedding $SU(3)_C\subset G(2)$ implies for the QCD $g_3$ coupling 
\begin{equation}
	g_3(M_\chi)=g_{G(2)}(M_\chi),
	\label{eq:G2_matching}
\end{equation}
with no additional normalization factors.
The final hierarchy is
\begin{equation}
	M_Z<M_\Omega\simeq M_\Theta<M_\chi<M_{E_6}\lesssim M_{E_7}.
\end{equation}
The scale \(M_\Omega\) is fixed by the one-loop crossing
using
\begin{equation}
	\alpha_{\rm em}^{-1}(M_Z)=127.955,
	\qquad
	\sin^2\theta_W(M_Z)=0.23122,
	\qquad
	\alpha_3(M_Z)=0.1179.
\end{equation}
This gives
\begin{equation}
	\alpha_1^{-1}(M_Z)=59.02,
	\qquad
	\alpha_2^{-1}(M_Z)=29.59,
	\qquad
	\alpha_3^{-1}(M_Z)=8.48,
\end{equation}
and hence
\begin{equation}
	M_\Omega\simeq1.03\times10^{13}\ {\rm GeV},
	\qquad
	\alpha_A^{-1}(M_\Omega)
	=
	\alpha_2^{-1}(M_\Omega)
	\simeq42.41.
\end{equation}

\subsection{Beta functions in the exceptional regime}
Above \(M_\chi\), the gauge group is
\begin{equation}
	G(2)\times SU(3)_A\times U(1)_X.
\end{equation}
The final Class A/B parent representations are not propagated as complete light multiplets.  Therefore the beta functions are not computed by inserting complete \(\overline{\mathbf{27}}\), \(\mathbf{351}\), \(\mathbf{2430}\), or \(\mathbf{2925}\) representations.  Instead, the running should be controlled by the projected effective spectrum: the \(G(2)\)-breaking scalar \(\chi\), the \(SU(3)_A\)-breaking scalar \(\Omega\), the possible active Higgs ancestor and a controlled set of projected spectators.

In an ordinary four-dimensional bulk theory, the branches from which the SM fields descend must contribute to both the beta functions and the coupled-channel unitarity matrix.  It is
therefore useful to test the selected \(G(2)\times SU(3)_A\) ancestor branches directly, independently of the complete \(E_6\) parent representations.
Using the Milagre--Lavoura coupled-channel estimate \cite{MILAGRE2024116542} described before one finds, for three families and \(g_A\simeq0.54\),
\begin{equation}
(a_0)_{G(2)}^{A}\simeq0.28,
\qquad
(a_0)_{SU(3)_A}^{A}\simeq0.74.
\end{equation}
For the representative Class B active branches one obtains
\begin{equation}
(a_0)_{G(2)}^{B}\simeq0.20,
\qquad
(a_0)_{SU(3)_A}^{B}\simeq (0.71_{u^c\subset (\mathbf7,\mathbf{15}_A)},0.36_{u^c\subset (\mathbf7,\mathbf3_A)}).
\end{equation}
The \(G(2)\) channel is perturbatively acceptable, but the \(SU(3)_A\)
channel already exceeds the perturbative bound $|{\rm Re}\,a_0|\leq {1\over2}$ in the $u^c\subset (\mathbf7,\mathbf{15}_A)$ case.
Thus the selected SM-ancestor branches are far less dangerous than complete large \(E_6\) parents, but they still cannot be treated as an ordinary
three-family perturbative bulk spectrum over the full interval $M_\Omega<\mu<M_{E_6}$.
This conclusion is reinforced by the beta functions.  Counting the same active branches gives
\begin{equation}
b_A^{A,\mathrm{act}}\simeq -228,
\qquad
b_A^{B,\mathrm{act}}\simeq (-199_{u^c\subset (\mathbf7,\mathbf{15}_A)},-65_{u^c\subset (\mathbf7,\mathbf3_A)}), 
\end{equation}
so that the full product branches cannot be all treated as active perturbative bulk multiplets.  In the projected-running interpretation they are zero-mode ancestors rather than complete
propagating multiplets, and their contributions are removed from both \(b_A\) and \(b_{G(2)}\) above the projection/lifting threshold.


\paragraph{$G(2)$ coefficient $b_{G(2)}$.}
For scalars, only $\chi$ is $G(2)$-charged, which is responsible for $G(2)\to SU(3)_C$; here we treat $\chi$ as a real scalar in the $\mathbf7$ of $G(2)$ which contributes as one-half of a complex $\mathbf7$: 
\begin{equation}
	\sum_s T_{G(2)}(R_s)=\frac12\,T_{G(2)}(\mathbf7)=\frac12,
\end{equation}
with $T_{G(2)}(\mathbf7)=1$. Using
\begin{equation}
	C_2(G(2))=4,
\end{equation}
one obtains
\begin{equation}
	b_{G(2)}
	=
	\frac{11}{3}C_2(G(2))
	-
	\frac{1}{3}\left(\frac12\,T_{G(2)}(\mathbf7)\right).
\end{equation}
Therefore
\begin{align}
	b_{G(2)}
	&=
	\frac{44}{3}
	-
	\frac{1}{6}
	\nonumber\\
	&=
	\frac{29}{2} = 14.5.
	\label{eq:bG2_final_projected}
\end{align}
The positive sign reflects the strong asymptotic freedom of the projected \(G(2)\) sector.

\paragraph{$SU(3)_A$ coefficient $b_A$.}
For \(SU(3)_A\), we include one $SU(3)_A$-charged adjoint scalar $\Omega\sim\mathbf8_A$ with $T(\mathbf8)=3$:
\begin{equation}
	\sum_s T(R_s)=\frac12\,T(\mathbf8)=\frac32.
\end{equation}
Using $C_2(SU(N))=N$, one has
\begin{equation}
	b_A
	=
	11
	-
	\frac23\sum_f T_A(R_f)
	-
	\frac12.
\end{equation}
To soften $b_A$ slope and obtain a coherent meeting point with $b_{G(2)}$, one should notice that the Class B solution naturally contains \(G(2)\)-singlet, \(SU(3)_A\)-charged SM leptonic branches,
\begin{equation}
L:\;(\mathbf1,\mathbf8_A)_{-3},
\qquad
e^c:\;(\mathbf1,\mathbf8_A)_{+3}.
\end{equation}
Since these branches are \(G(2)\)-singlets, they contribute to \(b_A\) but not to \(b_{G(2)}\). 
This pair has
\begin{equation}
I_A^{8+8}
=
T(\mathbf8)+T(\mathbf8)
=
6.
\end{equation}
Therefore
\begin{equation}
b_A
=
11-\frac23(6)-\frac12
=
\frac{13}{2}
= 6.50.
\end{equation}
The Higgs \(H=(\mathbf1,\mathbf8_A)_0\) is a \(G(2)\)-singlet so, if its full \(SU(3)_A\) adjoint parent fragment remains active, it only slightly shifts
\begin{equation}
\Delta b_A=-{1\over 6}T_A(\mathbf8)=-{1\over2},
\qquad
\Delta b_{G(2)}=0.
\end{equation}
The scale \(M_\chi\) of this exceptional regime is not fixed by the coupling intersection.  Once \(M_\chi\) is specified, the
intersection of \(\alpha_A\) and \(\alpha_{G(2)}\) determines the effective \(E_6\) completion scale, according to
\begin{equation}
M_{E_6}
=
M_\chi
\exp\left[
\frac{
	2\pi\left(
	\alpha_A^{-1}(M_\chi)
	-
	\alpha_{G(2)}^{-1}(M_\chi)
	\right)
}{
	b_{G(2)}-b_A
}
\right].
\end{equation}
 Conversely, if one demands a target
\(E_6\)-meeting scale, the same equation fixes the required \(M_\chi\).
For the Class-B spectator slope \(b_A=13/2\) and
\(b_{G(2)}=29/2\), we adopt the benchmark choice \(M_\chi=10^{14}\,\mathrm{GeV}\) to separate by about one order of magnitude in energy the breaking scales of $SU(3)_A$ and $G(2)$, obtaining
\begin{equation}
M_{E_6}\simeq6.9\times10^{15}\,\mathrm{GeV}.
\end{equation}
Demanding \(M_{E_6}=10^{16}\,\mathrm{GeV}\) instead gives
\begin{equation}
M_\chi\simeq1.5\times10^{14}\,\mathrm{GeV}.
\end{equation}
Thus \(M_\chi\) and \(M_{E_6}\) are linked by the running, but one of them must be chosen as a physical threshold input.

If we keep also the branch $\Theta\sim(\mathbf1,\mathbf{10}_A)_{+3}$, for \(SU(3)_A\) active, with $T_A(\mathbf{10})=\frac{15}{2}$, the resulting shift is
\begin{equation}
	\Delta b_A^\Theta
	=
	-\frac13 T_A(\mathbf{10})
	=
	-\frac13\cdot\frac{15}{2}
	=
	-\frac52.
	\label{eq:Theta_shift_bA}
\end{equation}
This lowers the \(SU(3)_A\)-\(G(2)\) meeting scale to approximately
\begin{equation}
	M_{E_6}^{\Theta\text{-active}}
	\simeq1.5\times10^{15}\,\mathrm{GeV},
	\qquad
	\alpha_{E_6}^{-1}\simeq45.6,
	\qquad
	g_{E_6}\simeq0.525.
	\label{eq:Theta_active_meeting}
\end{equation}
An active \(\Theta\) is therefore not catastrophic, but it noticeably lowers
the exceptional meeting scale.  For the cleaner
\(M_{E_6}\sim10^{16}\,\mathrm{GeV}\) benchmark, \(\Theta\) should be regarded
as threshold-localized near \(M_\Theta\simeq M_\Omega\), with its effect absorbed into finite matching corrections rather than included as a long-distance contribution to the \(SU(3)_A\) beta function. In this case it is not included as a long-distance contribution to the \(SU(3)_A\) or \(U(1)_X\) beta functions between \(M_\Omega\) and \(M_{E_6}\).\\

The \(U(1)_X\) factor is hereafter treated in the threshold-localized approximation:
\begin{equation}
	b_X=0.
\end{equation}
This does not mean that a long-lived unbroken \(U(1)_X\) gauge theory with no charged states exists, rather the \(X\)-charged exotic branches and the \(\Theta\) sector are localized near $M_\Theta\simeq M_\Omega$ and are integrated out as threshold corrections.  

\subsection{Intermediate Regime: $SU(3)_C\times SU(3)_A \times U(1)_X$ \; (for $M_ \Omega<\mu<M_\chi$)}
Below $M_\chi$, the $G(2)$ gauge group is broken to $SU(3)_C$; the matter charged under color is obtained by decomposing the $G(2)$ fundamental
$\mathbf7\to \mathbf3\oplus \bar{\mathbf3}\oplus \mathbf1$ \cite{Slansky1981}.
If some \(X\)-charged spectators were kept light above \(M_\Omega\), then
\begin{equation}
	b_X=
	-\frac{2}{3}\sum_{\rm Weyl\ f}X_f^2
	-\frac{1}{3}\sum_{\rm complex\ s}X_s^2
\end{equation}
in the present inverse-coupling convention would no longer vanish; equivalently the usual beta coefficient for \(g_X\) would be positive and \(\alpha_X^{-1}\) would decrease toward the ultraviolet. This is the Landau-pole-prone regime \cite{ModernPP} that the projected construction avoids.

The \(SU(3)_A\) coefficient is the only place where the projected spectator content matters. If all large parent branches were removed, the running would be close to the pure-gauge value, up to the real adjoint scalar \(\Omega\). If instead, as discussed before complete parent multiplets were retained, the theory would become non-perturbative and would violate partial-wave unitarity. The final benchmark therefore uses an intermediate projected effective slope, that can remain the same as long as the $SU(3)_A$-charged spectrum is unchanged; in the setup with $ \Omega$ still present and the Class B $G(2)$-singlet spectator pair, one obtains again
\begin{equation}
	b_A=\frac{13}{2}\qquad (M_ \Omega<\mu<M_\chi).
\end{equation}
The color coefficient $b_{3C}$ can be derived decomposing the fermion content into $SU(3)_C$ representations (the explicit $SU(3)_C$ counting is given in Appendix~\ref{app:colorcount}).

\subsection{SM Regime ($\mu<M_ \Omega$)}
Below $M_ \Omega$, the theory reduces to the SM. In GUT normalization for $U(1)_Y$ \cite{Slansky1981},
\begin{equation}
	(b_1,b_2,b_3)=\left(-\frac{41}{10},\;\frac{19}{6},\;7\right),
\end{equation}
consistent with our sign convention.\\

\begin{figure}[t]
	\centering
	\includegraphics[width=0.95\linewidth]{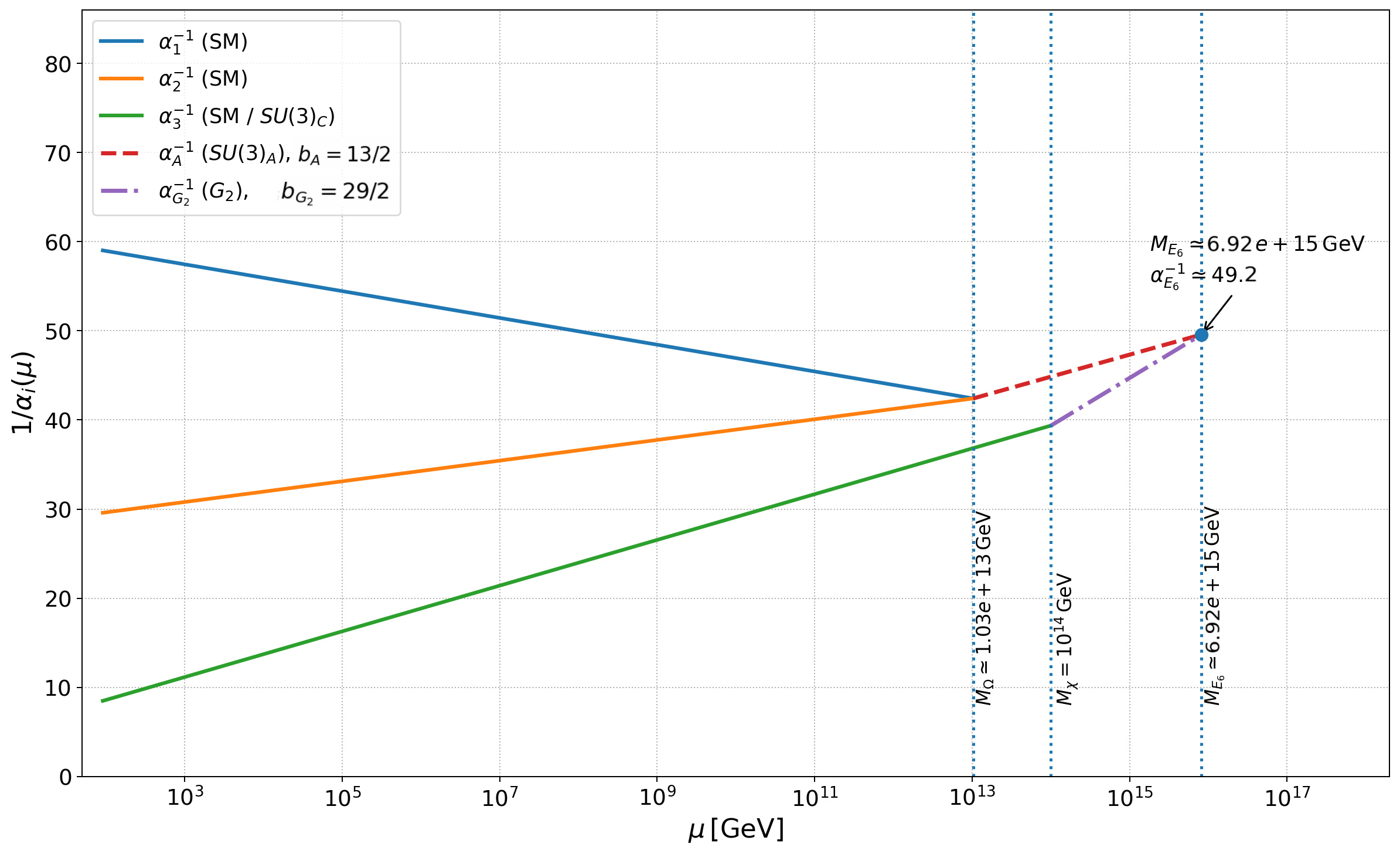}
	\caption{One-loop running of gauge couplings with exact piecewise solutions. The legend labels the inverse gauge couplings in the SM and in the additional $G(2)$ and $SU(3)_A$ sectors. The blue dot marks the $E_6$ meeting point (values shown). Vertical dashed lines indicate the approximate matching thresholds $M_ \Omega$, $M_\chi$ and $M_{E_6}$for the gauge phase transitions, where the conditions are imposed as $\alpha_1=\alpha_A=\alpha_2$ at $M_ \Omega$ and $\alpha_{G(2)}=\alpha_3$ at $M_\chi$.} 
	\label{fig:runningB}
\end{figure}

\subsection{Intermediate exceptional unification}

The benchmark scenario is defined by a comparatively low \(G(2)\)-breaking scale, with the VEV-defined scales
\begin{equation}
	M_\Omega\simeq M_\Theta\simeq 1.03\times10^{13}\,\mathrm{GeV},
	\qquad
	M_\chi\simeq 10^{14}\,\mathrm{GeV}.
\end{equation}
Above \(M_\chi\), the \(SU(3)_A\) and \(G(2)\) couplings run and intersect at
\begin{equation}
	\mu_{\rm meet}\simeq 6.9\times10^{15}\,\mathrm{GeV},
	\qquad
	\alpha_{\rm meet}^{-1}\simeq49.2,
	\qquad
	g_{\rm meet}=\sqrt{4\pi\alpha_{\rm meet}}\simeq0.51,
\end{equation}
well below the Planck scale. The masses of the heavy gauge bosons generated at symmetry breaking are
\begin{equation}
	m_Y = g_A(M_\Omega)\,v_\Omega \simeq g_A(M_\Omega)M_\Omega,
	\qquad
	m_{\mathcal X}=g_{G(2)}(M_\chi)\,v_\chi\simeq g_{G(2)}(M_\chi)M_\chi.
\end{equation}
At the \(SU(3)_A\)-breaking scale one finds
\begin{equation}
	\alpha_A^{-1}(M_\Omega)\simeq42.4,
	\qquad
	g_A(M_\Omega)\simeq0.54,
\end{equation}
and therefore
\begin{equation}
	m_Y\simeq 5.6\times10^{12}\,\mathrm{GeV}.
\end{equation}
Placing the \(\mathbf{650}\) breaking at the meeting point, so that $\langle\Phi_{\mathbf{650}}\rangle\equiv M_{E_6}=\mu_{\rm meet}$, the \(E_6\) completion unifies \(G(2)\) and \(SU(3)_A\) with a common one-loop coupling
\begin{equation}
\alpha_{E_6}^{-1}\simeq49.2.
\end{equation}
At the \(G(2)\)-breaking scale,
\begin{equation}
	\alpha_3(M_\chi)=\alpha_{G(2)}(M_\chi)\simeq0.0254,
	\qquad
	g_{G(2)}(M_\chi)=\sqrt{4\pi\alpha_{G(2)}(M_\chi)}\simeq0.565.
\end{equation}
Thus the mass of the heavy dark-sector vector bosons is
\begin{equation}
	m_{\mathcal X}
	=
	g_{G(2)}(M_\chi)M_\chi
	\simeq
	5.7\times10^{13}\,\mathrm{GeV}.
\end{equation}
This scenario requires a modest mixing-angle suppression of proton decay
\cite{Weinberg1979Bviol,WilczekZee1979,SuperK2017,PDG2024}, as discussed later, but still allows a relatively low exceptional scale and an intermediate \(G(2)\) threshold, favoring the dark-glueball phenomenology.
\newline

%
%
%
Among the possible realizations of the renormalization–group evolution, in this configuration the symmetry breaking scales are hierarchically separated yet close enough to ensure a controlled intermediate regime in which $SU(3)_C$ and $SU(3)_A$ coexist. The resulting one–loop running yields a dynamical intersection of $\alpha_A$ and $\alpha_{G(2)}$ well below the reduced Planck scale. The separation $M_ \Omega < M_\chi < \mu_{\rm meet}$ naturally orders the symmetry breaking sequence, preventing accidental degeneracies of phase transitions and simplifying the cosmological evolution. 

The heavy vector boson masses are automatically separated in a manner consistent with proton–decay constraints \cite{Masi2024,LuciniTeper2001,MorningstarPeardon1999} and with the decoupling of ultra–heavy degrees of freedom from low–energy phenomenology. Importantly, the meeting of $\alpha_A$ and $\alpha_{G(2)}$ arises dynamically from the renormalization group flow and not from a forced identification of symmetry–breaking scales with SM crossing points. 
Throughout we interpret the one-loop meeting as evidence of consistency of the exceptional embedding and use it primarily to motivate benchmark scales.

$G(2)$ breaks first at the higher scale $M_\chi$ (producing the 6 heavy $\mathcal X$-bosons), whereas $SU(3)_A$ breaks later at the lower scale $M_ \Omega$ (producing the 4 heavy $Y$-bosons). So $G(2)$ breaking precedes $SU(3)_A$ one and $\mathcal X$-bosons are more massive relics. It is perfectly consistent that the heavier $\mathcal X$-bosons (from $G(2)\to SU(3)_C$) end up stable/long-lived as dark glueballs, while the lighter $Y$-bosons decay promptly and never join the dark sector. In fact stability is set by symmetries and couplings, not by mass ordering. 
The projected-running approximation should be understood as a short-threshold approximation, not as the statement that the SM ancestors leave no imprint. The \(G(2)\times SU(3)_A\) branches are integrated out, projected or localized at a scale \(M_{\rm lift}\) close to \(M_\Omega\simeq M_\Theta\). Their effect on the subsequent bulk running should be then encoded in threshold corrections.

Concluding, $E_6$ embedding, equipped with $E_7$ uplift, provides sufficient flexibility to accommodate unification, DM and low-energy SM constraints within a single coherent framework.

\subsection{Uncertainties from two-loop running and threshold corrections}
\label{subsec:unif_uncertainties}

The one-loop running and sharp threshold matching used in Fig.~\ref{fig:runningB} provide a transparent analytic picture, but the inferred meeting scale and apparent convergence are subject to possible corrections from (i) two-loop running and (ii) threshold effects due to possible mass splittings within heavy multiplets around $M_ \Omega$ and $M_\chi$. In practice, unknown $\mathcal{O}(1)$ splittings among the heavy fields induce logarithmic threshold shifts in the matching conditions that can slightly move the meeting point. 

\paragraph{Parametric size of two-loop effects.}
At two loops, gauge couplings obey \cite{MachacekVaughn1983,MachacekVaughn1984}
\begin{equation}
	\mu\frac{d g_i}{d\mu} = \frac{b_i}{16\pi^2}g_i^3
	+ \frac{g_i^3}{(16\pi^2)^2}\left(\sum_j b_{ij} g_j^2 - \sum_a c_{ia} y_a^2 + \ldots\right),
\end{equation}
where the parentheses denote scalar quartic contributions $b g^2$ and model-dependent Yukawa structures $c y^2$.
Two-loop corrections could shift $\alpha_i^{-1}(\mu)$ by $\mathcal{O}(1)$ over many decades of running when additional gauge sectors and matter are present. Consequently, the apparent one-loop meeting point $(\mu_{\rm meet},\alpha^{-1}_{\rm meet})$ should be assigned a
conservative theory uncertainty.

\paragraph{Threshold matching and non-degenerate heavy spectra.}
Matching across a threshold scale $M_T$, \textit{e.g.} $M_ \Omega$ or $M_\chi$, receives finite corrections $\Delta_i^{\rm th}(\mu)$ from
integrating out heavy states with different masses $M_k$ not equal to $M_T$ \cite{Hall1981_EffectiveGUT}:
\begin{equation}
	\alpha_i^{-1}(\mu)\Big|_{\rm EFT} = \alpha_i^{-1}(\mu)\Big|_{\rm UV}
	+ \Delta_i^{\rm th}(\mu),\qquad
	\Delta_i^{\rm th}(\mu) \simeq -\frac{1}{2\pi}\sum_{k\in {\rm heavy}} b_i^{(k)}\ln\frac{M_k}{M_T},
	\label{eq:threshold_generic}
\end{equation}
with the usual subscript ${\rm EFT}$ for effective field theory and ${\rm UV}$ for ultraviolet completion.
Even modest splittings, $M_k/M_T \in [1/3,3]$, can generate $\Delta_i^{\rm th}\sim \mathcal{O}(0.1\!-\!1)$.
Because the meeting criterion is sensitive to differences between inverse couplings, the relevant
uncertainty is set by $\Delta_i^{\rm th}-\Delta_j^{\rm th}$.
The group-theory coefficients \(b_i^{(k)}\) are fixed by the ancestor branch representations, while the masses \(M_k\) and finite pieces depend on the projection or lifting mechanism.  Thus the representation tables
determine the possible size and sign of the threshold corrections, but not their unique numerical values.
Because the active-ancestor coefficients are very large, ($b_A^{A,\rm act}\simeq -228$, $b_A^{B,\rm act}\simeq (-199, -65)$),
the logarithmic interval over which these branches can propagate must be short.  For Class A, even a factor of two separation
\(M_\Omega/M_{T}=1/2\) would shift $\alpha_A^{-1}$
by approximately \(-25\). Therefore the viable running scenario requires
\(M_{T}\) to lie close to \(M_\Omega\), so that the ancestor
branches appear as threshold corrections rather than as long-distance
bulk-running degrees of freedom.\\

We therefore interpret Fig.~\ref{fig:runningB} as demonstrating that the model admits a consistent
high-scale embedding, while allowing for a theory band \cite{EllisWellsZheng2015_ThresholdUnification}
\begin{equation}
	\alpha^{-1}_{\rm meet} \to \alpha^{-1}_{\rm meet}\pm \delta\alpha^{-1},\qquad
	\log_{10}\mu_{\rm meet}\to \log_{10}\mu_{\rm meet}\pm \delta_{\log\mu},
\end{equation}
with conservative estimates from the combined two-loop and threshold effects. A full two-loop analysis is left for a future work.

\subsection{Observability of ultra--heavy gauge bosons and exotic Higgs states}

All three mass scales lie many orders of magnitude above the reach of the LHC or any planned
collider: 
\begin{equation}
	M_Y \sim M_{Z'}\sim 10^{13}\,\text{GeV}, \qquad
	M_{\mathcal X} \sim 10^{14}\,\text{GeV}, \qquad
	M_{\mathcal{E}} \gtrsim 10^{15\div16}\,\text{GeV}. 
\end{equation}
At energies well below the heavy thresholds, the virtual effects of ultra-heavy gauge bosons or scalars are
captured systematically by the SM Effective Field Theory (SMEFT): integrating out states of mass
$M\sim \Lambda$ generates higher-dimensional operators suppressed by $1/M^2$, so that electroweak-scale observables receive relative corrections of order \cite{Weinberg1979Bviol,WilczekZee1979,NathFileviezPerez2007}  
\begin{equation}
\delta \sim v_{EW}^2/M^2
\end{equation} 
up to coupling and mixing-angle factors.
This scaling follows directly from the low-momentum expansion of heavy propagators and the appearance of Higgs
vacuum insertions after electroweak symmetry breaking, and it underlies the standard treatment of oblique and
vertex corrections in extended gauge sectors and in SMEFT \cite{Grzadkowski2010Warsaw,BrivioTrott2019,Barbieri2004,PeskinTakeuchi1990,PeskinTakeuchi1992}.
For the present benchmark scenario, $M\sim 10^{13\div14}\,$GeV implies $v_{EW}^2/M^2\lesssim 10^{-24}\div10^{-24}$, 
rendering any collider or precision effects completely negligible unless additional
model-dependent enhancements (e.g.\ large gauge mixing) are introduced \cite{HanSkiba2005,BrivioTrott2019}.\\

In details, for what concerns indirect low--energy effects, heavy gauge vectors with mass $M_V$ generate effective four--fermion operators when integrated
out at energies $E\ll M_V$. If a heavy vector $V_\mu$ couples to a fermion current
$J^\mu=\bar f\gamma^\mu T f$ (where $T$ is the generator acting on fermions) with strength $g_V$, integrating out $V_\mu$ induces the dimension--six operator \cite{BuchmullerWyler1986,Grzadkowski2010Warsaw,BrivioTrott2019}
\begin{equation}
	\mathcal L_{\rm eff} =
	\frac{g_V^2}{M_V^2}
	(\bar f \gamma^\mu f)(\bar f' \gamma_\mu f') .
\end{equation}
For $M_V  =M_\chi\sim 10^{14}$~GeV and $g_V\sim 0.5\div0.6$, the coefficient $g_V^2/M_V^2\sim 10^{-28}\,\mathrm{GeV}^{-2}$,
which is many orders of magnitude below current experimental sensitivity. If couplings to light fermions arise only through small mixing angles
$\theta_f$, the effective coefficient scales as
\begin{equation} 
	g_V^2\theta_f^2/M_V^2,
	\end{equation}
leading to an additional strong suppression.
Consequently, electroweak precision observables receive completely negligible corrections \cite{Barbieri2004,HanSkiba2005} for the benchmark scales considered in the present GUT.\\

In addition, the presence of several high symmetry-breaking scales introduces the
usual hierarchy problem associated with scalar masses \citep{ModernPP}.
Loop corrections from heavy particles generate contributions to the SM Higgs mass of the order $\delta m_H^2
\sim
\frac{\lambda}{16\pi^2} M_{\rm heavy}^2$,
where $M_{\rm heavy}$ represents the mass of heavy bosons. Several mechanisms can mitigate the hierarchy problem \cite{PESKIN2025116971}, starting from a portal suppression for the SM Higgs--extra Higgs quadratic interactions. At one loop, integrating out one scalar Higgs $h_\sigma$ induces a threshold correction to the SM Higgs mass parameter of order $\delta m_H^2 \sim \frac{\kappa_{H\sigma}}{16\pi^2} m_{h_\sigma}^2$: requiring the correction not to exceed the physical Higgs mass, $|\delta m_H^2| \lesssim m_{h_{125}}^2$,
gives the naturalness bound
\begin{equation}
	\kappa_{H\sigma} \lesssim 16\pi^2\,\frac{m_{h_{125}}^2}{m_{h_\sigma}^2}.
	\label{eq:kappa_naturalness}
\end{equation}
Applying this estimate to the heavy scalar sectors of the benchmark model yields to $\kappa_{H\sigma}<10^{-21}$.\\


Concluding, the $E_6$ embedding is safe with respect to electroweak precision tests. All exotics decouple, while the SM Higgs might emerge naturally from $ \Omega$ in one scenario, as the unique light remnant of the high-energy symmetry breaking.
Their only observable imprints are indirect footprints: cosmological effects on inflationary dynamics and reheating and possible gravitational wave signatures from phase transitions \citep{ATLAS_Higgs2012,CMS_Higgs2012,EinsteinTelescope2011,CosmicExplorer2019,LIGOScientific2016GW150914}. 
First--order phase transitions at the three VEVs scales may generate stochastic gravitational waves.
However, their characteristic frequencies should lie
far above the sensitivity bands of LISA, Einstein Telescope or Cosmic Explorer, as presented in Section 12.7.

\section{Isolation and Conservation of the \texorpdfstring{$G(2)$}{G2} Dark Sector}

A central requirement for this GUT proposal is that the dark sector identified with the
massive $G(2)/SU(3)_C$ gauge bosons described in \cite{Masi2024} remains dynamically isolated from the visible sector, while all other heavy gauge bosons arising from the extended unification chain decay promptly.
In this section we show that this separation is automatic, technically natural and stable under radiative corrections. 

\paragraph{Orthogonality of gauge charges.}

As stressed before, the exceptional embedding 
\begin{equation}
	E_6 \;\supset\; G(2) \times SU(3)_A
\end{equation}
is \emph{special} rather than regular.
As a consequence, the generators of $G(2)$ are orthogonal to those of $SU(3)_A$ with respect to the
$E_6$ Killing form \cite{WeinbergQTF2,Slansky1981,Dynkin1957},
\begin{equation}
	\mbox{Tr}(T^a_{G(2)} T^b_{SU(3)_A}) = 0 .
\end{equation}
This implies that the massive vectors $\mathcal{X}_\mu \in G(2)/SU(3)_C$ are neutral under $SU(3)_A$, $U(1)_X$ and carry no hypercharge and no weak isospin, therefore cannot couple directly to any SM current.
Their only gauge interactions are with themselves and with the $SU(3)_C$ gluons inherited from the
unbroken $SU(3)\subset G(2)$ \cite{Masi2024}.

On the other hand, $ \Omega\sim(\mathbf 1,\mathbf 8)$ breaks $SU(3)_A\to SU(2)_L\times U(1)_A$ and carries no $G(2)$ charge and, because of the absence of bifundamental Higgs, the broken vectors of one factor are singlets under the other factor. Concretely, the six $G(2)/SU(3)_C$ vectors (the dark glueballs constituents) are neutral under $SU(3)_A$, whereas the four $SU(3)_A/(SU(2)_L\times U(1)_A)$ vectors are neutral under $G(2)$ and cannot carry \textit{dark color} and they do not hadronize into the dark glue sector \cite{Georgi1999,Slansky1981}.

This orthogonality is the group–theoretic origin of the \textit{darkness} of the $G(2)$ broken sector: the heavy $G(2)$ vectors are \emph{automatically} neutral under the visible gauge group. This is a fundamental and original feature of the present GUT model.

\paragraph{Absence of renormalizable gauge mixing.}

Gauge kinetic mixing between two non–Abelian factors is forbidden at dimension four \cite{Slansky1981}.
In particular, terms of the form \cite{WeinbergQTF2,Barbieri2004}
\begin{equation}
	F^{a}_{\mu\nu}(G(2))\,F^{\mu\nu}_{b}(SU(3)_A)
\end{equation}
are identically zero because the indices belong to different simple algebras.
Mass mixing between $G(2)$ and $SU(3)_A$ vectors requires a bifundamental VEV and we do not have one by construction. The lowest cross terms you could imagine are higher-dim, e.g. \cite{BuchmullerWyler1986,Grzadkowski2010Warsaw,BrivioTrott2019}
\begin{equation}
	\frac{1}{\Lambda^2}\,\mathrm{Tr}\!\big(F_{G(2)}^{\mu\nu} \,\chi\,\chi^\dagger\big)\,\mathrm{Tr}\!\big(F_{A,\mu\nu}\, \Omega\, \Omega^\dagger\big),
\end{equation}
but these vanish because $\chi$ and $ \Omega$ live in disjoint factors. 
There is no operator that endows the $SU(3)_A$ vectors with dark $G(2)$ color, even radiatively. 
So, in the minimal \(E_6\)-core sector, no bifundamental scalar connecting
\(G(2)\) and \(SU(3)_A\) is present. 

In the refined \(E_7\) uplift, the abelian-breaking field can be chosen from a
\(G(2)\)-singlet branch of the \(\mathbf{2925}\) of \(E_6\). In class A/B solutions one has
\begin{equation}
\Theta\subset\mathbf{2925}_{+3},
\qquad
\Theta\sim
(\mathbf1_{G(2)},\mathbf{10}_A)_{+3}.
\end{equation}
It contains an \(SU(2)_L\)-singlet component satisfying
\begin{equation}
A_\Theta+X_\Theta=0,
\end{equation}
and hence preserves the hypercharge generator while breaking the orthogonal
abelian combination in $U(1)_A\times U(1)_X\longrightarrow U(1)_Y$.
This realization is \(G(2)\)-neutral already before \(G(2)\to SU(3)_C\). Therefore
the \(\Theta\) VEV cannot induce renormalizable mass mixing between the
\(G(2)/SU(3)_C\) broken vectors and the electroweak or abelian gauge bosons.
Its communication with the \(G(2)\)-origin dark sector is restricted to
higher-dimensional operators and scalar portals such as $
(\Theta^\dagger\Theta)(\chi^\dagger\chi)$,
which can be naturally suppressed. The abelian-breaking sector is therefore
sequestered from the \(G(2)\) dark sector at the level of gauge quantum numbers.

\paragraph{Higgs portals and scalar mediation.}

After $E_6\to G(2) \times SU(3)_A$, the two sectors have no renormalizable gauge mixing.
The leading renormalizable communication arises from scalar portals, e.g.
$V \supset \kappa_{\chi \Omega}(\chi^T\chi)\mbox{Tr}( \Omega^\dagger \Omega)$.
The heavy $G(2)/SU(3)_C$ dark vectors $\mathcal{X}_\mu$ couple at tree level to the $G(2)$-breaking radial mode $h_\chi$
through the $\chi$ kinetic term \cite{BrivioTrott2019}:
\begin{equation}
	\mathcal L \supset (D_\mu\chi)^\dagger(D^\mu\chi)
	\;\Rightarrow\;
	\mathcal L \supset g_{G(2)}^2\,v_\chi\,\mathcal{X}_\mu \mathcal{X}^\mu\,h_\chi,
\end{equation}
where $m_{\mathcal X}^2=g_{G(2)}^2v_\chi^2$.
Portal interactions induce CP-even scalar mixing: in a minimal two-field approximation,
the off-diagonal entry scales as $m_{\chi \Omega}^2\sim \kappa_{\chi \Omega}v_\chi v_ \Omega$, giving \cite{HanSkiba2005}
\begin{equation}
	\theta_{\chi \Omega}\simeq \frac{m_{\chi \Omega}^2}{m_{h_\chi}^2-m_{h_ \Omega}^2}
	\sim \frac{\kappa_{\chi \Omega}}{2\lambda_\chi}\frac{v_ \Omega}{v_\chi},
	\qquad
	m_{h_\chi}^2 = 2\lambda_\chi v_\chi^2,\ \ m_{h_ \Omega}^2 = 2\lambda_ \Omega v_ \Omega^2.
\end{equation}
Consequently, the effective coupling of $\mathcal{X}_\mu$ to the SM-like Higgs $H$ is suppressed by the
small admixture of $h_\chi$ in $H$ defined by the angle $\theta_{H\chi}$ \cite{WeinbergQTF2}:
\begin{equation}
	\mathcal L_{\rm eff}\supset g_{G(2)}^2\,v_\chi\,\sin\theta_{H\chi}\;\mathcal{X}_\mu \mathcal{X}^\mu\,H,
	\qquad \sin\theta_{H\chi}\lesssim \sin\theta_{\chi \Omega}\ll 1,
\end{equation}
so any portal-induced visible--dark energy transfer or late decays are parametrically suppressed.


\paragraph{Prompt decays of non-dark heavy vectors and cosmological cleanness.}

The other heavy gauge bosons of the theory are:
\begin{itemize}
	\item \(Y_\mu\): from \(SU(3)_A/(SU(2)_L\times U(1)_A)\) breaking,
	\item \(Z'\): from \(U(1)_A\times U(1)_X\to U(1)_Y\),
	\item \(\mathcal E_\mu\): from \(E_6/(G(2)\times SU(3)_A)\),
	\item \(\mathcal E^{(7)}_\mu\): from \(E_7/E_6\), if the full uplift is retained.
\end{itemize}
Unlike the $\mathcal{X}_\mu$, these states carry electroweak or fermionic charges and possess renormalizable
couplings to SM fields. Their decay widths scale as \cite{GoodsellLieblerStaub2017TwoBodyWidths}
\begin{equation}
	\Gamma_i \sim \frac{g_i^2}{16\pi} m_i,
\end{equation}
which implies lifetimes far shorter than the Hubble time at their mass scales \cite{KolbTurner1990}.
The mass in the derived benchmark model for $Y$ bosons $m_Y=g_A v_ \Omega\approx 6\times10^{12}\,\mathrm{GeV}$ implies $\Gamma_Y\sim \mathcal O(10^{10})\,\mathrm{GeV}$ and $\tau_Y\sim \mathcal O(10^{-35})\,\mathrm{s}$. 
The ultra-heavy $E_6$--step vectors (and the $\mathcal E^{(7)}_\mu$) are heavier and decay even faster, so that these lifetimes are orders of magnitude shorter than any cosmological timescale and occur before the dark $G(2)$ confinement epoch, so they cannot accumulate or hadronize into the dark glue sector \cite{GurseyRamondSikivie1976,AchimanStech1978,HewettRizzo1989}. They therefore decay promptly into visible matter and leave no relic abundance.\\


The only candidate long-lived non-SM relics in this sector are the $G(2)$-origin heavy dark vectors states forming dark glueballs: they confine into dark glueballs at the scale $\Lambda_{G(2)}$, exactly as described in \cite{Masi2024}. The other broken vectors are gone from the primordial plasma essentially immediately: no stable relic from $E_6$ or $SU(3)_A$ breaking survives. Thus the cosmological dark sector is the confined $G(2)$ \text{glueballs only} \cite{Masi2024}, as discussed in Section 12.6.

\paragraph{Optional dark parity.}

As a precautional protection for the lightest $J=0$ glueball state, one may also impose a discrete symmetry such as
\begin{equation}
	\mathbb Z_2^{\rm dark}:\qquad \mathcal{X}_\mu \to -\mathcal{X}_\mu ,
\end{equation}
under which all visible fields are even. This parity forbids any higher–dimensional operator that could mediate $\mathcal X$ decay.
In \citep{Masi2021} few alternatives have been also proposed, such as an accidental symmetry, \textit{i.e.} a conserved additive gluon number $\Gamma$ to prevent the decay into mesons (like the baryon number $B$ for protons) or a $G$--parity conservation for a generic Yang--Mills theory, to prevent decay into $G$--even SM particles, unlike pions. However, a dark parity is not strictly required: the special embedding already ensures stability at the renormalizable level.

\paragraph{\(\Theta\) and the dark sector}
\label{subsec:Theta_Zprime_dark_seclusion}


The leading scalar portal connecting the abelian-breaking sector to the
\(G(2)\)-breaking sector is $V\supset
\lambda_{\Theta\chi}
(\Theta^\dagger\Theta)(\chi^T\chi)$: this coupling is allowed by gauge symmetry, even if $\Theta$ is a $G(2)$-singlet, and the model must impose either
\begin{equation}
|\lambda_{\Theta\chi}|\ll1,
\end{equation}
or a seclusion symmetry under which the leading \(\Theta\)-\(\chi\) portal is
forbidden or generated only at threshold.
After \(\chi\) and \(\Theta\) acquire VEVs, this portal induces scalar mixing between the radial modes \(\Theta_R=h_{\Theta}\) and \(\chi_R=h_\chi\). 
The off-diagonal scalar mass entry is parametrically
$m^2_{\Theta\chi}
\simeq
\lambda_{\Theta\chi}v_\Theta v_\chi$ and the mixing angle is approximately
\begin{equation}
\sin 2\alpha_{\Theta\chi}
\simeq
\frac{2\lambda_{\Theta\chi}v_\Theta v_\chi}
{m_{h_\chi}^2-m_{h_\Theta}^2}.
\end{equation}
A conservative seclusion requirement is therefore
\begin{equation}
|\sin\alpha_{\Theta\chi}|\ll1,
\end{equation}
which implies
\begin{equation}
|\lambda_{\Theta\chi}|
\ll
\frac{|m_{h_\chi}^2-m_{h_\Theta}^2|}
{2v_\Theta v_\chi},
\end{equation}
together with a small portal condition such as
\begin{equation}
|\lambda_{\Theta\chi}|\lesssim 10^{-2}\text{--}10^{-1}
\end{equation}
if one wants negligible scalar mixing without relying on large mass separation.

\section{FCNC Suppression and Proton Stability}

In unified theories, flavor physics and baryon-number violation are intimately related \cite{Hisano1993} \cite{Baek2004} \cite{IsidoriNirPerez2010} \cite{NathFileviezPerez2007}.
The same heavy states that mediate proton decay \cite{Weinberg1979Bviol,WilczekZee1979,SuperK2017,PDG2024} can, in principle, also induce FCNCs. In this section we show that the present
$E_6\to G(2) \times SU(3)_A$ embedding suppresses both effects in a correlated and natural way -- and that the \(E_7\) uplift preserves this suppression once the projected spectrum and the \(U(1)_X\) completion are treated consistently -- clarifying the role of Minimal Flavor Violation (MFV) as optional organizing principles \cite{FroggattNielsen1979,DAmbrosio2002}.

\subsection{Absence of tree-level FCNCs}

In the SM, FCNC processes such as $K^0$--$\bar K^0$ mixing or $b\to s\ell^+\ell^-$ transitions are suppressed by the Glashow--Iliopoulos--Maiani (GIM)
mechanism and arise only at loop level \citep{ModernPP}. Beyond the SM, dangerous FCNCs typically
originate from one of the following sources: multiple Higgs doublets coupling to the same fermion species, heavy gauge bosons with non-universal flavor couplings, mixing between light fermions and heavy vectorlike states \citep{PDG2024}.

In the present GUT, despite the presence of several scalar multiplets in the symmetry-breaking chain
($\Phi$, $\chi$, $ \Omega$ and $\Theta$), only a single $SU(2)_L$ Higgs doublet remains light and
develops an electroweak VEV. All additional electroweak doublets, if present in the parent exceptional multiplets, are taken to be either heavy or aligned with
the light Higgs direction. Thus each fermion species couples to a unique SM Higgs doublet and tree-level Higgs-mediated FCNCs are absent. This property is known as
\emph{Natural Flavor Conservation} \cite{GATTO1980221} and mirrors the situation in the SM. 

The abelian-breaking field may be chosen from the \(G(2)\)-singlet branch. The component acquiring a VEV is an \(SU(2)_L\)-singlet satisfying $A_\Theta+X_\Theta=0$,
so it breaks the orthogonal abelian combination in $U(1)_A\times U(1)_X\to U(1)_Y$ without introducing an additional light electroweak Higgs doublet. Therefore
the \(\Theta\) sector does not generate tree-level Higgs-mediated FCNCs. Its possible scalar portals, such as
$(\Theta^\dagger\Theta)(H^\dagger H)$ and $(\Theta^\dagger\Theta)(\chi^T\chi)$,
are flavor-blind at the renormalizable level.

Gauge interactions are also family-universal in the gauge basis. In the minimal \(E_6\) core this follows from placing the three families in identical
\(E_6\) representations. In the completed \(E_7\) picture the statement is replaced by projected family universality: each family is selected from the
same projected \(E_7\)-derived branch structure, with identical gauge quantum numbers. 
Heavy gauge bosons associated with the $G(2)$ and $SU(3)_A$ breaking steps do not carry intrinsic family indices. Consequently, neither the $\mathcal X$ bosons from
$G(2)/SU(3)_C$ nor the $Y$ bosons from $SU(3)_A/(SU(2)_L\times U(1)_A)$ generate tree-level FCNCs through gauge exchange \cite{Georgi1999,Slansky1981}.

The extra \(U(1)_X\) charges are species-dependent but family-universal. Hence the heavy abelian boson \(Z'\)
does not generate tree-level FCNCs. Its gauge couplings are diagonal in family space before Yukawa rotations and any flavor violation induced after mixing with vectorlike states is suppressed by the same small light--heavy mixing
angles that control the rest of the exotic sector.




\subsection{Vectorlike fermions and mixing-induced effects}

Exotic fermions appearing in the projected \(E_7\)-derived spectrum are removed from the low-energy spectrum through the vectorlike completion. In the completed
framework this lifting is naturally multi-threshold. Exceptional exotics may
receive masses near \(M_{E_6}\), \(G(2)\)-charged exotics near \(M_\chi\),
\(SU(3)_A\)-charged exotics near \(M_\Omega\), and purely abelian exotics near
\(M_\Theta\simeq M_\Omega\). After symmetry breaking, the fermion mass matrices in a $(f_{\rm SM},\,f_{\rm ex})$ basis take the schematic form \cite{WeinbergQTF2}
\begin{equation}
	\mathcal M_f =
	\begin{pmatrix}
		m_f^{\rm EW} & \Delta_f \\
		0 & M_f
	\end{pmatrix},
\end{equation}
where $m_f^{\rm EW}=y_f v_{EW}/\sqrt2$ is the usual electroweak mass with Yukawa coupling $y_f$, \(M_f\) denotes the appropriate high-threshold vectorlike mass generated by a high-scale singlet VEV, $\Delta_f \sim y_{\rm mix} v_{EW}/\sqrt2$ encodes mixing between light and heavy states.

For $M_f \gg m_f^{\rm EW}\sim\Delta_f$, diagonalization induces small corrections to the light-fermion couplings,
suppressed by a small mixing angle
\begin{equation} 
\theta_f\simeq \Delta_f/M_f\ll 1.
\end{equation}
These mixings represent the only potential source of both loop-level FCNCs and baryon-number violating interactions involving light fermions.
The physical light fermions therefore contain only a tiny admixture of heavy states.
All corrections to SM couplings scale as $\mathcal O(\theta_f^2)$,
and any flavor-changing or baryon-number violating effects induced by heavy
sector interactions are suppressed by powers of $\theta_f$.
Given $M_f\sim 10^{14-15}\,\mathrm{GeV}$, these effects are negligibly small.

\subsection{Proton decay}
In the special embedding $E_6\to G(2) \times SU(3)_A$, proton decay \cite{Weinberg1979Bviol,WilczekZee1979,SuperK2017,PDG2024} is naturally suppressed for several reasons:
i) the $SU(3)_A$ coset vectors $Y$ carry no color and do not generate baryon-violating dimension-6 operators associated with colored leptoquark gauge boson exchange;
ii) the heavy \(Z'\) introduced by the \(E_7\)
uplift is also colorless and transforms trivially under \(G(2)\) as $Z'_\mu\sim\mathbf1_{G(2)}$, being neither a dark gluon, nor a \(G(2)\)-coset vector and not a
leptoquark gauge boson;
iii) the $E_6$ and $E_7$ coset vectors are made very heavy and/or prevented from coupling to light quark–lepton currents by symmetry, leaving only negligible higher-order contributions \cite{GurseyRamondSikivie1976,AchimanStech1978,HewettRizzo1989,Slansky1981,Dorsner}, as described below. 

In the \(E_7\) uplift the abelian-breaking scalar can be chosen from a \(G(2)\)-singlet branch, $\Theta\subset\mathbf{2925}_{+3}$ with $\Theta\sim(\mathbf1_{G(2)},\mathbf{10}_A)_{+3}$.
Thus the \(\Theta\) VEV breaks only the orthogonal abelian combination in $U(1)_A\times U(1)_X\to U(1)_Y$
and does not introduce a new \(G(2)\)-charged bridge between the dark \(G(2)/SU(3)_C\) vectors and light quark--lepton currents. 

The only potentially dangerous gauge bosons at the lower exceptional threshold are the broken \(G(2)/SU(3)_C\) vectors \(\mathcal X\): these \(\mathbf3_C\oplus\overline{\mathbf3}_C\) vectors can mediate baryon-violating operators only if they have non-negligible couplings to both light quark and light lepton zero modes. In the original incomplete \(E_6\) core this can
be arranged by placing the light leptons in \(G(2)\)-singlet branches: \(\mathcal X\) can couple to leptons only through the $G(2)$-charged $(\mathbf7,\mathbf3)$ sector, while the light leptons can be arranged to lie in the $G(2)$-singlet $(\mathbf1,\bar{\mathbf6})$ sector. 
In the \(E_7\) uplift the preferred projected spectrum may select some light leptons from color-singlet components of \(G(2)\)-charged branches $\mathbf7_{G(2)}\to\mathbf1_C\oplus\mathbf3_C\oplus\overline{\mathbf3}_C$. Therefore the correct general condition is not simply that leptons are \(G(2)\)-singlets, but that the projected light quark--lepton current coupled to the broken \(G(2)/SU(3)_C\) generators is absent or mixing-suppressed.

Still proton decay might be mediated by heavy colored vectors $ \mathcal X$, even if leptons are neutral under $G(2)$ at the gauge level and the $ \mathcal X$ bosons do not couple directly to quark--lepton currents. Such couplings arise only after mixing between light fermions and heavy $G(2)$-charged fermions.

The broken \(G(2)/SU(3)_C\) generators connect the colored and color-singlet components
inside the same \(G(2)\) multiplet.  At the level of a parent \(G(2)\) current one has
schematically
\begin{equation}
	\mathcal L_{\mathcal X}
	=
	g_{G(2)}
	\mathcal X_\mu^I
	\overline\Psi\,\gamma^\mu T^I_{\mathcal X}\Psi ,
	\qquad
	T^I_{\mathcal X}\in \mathfrak g_2/\mathfrak{su}(3)_C .
	\label{eq:Xdark_parent_current}
\end{equation}
Here \(I=1,\ldots,6\) labels the six broken generators of
$G(2)$.  The matrices
\(T^I_{\mathcal X}\) are the corresponding \(G(2)\) generators in the
representation of the fermion multiplet \(\Psi\).  In the fundamental
\(\mathbf7\), they are the off-diagonal generators connecting the
\(\mathbf3_C\oplus\overline{\mathbf3}_C\) colored components with the
\(\mathbf1_C\) singlet component.
After the branching $\mathbf7_{G(2)}\to\mathbf1_C\oplus\mathbf3_C\oplus\overline{\mathbf3}_C$, this current may contain
terms of the form $\mathcal L_{\mathcal X}
\supset
g_{G(2)}\,
\mathcal X_\mu\,
\overline q\,\gamma^\mu \ell
+
\mathrm{h.c.}$,
where \(q\) denotes a colored component and \(\ell\) a color-singlet component of the same
\(G(2)\)-origin multiplet. The broken \(G(2)\) vectors are singlets
under \(SU(3)_A\), hence they preserve the \(SU(3)_A\) representation and the abelian
charges carried by the field.  A direct light-light \(\mathcal X q\ell\) current exists
only if the projected light quark and lepton components have nonzero overlap inside the
same \(G(2)\) branch and compatible \(SU(3)_A\) sector. 
In details, after the chiral projection, the coupling of \(\mathcal X_\mu\) to two light fields is
not simply the parent gauge coupling, it is weighted by a projected Clebsch/overlap coefficient:
\begin{equation}
	g_{G(2)}\,\mathcal C^I_{ql}\,
	\mathcal X_\mu^I\,\overline q\gamma^\mu l,
	\qquad
	\mathcal C^I_{ql}
	=
	\langle q_{\rm light}|T^I_{\mathcal X}|l_{\rm light}\rangle .
	\label{eq:projected_X_Clebsch}
\end{equation}
The symbol \(\mathcal C^I_{ql}\) therefore denotes a \emph{single-vertex} group-theoretic
projector coefficient \cite{Slansky1981,BairdBiedenharn1972,LieArt}.
The proton-stability requirement is precisely
\begin{equation}
	\mathcal C^I_{q\ell}
	\equiv
	\langle q_{\rm light}|T^I_{\mathcal X}|\ell_{\rm light}\rangle
	=
	0
	\qquad
	\text{at zeroth order in light-heavy mixing}.
	\label{eq:direct_X_Clebsch_zero}
\end{equation}
A full numerical proton-decay prediction would require constructing the
explicit current matrices and projector coefficients for the chosen Class A or Class B
branch.  In the present work we give an operator-level bound and identify the required
condition on the projector. We can verify that the zeroth-order condition of Eq.~\eqref{eq:direct_X_Clebsch_zero} is satisfied by the representation assignments of Tables~\ref{tab:ClassA} and~\ref{tab:ClassB}, without recourse to any uncomputed numerical \(G(2)\) Clebsch--Gordan coefficient.

The six broken generators \(T^I_{\mathcal X}\in\mathfrak g_2/\mathfrak{su}(3)_C\) belong to the \(G(2)\) factor alone: on any state transforming as \((\mathbf7,\mathbf R_A)_X\) under \(G(2)\times SU(3)_A\times U(1)_X\), they act as \(T^I_{\mathcal X}\otimes\mathbb 1_{SU(3)_A}\otimes\mathbb 1_{U(1)_X}\), and hence commute with every generator of \(SU(3)_A\) and of \(U(1)_X\). For two light fields \(q\in(\mathbf7,\mathbf R_A)_{X_q}\) and \(\ell\in(\mathbf7,\mathbf R_A')_{X_\ell}\), the projector of Eq.~\eqref{eq:projected_X_Clebsch} therefore factorizes exactly as
\begin{equation}
	\mathcal C^I_{q\ell}
	=
	\langle q|T^I_{\mathcal X}|\ell\rangle
	=
	\big\langle \mathbf7_q\big|T^I_{\mathcal X}\big|\mathbf7_\ell\big\rangle_{G(2)}
	\;\times\;
	\big\langle \mathbf R_A\big|\,\mathbb 1\,\big|\mathbf R_A'\big\rangle_{SU(3)_A}
	\;\times\;
	\delta_{X_q,X_\ell},
	\label{eq:ClassA_Cql_factorization}
\end{equation}
which vanishes identically whenever \(\mathbf R_A\neq\mathbf R_A'\) or \(X_q\neq X_\ell\), \emph{regardless of the value of the \(G(2)\)-only matrix element} -- a consequence of orthogonality of inequivalent irreducible representation spaces for the \(SU(3)_A\times U(1)_X\) symmetry case, requiring no input beyond the representation content already tabulated.

From Class A, Table~\ref{tab:ClassA}, the light quark fields occupy \(\mathbf R_A\in\{\overline{\mathbf3}_A,\mathbf3_A,\overline{\mathbf6}_A\}\) (for \(Q_L,u^c,d^c\) respectively), while the light lepton fields occupy \(\mathbf R_A=\mathbf8_A\) (for \(L,e^c\)). Since \(\dim\overline{\mathbf3}_A=\dim\mathbf3_A=3\), \(\dim\overline{\mathbf6}_A=6\) and \(\dim\mathbf8_A=8\) are pairwise distinct, \(\mathbf8_A\) is inequivalent to any of the three quark representations, and Eq.~\eqref{eq:ClassA_Cql_factorization} gives
\begin{equation}
	\mathcal C^I_{q\ell}=0
	\qquad
	\text{exactly, for every } q\in\{Q_L,u^c,d^c\},\ \ell\in\{L,e^c\},\ I=1,\ldots,6 .
	\label{eq:ClassA_Cql_vanish}
\end{equation}
Explicitly, for \(Q_L\in(\mathbf7,\overline{\mathbf3}_A)_{+1}\) and \(L\in(\mathbf7,\mathbf8_A)_{-3}\) in Table~\ref{tab:ClassA}, both the \(SU(3)_A\) factor (\(\overline{\mathbf3}_A\neq\mathbf8_A\)) and the \(U(1)_X\) factor (\(X_{Q_L}=+1\neq-3=X_L\)) independently forbid a nonzero overlap: the suppression is doubly protected, by two commuting and independently exact symmetries, rather than relying on either alone.

From Class B, Table~\ref{tab:ClassB}, the conclusion is even more immediate: the lepton fields \(L,e^c\) there sit in \((\mathbf1,\mathbf8_A)\), \textit{i.e.}\ they are genuine \(G(2)\) \emph{singlets}, so that \(T^I_{\mathcal X}|\ell\rangle=0\) identically for every broken generator, independently of any \(SU(3)_A\) or \(U(1)_X\) bookkeeping.

Therefore the \(E_7\) uplift, by assigning the lepton sector to a different \(SU(3)_A\) (and, in Class B, \(G(2)\)) representation than the quarks, simultaneously repairs the hypercharge normalization \emph{and} enstablish the leptoquark-current protection: the two problems share a common representation-theoretic origin and a common cure.
Beyond this zeroth-order selection rule, only the mixing-suppressed residual discussed below, controlled by the mixing angle \(\vartheta_{\rm LQ}\), can populate the would-be leptoquark vertex; a complete numerical prediction for that residual still requires the explicit heavy-sector mixing matrices, which remains a next-stage calculation.
The relevant heavy gauge mediators are summarized in Table~\ref{tab:proton_decay_mediators}.

\begin{table}[t]
	\centering
	\small
	\renewcommand{\arraystretch}{1.25}
	\setlength{\tabcolsep}{4pt}
	\begin{tabularx}{\textwidth}{>{\raggedright\arraybackslash}p{0.23\textwidth}
			>{\raggedright\arraybackslash}p{0.24\textwidth}
			>{\centering\arraybackslash}p{0.10\textwidth}
			>{\raggedright\arraybackslash}X}
		\hline
		Mediator sector & Representation after breaking & Color & Proton-decay role \\
		\hline
		\(\mathcal X_\mu\in G(2)/SU(3)_C\)
		&
		\(\mathbf3_C\oplus\overline{\mathbf3}_C\)
		&
		yes
		&
		Primary leptoquark-like concern.  These are the broken dark \(G(2)\) vectors connecting colored and color-singlet components inside compatible \(G(2)\) branches.
		\\[1mm]
		
		\(Y_\mu\in SU(3)_A/(SU(2)_L\times U(1)_A)\)
		&
		\(SU(2)_L\) doublets
		&
		no
		&
		Not direct gauge leptoquarks.  They are relevant for electroweak-ancestor transitions, flavor effects and exotic-light mixing, but not as the leading proton-decay mediators.
		\\[1mm]
		
		\(\mathcal{E_\mu}\in E_6/(G(2)\times SU(3)_A)\)
		&
		\((\mathbf7,\mathbf8_A)\)
		&
		yes after \(G(2)\) breaking
		&
		Possible baryon-violating currents after projection, but the vectors are heavier, with masses of order \(M_{E_6}\).
		\\[1mm]
		
		\(E_7/(E_6\times U(1)_X)\) vectors
		&
		\(27_{+2}\oplus\overline{27}_{-2}\)
		&
		branch-dependent
		&
		Possible GUT-like currents, but they live at the highest scale \(M_{E_7}\) and are further suppressed by the same projection/mixing mechanism.
		\\[1mm]
		
		\(Z'\)
		&
		orthogonal abelian combination
		&
		no
		&
		No tree-level \(\Delta B\) gauge current.  Its constraints are instead abelian, flavor and threshold-matching constraints.
		\\
		\hline
	\end{tabularx}
	\caption{Gauge sectors relevant for proton-decay estimates.  The primary leptoquark-like vectors in the present construction are the broken \(G(2)/SU(3)_C\) dark vectors \(\mathcal X_\mu\).  The \(Y_\mu\) vectors from \(SU(3)_A\) breaking are color singlets and therefore are not the dominant gauge-mediated proton-decay mediators.}
	\label{tab:proton_decay_mediators}
\end{table}

Let \(f_{\rm light}^{(0)}\) be a projected chiral field and \(F_{\rm heavy}^{(0)}\) a heavy state with the same low-energy quantum numbers.  A typical mass matrix gives
\begin{equation}
	f_{\rm light}
	=
	f_{\rm light}^{(0)}
	+
	\theta_f F_{\rm heavy}^{(0)}
	+
	\cdots,
	\qquad
	\theta_f\simeq\frac{\Delta_f}{M_f}
	\lesssim
	\frac{v_{\rm EW}}{M_f}.
	\label{eq:mixing_angle_darkX}
\end{equation}
then the effective projected Clebsch at one leptoquark vertex becomes schematically
\begin{equation}
	\mathcal C^I_{q\ell,\rm eff}
	\simeq
	\vartheta_{\rm LQ}\,
	\widehat{\mathcal C}^I_{q\ell},
	\label{eq:effective_X_Clebsch_mixing}
\end{equation}
where \(\widehat{\mathcal C}^I_{q\ell}\) is the parent-theory Clebsch coefficient, expected
to be of order unity unless further group-theoretic zeros occur, while \(\vartheta_{\rm LQ}\) is the effective light-heavy overlap controlling one \(\mathcal X q\ell\) vertex for a common mixing parameter $\theta_q\sim\theta_\ell\simeq \vartheta_{\rm LQ}$
of the same order as the dominant \(\theta_f\). Here we adopt the \emph{mixing-angle convention} to study leptoquark suppression, being $\vartheta_{\rm LQ}$ the mixing angle that measures the amount by which a light lepton or quark state overlaps with a heavy
\(G(2)\)-charged component connected by the broken \(\mathcal X\) generators.
Then the effective quark--lepton coupling of $\mathcal X$ to light fields is
\begin{equation}
	g_{\rm eff} \;=\; g_{G(2)}\,\vartheta_{\rm LQ}.
\end{equation}
In the absence of projection or mixing
suppression, one would have
\begin{equation}
	C_6^{\rm unsup}
	\sim
	\frac{g_{G(2)}^2}{m_{\mathcal X}^2}
	\left(\mathcal C_1\mathcal C_2\right),
	\label{eq:C6_unsup_darkX}
\end{equation}
where \(\mathcal C_1\) and \(\mathcal C_2\) are the two single-vertex Clebsch coefficients
appearing at the two ends of the exchanged \(\mathcal X\) propagator \cite{langacker1981Unification,NathFileviezPerez2007} (omitting the $I$ apex for simplicity).  For an unsuppressed ordinary leptoquark vector, these factors are \(O(1)\).
In the projected model, however, the direct light-light Clebsches vanish as in
Eq.~\eqref{eq:direct_X_Clebsch_zero}.  With the mixing-angle convention
\begin{equation}
	\mathcal C_1\sim\vartheta_{\rm LQ}\widehat{\mathcal C}_1,
	\qquad
	\mathcal C_2\sim\vartheta_{\rm LQ}\widehat{\mathcal C}_2,
\end{equation}
the Wilson coefficient becomes \cite{Weinberg1979Bviol,WilczekZee1979,NathFileviezPerez2007}
\begin{equation}
	C_6
	\simeq
	\frac{g_{\rm eff}^2}{m_{\mathcal X}^2}
	=
	\frac{g_{G(2)}^{\,2}\,\vartheta_{\rm LQ}^{\,2}}{m_{\mathcal X}^2}
	\left(\widehat{\mathcal C}_1\widehat{\mathcal C}_2\right)
	=
	\frac{4\pi\,\alpha_{G(2)}\,\vartheta_{\rm LQ}^{\,2}}{m_{\mathcal X}^2}
	\left(\widehat{\mathcal C}_1\widehat{\mathcal C}_2\right).
	\label{eq:C6mix_darkX}
\end{equation}
Since partial widths scale as $\Gamma\propto |C_6|^2$, 
the proton lifetime scales as \cite{NathFileviezPerez2007,SuperK2017}
\begin{equation}
	\tau_p \;\propto\; \frac{1}{|C_6|^2}\;\propto\;
	\frac{m_{\mathcal X}^4}
	{g_{G(2)}^4\,\vartheta_{\rm LQ}^{\,4}}
	\frac{1}{|\widehat{\mathcal C}_1\widehat{\mathcal C}_2|^2}.
	\label{eq:theta4_lifetime_scaling_darkX}
\end{equation}
showing a quartic sensitivity to the mixing angle.
For the benchmark channel \(p\to e^+\pi^0\), a standard chiral reduction gives \cite{Aoki2017ProtonDecay,ClaudsonWiseHall1982,Aoki2008NucleonDecayMatrix}
\begin{equation}
	\Gamma(p\to e^+\pi^0)
	\simeq
	\frac{m_p}{64\pi f_\pi^2}
	\left(1-\frac{m_{\pi^0}^2}{m_p^2}\right)^2
	A_R^2
	|\alpha_H|^2
	(1+D+F)^2
	|C_6|^2 .
	\label{eq:p_to_epi_width_darkX}
\end{equation}
Here \(A_R=A_LA_S\) is the short- and long-distance renormalization factor,
\(f_\pi\simeq0.13\,\mathrm{GeV}\), and \(D\simeq0.80\), \(F\simeq0.47\) are baryon
chiral Lagrangian parameters.  The lattice hadronic matrix element is of order
\begin{equation}
	|\alpha_H|\sim10^{-2}\,\mathrm{GeV}^3 .
\end{equation}
These hadronic, nuclear, flavor and Clebsch factors are important for a final
channel-by-channel prediction, but for the present operator estimate they are absorbed into
an \(O(1)\) factor.

For a compact numerical normalization,
one may write schematically \cite{NathFileviezPerez2007,Aoki2017ProtonDecay}
\begin{equation}
	\tau(p\to e^+\pi^0)
	\sim
	10^{35}\,{\rm yr}\,
	\left(\frac{m_{\mathcal X}}{10^{16}\,{\rm GeV}}\right)^4
	\left(\frac{0.03}{\alpha_{G(2)}}\right)^2
	\vartheta_{\rm LQ}^{-4}
	\left(\frac{1}{|\widehat{\mathcal C}_1\widehat{\mathcal C}_2|^2}\right)
	\mathcal H^{-1},
	\label{eq:tauScaling}
\end{equation}
where \(\mathcal H=O(1)\) parametrizes the remaining hadronic, flavor and
renormalization uncertainty.  
Using the current Super-Kamiokande bound on the dominant gauge-mediated channel \cite{SuperK2017,PDG2024},
\begin{equation}
	\tau(p\to e^+\pi^0)\gtrsim 2.4\times 10^{34}\,{\rm yr} \qquad {\rm (90\% C.L.)},
\end{equation}
we can translate Eq.~\eqref{eq:tauScaling} into an upper bound on $\vartheta_{\rm LQ}$.
Adopting the schematic normalization used above,
the experimental constraint implies 
\begin{equation}
	\vartheta_{\rm LQ}
	\lesssim
	\left[
	\frac{10^{35}\,{\rm yr}}{2.4\times10^{34}\,{\rm yr}}
	\right]^{1/4}
	\left(\frac{m_{\mathcal X}}{10^{16}\,{\rm GeV}}\right)
	\left(\frac{0.03}{\alpha_{G(2)}}\right)^{1/2}
	\left(\frac{1}{|\widehat{\mathcal C}_1\widehat{\mathcal C}_2|^2}\right)^{1/4}
	\mathcal H^{-1/4}.
	\label{eq:thetaBoundGeneral_darkX}
\end{equation}
For order-one Clebsches and \(\mathcal H\simeq1\), this simplifies to
\begin{equation}
	\vartheta_{\rm LQ}
	\lesssim
	1.4
	\left(\frac{m_{\mathcal X}}{10^{16}\,{\rm GeV}}\right)
	\left(\frac{0.03}{\alpha_{G(2)}}\right)^{1/2}.
	\label{eq:thetaBoundGeneral_simple_darkX}
\end{equation}
For the benchmark scenario with $
m_{\mathcal X}\simeq g_{G(2)}M_\chi
\simeq5.6\times10^{13}\,\mathrm{GeV}$ and $\alpha_{G(2)}\simeq0.025\text{--}0.03$,
this gives
\begin{equation}
	\vartheta_{\rm LQ}
	\lesssim
	(0.8\text{--}1.0)\times10^{-2},
	\label{eq:thetaBoundB_darkX}
\end{equation}
showing that a percent-level (or smaller) leptoquark mixing is sufficient to reconcile a comparatively low
$M_X$ with proton-decay limits \cite{Weinberg1979Bviol,WilczekZee1979,SuperK2017,PDG2024}. 
Thus, proton stability is controlled by the \emph{same mixing angles} that govern the coupling of heavy states to light fermions.
Conversely, if the \(\mathcal X\) vectors had unsuppressed leptoquark couplings to the light projected
fields, Eq.~\eqref{eq:tauScaling} would give
\begin{align}
	\tau_{\rm unsup}^{\mathcal X}
	&\sim
	10^{35}\,{\rm yr}
	\left(\frac{5.6\times10^{13}}{10^{16}}\right)^4
	\left(\frac{0.03}{0.03}\right)^2
	\nonumber\\
	&=
	10^{35}\,{\rm yr}
	\left(5.6\times10^{-3}\right)^4\simeq 10^{26}\,{\rm yr},
	\label{eq:unsuppressed_lifetime_darkX}
\end{align}
up to the \(O(1)\) hadronic and Clebsch factor \(\mathcal H^{-1}\).  This is below the experimental lower bound.  Thus an unsuppressed direct \(\mathcal X q\ell\) current is not allowed, as expected.

The expected overlap from high-scale vectorlike lifting is much smaller: taking $M_f\gtrsim 10^{13}\,\mathrm{GeV}$, one finds
\begin{equation}
	\vartheta_{\rm LQ}
	\lesssim
	\frac{v_{\rm EW}}{M_f}
	\simeq 2.5 \times 10^{-11},
\end{equation}
and therefore
\begin{equation}
	\vartheta_{\rm LQ}^{\,4}
	\lesssim
	3.8 \times 10^{-43}.
	\label{eq:theta4_less_10minus44_darkX}
\end{equation}
If the vectorlike states are lifted closer to \(M_{E_6}\sim10^{16}\,\mathrm{GeV}\), the
suppression becomes even stronger,\textit{ i.e.} $\vartheta_{\rm LQ}\lesssim 10^{-14}$.
So, they may in principle generate ordinary GUT-like dimension-six operators, but their larger masses and the same projection/mixing suppression make them extremely
less dangerous than the \(G(2)/SU(3)_C\) dark vectors. \\

It is instructive to contrast the present construction with more conventional GUTs based on regular embeddings, such as $SU(5)$ or $SO(10)$.
In minimal $SU(5)$, proton decay\cite{Weinberg1979Bviol,WilczekZee1979,SuperK2017,PDG2024} arises at tree level from leptoquark gauge bosons
that couple directly to quark--lepton currents with unsuppressed strength, leading to
stringent lower bounds on the unification scale.
In $SO(10)$ models, although the embedding of a full fermion family is more natural,
dangerous baryon- and lepton-number violating operators typically persist unless
additional symmetries or carefully tuned Higgs sectors are introduced.

By contrast, in the special \(E_6\to G(2)\times SU(3)_A\) framework, the
dangerous leptoquark currents are not generic gauge currents of the low-energy
spectrum. In the \(E_7\) completion this statement becomes a projection
condition: the projected light spectrum must have negligible overlap with the
broken \(G(2)/SU(3)_C\) currents that connect light quarks to light leptons.
Any residual proton-decay amplitude is then controlled by
\(\vartheta_{\rm LQ}\) and by the heavy vector mass \(m_{\mathcal X}\). These negligible controlled mixings explain the longevity of the proton and the absence of observable FCNCs, without affecting the gauge unification pattern or the dark
$G(2)$ sector responsible for glueball DM \cite{Masi2024}. Therefore proton stability
and flavor protection are not unrelated assumptions, but two consequences of
the same projected and multi-threshold exceptional structure.

\section{Cosmological Features}
\label{sec:defects}

The multi-stage symmetry-breaking pattern of the present
framework has important implications for early-Universe cosmology. In particular, the formation and subsequent evolution of topological defects must be examined to ensure consistency with observational bounds \cite{Witten1984,Hogan1986,BezrukovShaposhnikov2008,KolbTurner1990,Starobinsky1980,Linde1982,AlbrechtSteinhardt1982,Lyth1997Bound}.
In this section we provide a coherent account of defect production, the timing of
inflation \cite{Guth1981,Linde1983}, reheating and baryogenesis \cite{FukugitaYanagida1986} and gravitational waves from phase transitions \citep{Cutting_2018,PhysRevLett.115.181101,Zhou_2020,2021JHEP...05..160Z}.

\subsection{Topological classification of defects}

For a gauge symmetry breaking $G \longrightarrow H$
the vacuum manifold is $\mathcal{M} = G/H$. Topological defects are classified by homotopy groups of $\mathcal{M}$, so that \cite{WeinbergQTF2,Kibble1976,VilenkinShellard2000}:
\begin{align}
	\pi_2(\mathcal{M}) \neq 0 &\quad \Rightarrow \quad \text{magnetic monopoles}, \\
	\pi_1(\mathcal{M}) \neq 0 &\quad \Rightarrow \quad \text{cosmic strings}. 
\end{align}

For compact connected Lie groups one has \cite{WeinbergQTF2,NakaharaTopology}
\begin{equation}
	\pi_2(G/H) \simeq \ker\!\left[\pi_1(H) \to \pi_1(G)\right],
\end{equation}
and since $\pi_2(G)=0$ for any Lie group, the monopole content is entirely determined by the fundamental groups of $G$ and $H$ \cite{Kibble1976,Preskill1979}.

In our breaking chain all groups are taken in their simply connected form.

\paragraph{Zero transition: $E_7\to E_6\times U(1)_X$}
The \(E_7\) stage introduces possible high-scale defects associated with
\begin{equation}
E_7\to E_6\times U(1)_X.
\end{equation}
For simply connected \(E_7\), up to possible discrete quotients in the global form of the embedded subgroup, the vacuum manifold
\begin{equation}
\mathcal M_{E7}=\frac{E_7}{E_6\times U(1)_X}
\end{equation}
has
\begin{equation}
\pi_2(\mathcal M_{E_7})\simeq\pi_1(E_6\times U(1)_X)=\mathbb Z.
\end{equation}
Thus \(E_7\)-scale monopoles may form. These are harmless if the \(E_7\) transition occurs before or during inflation, or if the reheating temperature is below the corresponding restoration scale. In fact the \(E_7\)
transition is assumed to occur at a UV scale
\begin{equation}
M_{E_7}\gtrsim M_{E_6}.
\end{equation}
If inflation occurs after this transition, the corresponding monopoles are
inflated away. Alternatively, if reheating satisfies
\begin{equation}
T_{\rm reh}<M_{E_7},
\end{equation}
they are never thermally regenerated. Therefore the \(E_7\)-scale monopoles can
be consistently decoupled from the later cosmology.

\paragraph{First transition: $E_6 \to G(2) \times SU(3)_A$.}
Since $\pi_1(E_6)=0$, $\pi_1(SU(3))=0$ and $\pi_1(H=SU(3)\times G(2))=0$,
one finds \cite{WeinbergQTF2}
\begin{equation}
	\pi_2(E_6/H)=0.
\end{equation}
Therefore no monopoles are generated at this stage.

\paragraph{Second transition: $G(2) \to SU(3)_C$.}
Again $\pi_1(G(2))=0$ and $\pi_1(SU(3))=0$, so 
\begin{equation}
	\pi_2(G(2)/SU(3))=0.
\end{equation}
No monopoles and no topologically stable strings arise from this breaking.

\paragraph{Third transition: $SU(3)_A \to SU(2)_L \times U(1)_A$.}
Here $\pi_1(SU(3))=0$ and $\pi_1(SU(2)\times U(1))=\mathbb{Z}$, hence 
\begin{equation}
	\pi_2\!\left(SU(3)/(SU(2)\times U(1))\right)
	\simeq \mathbb{Z}.
\end{equation}
Magnetic monopoles are therefore generically produced at this transition.
However	$\pi_1(G/H)=0$, so no stable cosmic strings arise from the gauge breaking itself \cite{Kibble1976,VilenkinShellard2000}.

\paragraph{Fourth transition: $U(1)_A\times U(1)_X\to U(1)_Y$}
The breaking
\begin{equation}
U(1)_A\times U(1)_X\to U(1)_Y
\end{equation}
has vacuum manifold schematically equivalent to \(U(1)\), and may produce local strings:
\begin{equation}
\pi_1\left(\frac{U(1)_A\times U(1)_X}{U(1)_Y}\right)=\mathbb Z.
\end{equation}
Their tension is of order
\begin{equation}
\mu_{\rm string}\sim 2\pi v_\Theta^2.
\end{equation}
 and because
\begin{equation}
v_\Theta\simeq M_\Theta\sim10^{13}\ {\rm GeV},
\end{equation}
these strings are high-scale objects. Their abundance must be removed or diluted by inflation, or the \(U(1)_X\)-breaking transition must not be thermally restored after reheating. The conservative cosmological assumption is that any such high-scale defects are inflated away or not regenerated after reheating.
Since the preceding \(SU(3)_A\to SU(2)_L\times U(1)_A\) transition produces monopoles carrying \(U(1)_A\) magnetic flux, the later breaking $U(1)_A\times U(1)_X\to U(1)_Y$
can confine part of this flux into strings. Thus the resulting strings may be isolated local strings, metastable strings, or string segments attached to
monopoles, depending on the flux matching between the \(SU(3)_A\) monopoles and the broken abelian generator. A complete treatment requires checking the
Aharonov--Bohm phases and flux quantization conditions for the projected \(E_7\)-derived matter spectrum. Recently, superconducting exceptional \(E_6\) strings were
analyzed in the \(E_6\to SO(10)\times U(1)_\psi\) chain context \cite{nym6-vpms}.

\subsection{Monopole mass scale and abundance}
The monopole problem in grand unification
was first analyzed by Preskill~\cite{Preskill1979}
(see also~\cite{Kibble1976,VilenkinShellard2000}).
The monopole mass is parametrically of order $M_M \sim \frac{4\pi}{g_A}\, v_ \Omega$,
where $v_ \Omega \sim v_\Theta$ is the aforementioned $SU(3)_A$ and $U(1)_X$ breaking scale and $g_A\sim g_\Theta$ the corresponding gauge coupling.
Using the benchmark values $v_ \Omega \simeq 10^{13}\,\text{GeV}$, $g_A \simeq 0.5$, one obtains $M_M \sim \mathcal{O}(10^{14})\ \text{GeV}$. 

The Kibble mechanism predicts approximately one monopole per correlated Hubble volume $H^3$ at the phase transition \cite{Kibble1976}. The initial monopole number density as a function of the transition temperature $T_ \Omega$ is therefore \cite{Preskill1979}
\begin{equation}
	n_M(T_ \Omega) \sim H_ \Omega^3,
\end{equation}
and $H_ \Omega \sim \frac{T_ \Omega^2}{\bar M_{\rm Pl}}$, with the reduced Planck mass $\bar M_{\rm Pl}=2.4\times 10^{18}\,$GeV. The initial monopole-to-entropy ratio is then \cite{KolbTurner1990}
\begin{equation}
	\frac{n_M}{s} \sim  \left(\frac{T_ \Omega}{\bar M_{\rm Pl}}\right)^3.
\end{equation}
For $T_ \Omega \sim M_ \Omega \sim 10^{13}\ \text{GeV}$, this gives 
\begin{equation}
	\frac{n_M}{s} \sim 7 \times 10^{-17}.
\end{equation}
The present monopole density is approximately 
\begin{equation}
	\Omega_M h^2 \sim 
	\frac{M_M}{\text{GeV}}
	\left(\frac{n_M}{s}\right)
	10^9.
\end{equation}
Substituting values, 
\begin{equation}
	\Omega_M h^2 \sim 10^{14} \times 7 \times 10^{-17} \times 10^9 
	\simeq 7 \times 10^6,
\end{equation}
which vastly exceeds observational bounds.
Therefore, without dilution, the monopoles from $SU(3)_A$ breaking would overclose the Universe.
The general requirement is that inflation \cite{Guth1981,Linde1983} occur after, or last sufficiently long after, the final dangerous defect-producing transition. In the \(E_6\)-core
the dangerous defects are the \(SU(3)_A\) monopoles, while in the \(E_7\) uplift one must also account for the strings associated with \(U(1)_A\times U(1)_X\to U(1)_Y\).
Inflationary dynamics are constrained by Planck data \cite{Planck2018} and reviewed in~\cite{LythRiotto1999}.

It is instructive to compare the defect evolution of the present model with that of more conventional GUTs. In minimal $SU(5)$ and $SO(10)$ models, symmetry breaking typically produces magnetic monopoles at the unification scale, and no subsequent gauge transition removes them. Avoiding monopole overclosure in such frameworks therefore requires inflation to occur at or below the GUT scale, often in tension with reheating and baryogenesis requirements \cite{FukugitaYanagida1986}. Since in the \(E_6\)-core monopoles arise at the \(SU(3)_A\to SU(2)_L\times U(1)_A\) transition and in the \(E_7\) uplift this is accompanied by a subsequent abelian string-forming transition, inflation
must dilute both the monopoles and any associated string network, occuring after or during these transitions.

This requirement is fully compatible with identifying the $E_6$ breaking Higgs as the inflaton, provided reheating temperature and symmetry restoration are appropriately arranged.
This allows inflation \cite{Guth1981,Linde1983} to occur after the exceptional breakings but before electroweak
symmetry breaking, diluting all dangerous relics without erasing the dark-sector dynamics.
In this sense, the exceptional-group structure provides a built-in solution to the monopole problem that does not rely on ad hoc assumptions and is compatible with both high-scale baryogenesis \cite{FukugitaYanagida1986} and the existence of a stable dark glueball sector.\\

In fact, the requirement that inflation occur after the breaking $G(2)\to SU(3)_C$ has an important additional implication: it naturally preserves the dark $G(2)$ sector responsible for glueball DM \cite{Masi2024}.
Once inflation ends, the $G(2)$ gauge symmetry is not restored because the reheating temperature satisfies $T_{\rm RH}\ll M_\chi$. Below the confinement scale, the massive vector bosons associated with the broken generators bind into colorless dark glueballs. As emphasized in \cite{Masi2024}, these glueballs are automatically stable or
cosmologically long-lived due to the absence of light states carrying $G(2)$ charge, as revised later. See \cite{Masi2021} for insights.


\subsection{Reheating constraints}

After inflation, the reheating temperature $T_{\rm RH}$ must be chosen to avoid
the regeneration of monopoles. This implies $T_{\rm RH} \;<\; T_ \Omega \sim M_ \Omega \simeq M_\Theta$. The most economical choice is to identify the gauge-singlet component of the $ \mathbf{650}$ Higgs,
denoted $\Phi$, as the inflaton. 
Near large field values the potential may be approximated as
 \cite{WeinbergQTF2}
\begin{equation}
	V(\Phi) = \frac{\lambda_\Phi}{4}\left(\Phi^2 - v_{\Phi}^2\right)^2 .
\end{equation}
After inflation, the inflaton oscillates about $v_\Phi$ and reheats via decays to heavy vectors/scalars and matter multiplets, which subsequently cascade to the SM sector \cite{GurseyRamondSikivie1976,AchimanStech1978,HewettRizzo1989,Slansky1981}.
Inflaton decay width into generic fermion pairs is
$\Gamma_\Phi \sim \frac{\bar{y}^2}{8\pi} m_\Phi$, with $\bar{y}$ encoding the Yukawa couplings and $m_\Phi=m_{h_\Phi}$. The reheating temperature is then \cite{KolbTurner1990,AllahverdiReheating2010}
\begin{equation}
	T_{\rm RH} \;\simeq\; 0.2\,\sqrt{\Gamma_\Phi \bar M_{\rm Pl}}
	\;\simeq\;
	0.2\,\bar{y}\,\sqrt{\frac{m_\Phi \bar M_{\rm Pl}}{8\pi}}.
	\label{eq:TRH_kappa}
\end{equation}
For $\bar{y} \sim 10^{-3}$ and $m_{\Phi} \sim 10^{15}\,\text{GeV}$ (with $\lambda_\Phi<1$), we obtain 
\begin{equation}
	T_{\rm RH} \sim 10^{11}\text{--}10^{13}\,\text{GeV}.
\end{equation}
which is compatible with the condition $T_{\rm RH} \le T_{\rm RH}^{\max} \approx v_ \Omega$ that ensures monopoles are not regenerated after inflation \cite{Guth1981,Linde1982}.
Conversely, imposing this monopole-safe bound we obtain 
\begin{equation}
	\bar{y} \;\lesssim\; 
	y_{\max}\;\equiv\;
	\frac{T_{\rm RH}^{\max}}{0.2}\,\sqrt{\frac{8\pi}{m_\Phi \bar M_{\rm Pl}}}
	\;\approx\;
	1.6\times10^{-2}\left(\frac{10^{14}\ \mathrm{GeV}}{m_\Phi}\right)^{1/2}
	\left(\frac{T_{\rm RH}^{\max}}{10^{13}\ \mathrm{GeV}}\right).
	\label{eq:kappa_max_numeric}
\end{equation}
For a $m_\Phi\sim 10^{15}\,$GeV inflaton, this bound produces a reasonable value $y_{\max}\sim 5.1\times10^{-3}$. \\

Then we can estimate possible bounds for the inflaton couplings with the other Higgs scalars. The aforementioned renormalizable portals $V \supset \kappa_{\Phi\chi}(\Phi^\dagger\Phi)(\chi^T\chi)
+\kappa_{\Phi \Omega}(\Phi^\dagger\Phi)\mbox{Tr}( \Omega^\dagger \Omega)+\kappa_{\Phi\Theta}(\Phi^\dagger\Phi)(\Theta^\dagger\Theta)$
generate trilinear couplings $g_{\Phi \sigma\sigma}=\kappa_{\Phi \sigma}v_\Phi$ with $\sigma=\chi, \Omega,\Theta$.
If the decays are kinematically open,\textit{ i.e.} $m_\Phi>2m_\sigma$, the partial widths scale as \cite{KolbTurner1990,AllahverdiReheating2010}
\begin{equation}
	\Gamma(\Phi\to \sigma\sigma)\simeq \frac{N_\sigma}{32\pi}\,\frac{(\kappa_{\Phi \sigma}v_\Phi)^2}{m_\Phi},\,
	\sqrt{1-\frac{4m_\sigma^2}{m_\Phi^2}}\,.
	\label{eq:Gamma_phi_SS}
\end{equation}

where $N_\sigma$ counts the real degrees of freedom accessible in the final state. We take
$N_\chi=7$ for a real $\mathbf7$ of $G(2)$ and $N_ \Omega=8$ for the adjoint $\mathbf8$ of $SU(3)_A$; the
\(\Theta\) multiplicity depends on the projected scalar branch retained in the effective theory. The reheating temperature fixes the total inflaton width as \cite{KolbTurner1990}
\begin{equation}
	\Gamma_{\rm tot}\simeq \left(\frac{\pi^2 g_*}{90}\right)^{1/2}\frac{T_{\rm RH}^2}{\bar M_{\rm Pl}}
	\approx 3.3\,\frac{T_{\rm RH}^2}{\bar M_{\rm Pl}}
	\qquad (g_*\simeq 100).
	\label{eq:Gamma_tot_TR}
\end{equation}
Requiring that a single scalar channel does not saturate the total width ($\Gamma(\Phi\to \sigma\sigma)< \Gamma_{\rm tot}$), yields
\begin{equation}
	\kappa_{\Phi \sigma}<
	\left[\frac{32\pi}{N_\sigma}\frac{m_\Phi\,\Gamma_{\rm tot}}{v_\Phi^2}\right]^{1/2}\left(1-\frac{4m_\sigma^2}{m_\Phi^2}\right)^{-1/4}
	\label{eq:kappa_max_eps1}
\end{equation}
When $
\kappa_{\Phi \sigma}^{\max}=
\left[\frac{32\pi}{N_\sigma}\frac{m_\Phi\,\Gamma_{\rm tot}}{v_\Phi^2}\right]^{1/2}$
for branching ratio ${\rm BR}(\Phi\to \sigma\sigma)\equiv \frac{\Gamma(\Phi\to\sigma\sigma)}{\Gamma_{\rm tot}}\to 1$,
in the benchmark scenario with $m_\Phi\sim v_\Phi=10^{15}\,\mathrm{GeV}$, this translates into
\begin{align}
	T_{\rm RH}=10^{11}\ \mathrm{GeV}:&\quad
	\kappa_{\Phi\chi} \leq 6\times10^{-6},\qquad
	\kappa_{\Phi \Omega} \leq 5\times10^{-6},\nonumber\\
	T_{\rm RH}=10^{13}\ \mathrm{GeV}:&\quad
	\kappa_{\Phi\chi} \leq 6\times10^{-4},\qquad
	\kappa_{\Phi \Omega} \leq 5\times10^{-4}.
\end{align}
These are conservative weak bounds: any additional decay channels into visible matter reduce the allowed $\kappa_{\Phi \sigma}$, and imposing
a ``dark quarantine'' condition ${\rm BR}(\Phi\to \sigma\sigma)\ll1$ further tightens the bounds.

\subsection{Baryogenesis and leptogenesis}
In conventional GUT scenarios such as minimal $SU(5)$, baryogenesis may proceed through the CP-violating decays of superheavy gauge or scalar bosons \cite{FukugitaYanagida1986}. These decays typically generate baryon and lepton numbers while preserving $B-L$, implying that the produced asymmetry is erased by electroweak sphalerons which conserve $B-L$ but violate $B+L$ \cite{Morrissey_2012}. In the
present special embedding, the couplings of the heavy gauge bosons associated with the coset $E_6/(G(2) \times SU(3)_A)$ to light quark--lepton
bilinears are suppressed by the structure of the subgroup embedding and by the vectorlike completion of exotic fermions. Consequently the traditional gauge-boson route to GUT baryogenesis is expected to be inefficient.

A natural alternative is leptogenesis \cite{PhysRevD.104.055007}.
In the \(E_7\)-uplifted benchmark, the heavy sterile right--handed neutrino \(N_i\) is selected from the projected \(\mathbf{2430}_{-3}\) sector with the same
quantum numbers of abelian-breaking scalar, although they belong to distinct fermionic and scalar parent multiplets:
\begin{equation}
N^c_i\subset\mathbf{2430}_{-3},
\qquad
(A,X)=(+3,-3),
\qquad
Y=\frac13(A+X)=0.
\end{equation}
The Majorana masses of these singlet states can originate
from the vacuum expectation value of the a scalar field: since \(N^c\subset\mathbf{2430}_{-3}\), the bilinear \(N^cN^c\) carries \(X=-6\) and the appropriate singlet has $S_N\sim\mathbf1_{+6}$. \(S_N\) is a SM-singlet Majorana scalar with $Y=0$: a pure \(\mathbf1_{+6}\) branch can participate in the \(E_7\)- or \(U(1)_X\)-breaking sector.
The Majorana interaction is then
\begin{equation}
	\mathcal L_N\supset
	\frac12\,y_N^{ij} S_N N_i^cN_j^c+\mathrm{h.c.},
	\qquad
	M_N^{ij}=y_N^{ij}\langle S_N\rangle .
\end{equation}
and it generates, for $\langle S_N\rangle$ around the $E_6$ scale, heavy Majorana masses $M_N \simeq y_N\, v_\Phi$, which naturally lie in the range $M_N \sim 10^{10} - 10^{14}\,\text{GeV}$ for Yukawa couplings of order \(10^{-5}\)--\(10^{-1}\), taken the breaking scale \(v_\Phi \sim 10^{15\div16}\) GeV. Additional neutral components present in the large \(\mathbf{2430}\) parent
are assumed to be projected out or lifted vectorlike at the high scale; if
they were kept light, the leptogenesis analysis would have to be enlarged to a
multi-neutrino and possibly type-III washout system.

In the \(G(2)\)-singlet \(\Theta_{2925}\) construction, the
abelian-breaking field is not used to generate the right-handed-neutrino
Majorana mass. Although the lower abelian charges allow one to write an
effective structure of the schematic form
\((\Theta^\dagger\Theta^\dagger)N^cN^c\), explicit \(E_6\) tensor-product
show that this operator is not generated as a simple
\(E_6\)-invariant coupling for \(\Theta\subset\mathbf{2925}_{+3}\). 

Their
out-of-equilibrium decays can produce a lepton asymmetry which is later partially converted into baryon number by electroweak sphalerons.
After the subsequent breaking of \(SU(3)_A\) to the electroweak group,
the neutrino Yukawa interaction becomes of the type \begin{equation}
	\mathcal{L}_Y \supset y_{\alpha i}\,\overline{L_\alpha} H N_i + h.c. ,
\end{equation} 
with $L_\alpha$ the left-handed lepton and $H$ the SM Higgs doublet, leading to the seesaw relation \citep{ModernPP}
\begin{equation}
	m_\nu \simeq -m_D^T M_N^{-1} m_D ,
\end{equation} where \(m_D = y v_{EW}\) is the Dirac neutrino mass matrix. The CP asymmetry generated in the decay of the lightest heavy neutrino
\(N_1\) is defined as \begin{equation}
	\varepsilon_1 =
	\frac{\Gamma(N_1\rightarrow \ell H)-\Gamma(N_1\rightarrow \bar{\ell}H^\dagger)}
	{\Gamma(N_1\rightarrow \ell H)+\Gamma(N_1\rightarrow \bar{\ell}H^\dagger)} \neq 0.
\end{equation}
For hierarchical heavy neutrinos, the Davidson--Ibarra bound \cite{DavidsonIbarra2002} gives
\begin{equation}
	|\varepsilon_1|
	\lesssim
	\frac{3}{16\pi}\,
	\frac{M_1 (m_3-m_1)}{v^2}
	\simeq
	10^{-6}\left(\frac{M_1}{10^{10}\,\text{GeV}}\right).
\end{equation}
with $v=v_{EW}/\sqrt{2}\simeq 246/\sqrt{2}\simeq 174\,$ GeV in the Yukawa normalization convention used for seesaw leptogenesis and $m_i$ the SM neutrino masses \cite{DavidsonIbarra2002,FongNardiRiotto2012}.
The lepton asymmetry produced by $N_1$ decays can be written
schematically as \cite{FongNardiRiotto2012}
\begin{equation}
	Y_{B-L} \simeq
	\kappa\,\frac{\varepsilon_1}{g_*},
\end{equation}
where $\kappa$ is an efficiency factor accounting for washout effects and $g_*$ denotes again the effective number of relativistic degrees of freedom. Electroweak sphalerons subsequently convert this into a baryon
asymmetry according to 
\begin{equation}
	Y_B = \frac{28}{79} Y_{B-L}.
\end{equation}
The sphaleron conversion factor $28/79$ is the
usual SM result above the electroweak crossover \cite{FongNardiRiotto2012,Strumia2006Leptogenesis}.
Using representative values $\varepsilon_1 \sim 10^{-6}$, $\kappa \sim 10^{-2}$, $g_* \sim 100$,
one obtains \begin{equation}
	Y_B \sim 10^{-10},
\end{equation} which reproduces the observed baryon asymmetry of the Universe \cite{KolbTurner1990}. Thermal leptogenesis typically requires heavy neutrino masses $M_{N_1} \gtrsim 10^{9}\,\mathrm{GeV}$, together with a reheating temperature exceeding this scale \cite{FongNardiRiotto2012}. In the benchmark scenario the reheating temperature lies in the range which is sufficient to thermally produce the heavy neutrinos. Therefore thermal leptogenesis is viable in the benchmark scenario, provided
the reheating temperature is high enough to thermally populate the lightest heavy neutrino.
Due to the dynamical isolation of the $G(2)$ sector, the generation of the baryon asymmetry proceeds entirely within the visible sector and is not diluted by additional dark-sector chemical potentials. This simplifies the cosmological dynamics compared
with scenarios involving asymmetric dark matter \cite{PETRAKI_2013} or dark sphalerons \cite{Blennow_2011}.

Also non-thermal leptogenesis from inflaton decay can be taken into account \cite{FongNardiRiotto2012}: if the inflaton \(\Phi\) couples directly to the heavy neutrinos, its decay may produce a non-thermal neutrino abundance.
The resulting non-thermal baryon asymmetry can be estimated as \begin{equation}
	Y_B^{NT} \sim 
	\frac{28}{79}\,
	\frac{3}{2}\,
	{\rm BR}(\Phi\rightarrow N_1N_1)\,
	\varepsilon_1
	\frac{T_{\rm RH}}{m_\Phi}.
\end{equation}

For $m_\Phi \sim 10^{15}\,\text{GeV}$, $T_{\rm RH} \sim 10^{12}\,\text{GeV}$ and $\varepsilon_1 \sim 10^{-6}$,
the observed baryon asymmetry is obtained for a branching ratio of order 
\begin{equation}
	{\rm BR}(\Phi\rightarrow N_1N_1) \sim 10^{-1},
\end{equation}
that is already sufficient to account for the overall observed baryon asymmetry. Non-thermal leptogenesis is thus viable although it is less minimal than the thermal scenario.

\subsection{Estimating the portal \textbf{$\kappa_{\chi \Omega}$} bound}
To preserve the desired isolation of the $G(2)$ dark enesemble, \textit{i.e.} a proper dark sector \textit{quarantine}, it is sufficient to enforce a small $G(2)$--$SU(3)_A$ portal described by the Higgs $\kappa$ parameters.
For the estimated reheating temperatures $T_{\rm RH} \;\simeq\;10^{11}\text{--}10^{13}\ \text{GeV}$,
two qualitatively different cases may arise:
\begin{itemize}
	\item[(i)] $T_{\rm RH} \ll m_{\rm scalars}$ (that is the main case for $v_{ \Omega,\chi,\Phi}\gtrsim 10^{13-14}$~GeV): heavy Higgs scalars are not thermally populated and the portal acts through higher-dimensional operators after integrating them out. In this case the bounds on
	$\kappa$ become very weak.
	\item[(ii)] $T_{\rm RH} \gtrsim m_{\rm scalars}$ for at least one heavy Higgs scalar: some portals operate via renormalizable scatterings 	among relativistic states.
\end{itemize}

For the second case, considering, for example, the energy transfer between the visible $ \Omega$ bath and the $\chi$ bath through
$\kappa_{\chi \Omega}(\chi^T\chi)\mbox{Tr}( \Omega^\dagger \Omega)$, \textit{i.e.} a renormalizable $2\leftrightarrow 2$ transfer at $T\sim m_{\rm scalars}$, the thermally averaged cross section for relativistic scalars scales as \cite{KolbTurner1990,Hall_2010}
\begin{equation}
	\langle \sigma v\rangle_{ \Omega \Omega\leftrightarrow\chi\chi} \;\sim\; \frac{\kappa_{\chi \Omega}^2}{16\pi\,T^2}\,,
	\label{eq:sigmav_scale}
\end{equation}
up to $\mathcal O(1)$ group-theory and multiplicity factors.
The number density of a relativistic boson is $n\sim \zeta(3)T^3/\pi^2$, so the interaction rate per particle scales as \cite{KolbTurner1990,Hall_2010}
\begin{equation}
	\Gamma_{\rm trans}(T)\;\sim\; n\,\langle \sigma v\rangle \;\sim\; c\,\kappa_{\chi \Omega}^2\,T,
	\qquad c\sim 10^{-2}\text{--}10^{-1},
	\label{eq:Gamma_scale}
\end{equation}
where the constant $c$ lumps phase-space, multiplicity and group factors. Imposing that the interaction rate that transfers energy/number between sectors remains
smaller than the Hubble expansion rate, \textit{i.e.} $\Gamma_{\rm trans}(T_{\rm RH})<H(T_{\rm RH})$ gives
\begin{equation}
	\kappa_{\chi \Omega}\;\lesssim\;
	\left[\frac{1.66\,\sqrt{g_*}}{c}\right]^{1/2}\,
	\sqrt{\frac{T_{\rm RH}}{M_{\rm Pl}}}\,.
	\label{eq:kappa_bound_general}
\end{equation}
Taking $g_*\sim 10^2$ and $c\sim 0.03$ as a representative value yields
\begin{equation}
		\kappa_{\chi \Omega}\ \lesssim\ 6\times 10^{-3}\,
		\left(\frac{T_{\rm RH}}{10^{13}\ \mathrm{GeV}}\right)^{1/2}
		\ \simeq\
		6\times 10^{-3}
		\quad \text{for}\quad T_{\rm RH}=10^{13}\ \mathrm{GeV}.
	\label{eq:kappa_bound_numeric}
\end{equation}
So $\kappa_{\chi \Omega} < 10^{-(3\div2)}$ is the most stringent bound one can put in the most dangerous reheating scenario. The same scaling applies to $\kappa_{\Phi\chi}$ and $\kappa_{\Phi \Omega}$ if the corresponding fields are relativistic at $T_{\rm RH}$, with mild modifications from different multiplicities.


\subsection{Dark Gluons Cosmology}
The $G(2)$ sector is never thermally equilibrated
with the visible sector and must be populated non-thermally, as discussed in \cite{Masi2024,Masi2021} or via suppressed portal interactions.
The cosmological evolution should proceed as follows:

\begin{itemize}
	
	\item {\bf Inflation and $E_6$ breaking.}  
	The radial mode of the $E_6$-breaking Higgs $\Phi_{ \mathbf{650}}$
	drives inflation. At the end of inflation $E_6$ is already broken to $G(2) \times SU(3)_A$.
	
	\item {\bf Reheating.}  
	The inflaton oscillates and decays. The visible sector receives the dominant fraction of energy,
	while the $G(2)$ sector receives only a tiny fraction
	through either direct branching/decay or portal-mediated production and remains colder. We define the dark-to-visible temperature ratio as $\xi \equiv \frac{T_{\rm DM}}{T_{\rm SM}}$.
	
	\item {\bf Internal dark-sector thermalization.}  
	Even if initially underpopulated,
	the $G(2)$ gauge strong interactions rapidly thermalize the dark sector internally (\textit{dark freeze--out} \citep{Masi2021}) due to non-Abelian self-interactions. The two sectors remain thermally decoupled being portal couplings sufficiently small.
	
	\item {\bf $G(2)$ breaking and mass generation.}  
	At $T\sim M_\chi \sim 10^{14}\,\text{GeV}$,
	the Higgs $h_\chi$ acquires a VEV and breaks $G(2) \to SU(3)_C$, six gauge bosons acquiring masses: these heavy vectors constitute the primordial dark ensemble.
	
	\item {\bf Dark confinement and glueball formation.}  
	As the dark temperature decreases below the effective
	confinement scale, the massive $G(2)$ sector forms
	bound states ($J=0$ ``dark glueballs'' might be the stable ones) and number-changing reactions such as $3\to2$
	(``cannibal'' processes \citep{Bernal_2017,Bernal_2016, Bernal_2019}) in a \textit{strong interacting massive particle} (SIMP) scenario determine the final relic density \cite{Forestell2017Glueball,Soni:2016gzf,Heikinheimo_2017}.
	
	\item {\bf Late-time evolution.}  
	Portal interactions remain negligible and the dark glueballs behave as cold dark matter.
\end{itemize}

\paragraph{Direct dimension-five inflaton portal analysis.}
A tempting minimal possibility would be to populate the dark \(G(2)\)-origin sector directly from the inflaton through a dimension-five gauge-kinetic portal \cite{Pustyntseva2024ALPConstraints,GoodsellLieblerStaub2017TwoBodyWidths}:
\begin{equation}
	\mathcal L_5
	=
	\frac{c_5}{\Lambda}\,
	\Phi\,\mathrm{Tr}\, G_{\mu\nu} G^{\mu\nu},
	\label{eq:dim5_phi_GG_revised}
\end{equation}
where \( G_{\mu\nu}\) is the \(G(2)\)-origin field strength, $c_5$ a dimensionless parameter and $\Lambda$ the UV scale. The corresponding two-body width is, up to convention-dependent group factors \cite{Pustyntseva2024ALPConstraints,GoodsellLieblerStaub2017TwoBodyWidths}:
\begin{equation}
	\Gamma_5(\Phi\to G G)
	\simeq
	\frac{c_5^2}{8\pi}
	\frac{m_\Phi^3}{\Lambda^2}.
	\label{eq:Gamma5_phi_GG_revised}
\end{equation}
The resulting dark-sector temperature ratio after reheating is \cite{TenkanenVaskonen2016HiddenReheating}
\begin{equation}
	\xi \sim 
	\left(\frac{\Gamma_{5,\Phi\to G(2)}}{\Gamma_{\Phi\to \rm SM}}\right)^{1/4}.
\end{equation}
For $m_\Phi \sim 10^{15}\,\mathrm{GeV}$ and $\Lambda \equiv \bar M_{\rm Pl} = \frac{1}{\sqrt{8\pi G}} \simeq 2.4\times10^{18}\,\mathrm{GeV}$, one obtains
\begin{equation}
	\Gamma_5
	\simeq
	6.9\times10^6\,\mathrm{GeV}\,
	c_5^2
	\left(\frac{m_\Phi}{10^{15}\,\mathrm{GeV}}\right)^3
	\left(\frac{M_{\rm Pl}}{\Lambda}\right)^2 .
	\label{eq:Gamma5_numeric_revised}
\end{equation}
Thus \(c_5=O(1)\) with a Planckian cutoff would overpopulate the dark sector by many
orders of magnitude.
The smallness of the required branching fraction can be seen directly from entropy
conservation. 
The total inflaton width is fixed by reheating by the previously described formula $\Gamma_{\rm tot}\approx 3.3\,\frac{T_{\rm RH}^2}{M_{\rm Pl}}$. Assuming instantaneous reheating with $n_\Phi/s\simeq 3T_{\rm RH}/(4m_\Phi)$, the dark yield from inflaton two-body decays can be written as \cite{Hall_2010,Bernal_2017}
\begin{equation}
	Y_{\rm DM} \equiv \frac{n_{\rm DM}}{s}\simeq 2\,{\rm BR}_{\rm DM}\,\frac{n_\Phi}{s}
	\simeq \frac{3}{2}\,{\rm BR}_{\rm DM}\,\frac{T_{\rm RH}}{m_\Phi}.
	\label{eq:Y_from_inflaton}
\end{equation}
Taking $m_{\rm DM}\simeq m_{\mathcal X}\sim 10^{14}\,$GeV, the observed abundance requires \cite{KolbTurner1990,GondoloGelmini1991}
\begin{equation}
	Y_{\rm DM}\equiv \frac{n_{\rm DM}}{s_0}
	=\frac{\Omega_{\rm DM}h^2}{m_{\rm DM}}\frac{\rho_c/h^2}{s_0}
	\simeq \frac{4.4\times 10^{-10}}{m_{\rm DM}/{\rm GeV}}
	\;\Rightarrow\;
	Y_{\rm DM}\simeq 4.4\times 10^{-24}.
\end{equation}
with $s_0$ the entropy density today, $\rho_c$ is the cosmological critical density and $h$ the Hubble parameter. 
For an inflaton mass $m_\Phi \sim 10^{15}\,\mathrm{GeV}$ and $T_{\rm RH} \sim 10^{12}\,\mathrm{GeV}$ one finds from the previous formulas
\begin{equation}
	{\rm BR}_{\rm DM}\simeq 2.9\times 10^{-21},
	\qquad
	\Gamma_{\rm tot}\simeq1.4\times10^6\,\mathrm{GeV},
	\qquad
	\Gamma_{\rm DM}
	\simeq
    {\rm BR}_{\rm DM}\Gamma_{\rm tot}
	\simeq
	4.1\times10^{-15}\,\mathrm{GeV}.
	\label{eq:GammaD_required_revised}
\end{equation}
This would imply $\xi\sim {\rm BR}_{\rm DM}^{1/4}\sim 10^{-5}$, \textit{i.e.} a \textit{very cold} dark sector. 
Rewriting Eq.~\eqref{eq:Gamma5_numeric_revised} to put in evidence $c_5$ coupling and assuming $\Gamma_{\rm DM}=\Gamma_5$ gives
\begin{equation}
	c_5
	\simeq
	1.1\times10^{-11}
	\left(\frac{\Gamma_{\rm DM}}{10^{-15}\,\mathrm{GeV}}\right)^{1/2}
	\left(\frac{\Lambda}{M_{\rm Pl}}\right)
	\left(\frac{10^{15}\,\mathrm{GeV}}{m_\Phi}\right)^{3/2},
	\label{eq:c5_required_revised}
\end{equation}
that shows that $c_5 \sim 10^{-11}$ for $\Lambda \sim M_{\rm Pl}$. Conversely, an order-one coefficient would require
\begin{equation}
	\Lambda\sim10^{29}\text{--}10^{30}\,\mathrm{GeV},
	\label{eq:Lambda_transplanckian_revised}
\end{equation}
which lies far above the Planck scale.  Thus the direct dimension-five inflaton--dark-gauge operator is not a viable dominant production mechanism.  



\paragraph{Higher-dimensional inflaton transfer.}

If Eq.~\eqref{eq:dim5_phi_GG_revised} is forbidden, the leading inflaton-to-dark transfer
may arise from higher-dimensional sequestered operators.  A dimension-\(d\) effective
two-body amplitude gives the parametric scaling
\begin{equation}
	\Gamma_d
	\sim
	\frac{c_d^2}{8\pi}\,
	m_\Phi
	\left(\frac{m_\Phi}{\Lambda}\right)^{2(d-4)}.
	\label{eq:Gamma_d_revised}
\end{equation}
This expression is an EFT power-counting estimate; the precise coefficient depends on the
operator structure and group factors. For example, for \(d=8\),
\begin{equation}
	\Gamma_8
	\sim
	\frac{c_8^2}{8\pi}
	\frac{m_\Phi^9}{\Lambda^8}
	\simeq
	3.1\times10^{-14}\,\mathrm{GeV}\,
	c_8^2
	\left(\frac{m_\Phi}{10^{15}\,\mathrm{GeV}}\right)^9
	\left(\frac{M_{\rm Pl}}{\Lambda}\right)^8 .
	\label{eq:Gamma8_revised}
\end{equation}
Hence
\begin{equation}
	c_8
	\simeq
	0.18
	\left(\frac{\Gamma_{\rm DM}}{10^{-15}\,\mathrm{GeV}}\right)^{1/2}
	\left(\frac{\Lambda}{M_{\rm Pl}}\right)^4
	\left(\frac{10^{15}\,\mathrm{GeV}}{m_\Phi}\right)^{9/2}.
	\label{eq:c8_required_revised}
\end{equation}
A leading dimension-eight transfer with \(\Lambda\simeq \bar M_{\rm Pl}\) and \(c_8=O(0.1)\)
therefore gives the correct scale without invoking a trans-Planckian cutoff, provided the
lower-dimensional operators are forbidden.  Such operators should be understood as
sequestered reheating-era transfer operators, possibly involving the inflaton condensate,
rather than as the unsuppressed two-body decay.
\cite{KofmanLindeStarobinsky1994Reheating,KofmanLindeStarobinsky1997ReheatingTheory}.

\paragraph{Portal-mediated freeze-in}

Alternatively, the $G(2)$ sector may be populated
through scalar portals. Visible-sector scatterings $\text{SM SM} \to$ off-shell scalars $\to X X$ produce dark states via \textit{freeze-in} \citep{Masi2021,Bernal_2017,Hall_2010}. If the reheating range satisfies $T_{\rm RH}<M_\chi\sim 10^{14}\,\mathrm{GeV}$, on-shell production of DM from the thermal bath is Boltzmann suppressed by 
\begin{equation}
n^{\rm eq}\propto e^{-m_{h_{\chi}/T}}\sim e^{-100}\simeq 10^{-43}
\end{equation}
for $T = T_{\rm RH}\simeq 10^{12}\,\mathrm{GeV}$, so that these processes are exponentially small for $T\ll m_{h_\chi}$. Consequently, purely renormalizable portal freeze-in is necessarily subdominant unless $T_{\rm RH}\gtrsim M_\chi$ or unless production occurs through UV-sensitive higher-dimensional operators at the
highest temperatures.
The result is again a sequestered $G(2)$ sector whose relic abundance is determined by inflaton decay and subsequent internal strong dynamics.

\paragraph{The \(G(2)\)-breaking scalar as the parent of the dark sector.}
\label{subsec:chi_decay_DM}

The most natural parent of the dark ensemble is the \(G(2)\)-breaking scalar $h_\chi$, which can itself populate the dark sector through decays into the six heavy vectors $\mathcal{X}_\mu\in G(2)/SU(3)_C$ that ultimately constitute the dark ensemble described in \cite{Masi2024}.
A controlled energy fraction may first be transferred from the inflaton sector to the \(\chi\) sector through a scalar coupling, as described before.
A trilinear term gives, with the convention
\(\mathcal L\supset-\frac12 g_{\Phi\chi\chi}\Phi\chi^2+\mathrm{h.c.}\),
\begin{equation}
	\Gamma(\Phi\to\chi\chi)
	\simeq
	\frac{g_{\Phi\chi\chi}^2}{32\pi m_\Phi}
	\left(1-\frac{4m_{h_\chi}^2}{m_\Phi^2}\right)^{1/2}.
	\label{eq:Phi_to_chichi_width_revised}
\end{equation}
For \(m_{h_\chi}\ll m_\Phi/2\) and writing
\(\Gamma(\Phi\to\chi\chi)=\epsilon_{\chi\to {\rm DM}}^{-1}\Gamma_{\rm DM}\), where $\epsilon_{\chi\to {\rm DM}}$ is the energy conversion efficiency from $\chi$ to final dark gluons relics, one has 
\begin{equation}
	g_{\Phi\chi\chi}^{\rm eff}
	\simeq
	\left[
	32\pi m_\Phi\Gamma_{\rm DM}
	\right]^{1/2}\epsilon_{\chi\to {\rm DM}}^{-1/2}.
	\label{eq:mu_eff_with_efficiency}
\end{equation}
When $\epsilon_{\chi\to {\rm DM}}=1$ one obtains the limit
case in which all energy transferred to the $\chi$ sector is transformed into dark matter. Using the previously obtained values
one can estimate the coupling:
\begin{equation}
	g_{\Phi\chi\chi}^{\rm eff}
	\simeq
	20\,\mathrm{GeV}\epsilon_{\chi\to {\rm DM}}^{-1/2}.
	\label{eq:mu_eff_numeric_old_branch}
\end{equation}
This value is not a prediction of the scalar potential but the size of the effective trilinear needed to inject the required tiny energy fraction into the \(\chi\) sector,
as a function of the conversion efficiency into final dark glueballs.
If the effective trilinear originates from the quartic portal so that $ g_{\Phi\chi\chi}^{\rm eff}\sim \kappa_{\Phi\chi}v_\Phi$, one obtains
\begin{equation}
\kappa_{\Phi\chi}
\sim
2\times10^{-14}
\left(\frac{10^{15}\,\mathrm{GeV}}{v_\Phi}\right)\epsilon_{\chi\to {\rm DM}}^{-1/2}.
\end{equation}
This fitted value is nonetheless technically natural: since no bifundamental scalar connects the \(\Phi\) and \(\chi\) sectors (Section~8), no coupling in the theory can regenerate \(\kappa_{\Phi\chi}\) radiatively (except through insertions of itself or of the equally small portals \(\kappa_{\chi\Omega},\kappa_{\Phi\Omega}\)): a small tree-level value is therefore stable under quantum corrections, exactly as for the other portal quartics of the multi-Higgs potential. It is also comfortably consistent with the independent perturbativity bound of Sec.~12.3 -- \(\kappa_{\Phi\chi}\lesssim6\times10^{-6}\) at \(T_{\rm RH}=10^{11}\,\mathrm{GeV}\) and \(\lesssim6\times10^{-4}\) at \(T_{\rm RH}=10^{13}\,\mathrm{GeV}\) -- which sits eight or more orders of magnitude above the bound obtained here, so the fit does not approach the edge of perturbative or ``dark quarantine'' viability.

Once the \(\chi\) sector is populated, the radial mode can decay into the six massive dark
vectors, if the channel is open.
The decay $h_\chi\to \mathcal X\mathcal X$ is kinematically allowed if
\begin{equation}
	m_{h_\chi} > 2m_{\mathcal X}
	\quad\Longleftrightarrow \quad
	\sqrt{2\lambda_\chi}\,v_\chi > 2 g_{G(2)}v_\chi
	\quad\Longleftrightarrow \quad
	\lambda_\chi > 2 g_{G(2)}^2.
	\label{eq:gate}
\end{equation}
For $g_{G(2)}\sim 0.6$, this corresponds to $\lambda_\chi\gtrsim 0.7$. Thus, choosing
$\lambda_\chi\lesssim 2g_{G(2)}^2$ kinematically closes the direct \(h_\chi\to\mathcal X\mathcal X\) channel and avoids
overproduction from a post-inflationary \(h_\chi\) condensate \cite{Enqvist_2013}.

If we allow the decay, from the $\chi$ kinetic term one obtains a trilinear coupling $h_\chi \mathcal{X}_\mu \mathcal{X}^\mu$ of order $g_{G(2)}^2 v_\chi$. Parametrically, the width scales as
\begin{equation}
	\Gamma(h_\chi\to \mathcal X \mathcal X)\ \sim\ 
	\frac{N_\chi}{64\pi}\,\frac{m_{h_\chi}^3}{v_\chi^2}\,
	\sqrt{1-\frac{4m_{\mathcal X}^2}{m_{h_\chi}^2}},
	\qquad N_\chi=6,
	\label{eq:Gamma_chi_to_XX}
\end{equation}
up to $\mathcal{O}(1)$ group-theory factors fixed by the chosen generator normalization. This channel is
therefore potentially efficient and must be kinematically controlled by the initial \(\chi\)-sector energy fraction.

Assuming $h_\chi$ decays during the inflaton-dominated era (before completion of reheating), the post-reheating yield of ultra-heavy vectors produced in $h_\chi\to \mathcal X \mathcal X$ is estimated as \cite{KolbTurner1990,Hall_2010}
\begin{equation}
	Y_{\rm DM} \equiv \frac{n_{\mathcal X}}{s}\ \simeq\
	\frac{3}{4}\,\frac{T_{\rm RH}}{m_{\mathcal X}}\,
	\left.f_\chi\right|_{\rm decay}
	\times {\rm BR}(h_\chi\to \mathcal X\mathcal X).
\end{equation}
where $f_\chi=\rho_\chi/\rho_{\rm tot}$ represents the energy fraction carried by $h_\chi$ at the decay.
Taking ${\rm BR}(h_\chi\to \mathcal X \mathcal X)=1$ and matching $Y_{\rm DM}\simeq 4.4\times 10^{-24}$ gives
\begin{equation}
	\left.f_\chi\right|_{\rm decay}\le	\frac{2}{3} Y_{\rm DM}\,\frac{m_{\rm DM}}{T_{\rm RH}}
	\ \simeq\
	\begin{cases}
		2.9\times 10^{-21}, & T_{\rm RH}=10^{11}\,\mathrm{GeV},\\
		2.9\times 10^{-22}, & T_{\rm RH}=10^{12}\,\mathrm{GeV},\\
		2.9\times 10^{-23}, & T_{\rm RH}=10^{13}\,\mathrm{GeV}.
	\end{cases}
\end{equation}
Thus, unless the $\chi$ sector energy density at decay is extraordinarily suppressed, $h_\chi\to \mathcal X \mathcal X$ would overproduce ultra-heavy dark relics, motivating either kinematic closure of the decay or suppression of a
post-inflation $h_\chi$ condensate \cite{Enqvist_2013}.

\paragraph{\(\chi\)-misalignment and \(G(2)\)-breaking transition production.}

A displaced \(\chi\) field can also act as a spectator source.  If it begins coherent
oscillations, when $H\simeq m_{h_\chi}$, for a quadratic potential
\begin{equation}
	\rho_\chi^{\rm osc}
	\simeq
	\frac12m_{h_\chi}^2\chi_i^2
	\label{eq:rho_chi_osc_revised}
\end{equation}
where $\chi_i$ is the initial value. A coherently oscillating scalar redshifts as matter
\cite{Turner1983CoherentOscillations}.  If only a tiny initial displacement is present,
the \(\chi\)-condensate can seed the dark sector without overproducing it.  The abundance is
then controlled by the initial condition \(\chi_i\), the subsequent fragmentation/decay of
the condensate and the branching into dark vectors.

Another possibility is nonthermal production at the \(G(2)\)-breaking phase transition \cite{BakerKoppLong2020FilteredDM,GiudiceLeePomarolShakya2024FOPTDM}.
If the transition is first order, the time-dependent background can produce heavy vector
and scalar excitations: a fraction $\epsilon_{\rm DM}$ of the released vacuum energy $\Delta U_\chi$ is converted into long-lived dark sector quanta.
This mechanism is viable but model-dependent, because it depends on the finite-temperature
\(\chi\) potential, wall velocity, transition strength and coupling to the heavy vectors
\cite{BakerKoppLong2020FilteredDM,GiudiceLeePomarolShakya2024FOPTDM}.

\paragraph{Irreducible gravitational production.}

Even if all non-gravitational portals between the visible/inflaton sector and the
\(G(2)\)-origin sector are forbidden, a minimal gravitational contribution could remain \cite{ChungKolbRiotto1998NonthermalSupermassive,ChungKolbRiotto1999ReheatingProduction}. During
reheating, dark-sector quanta can be produced through graviton-mediated processes such as $
\Phi\Phi\to \mathcal X\mathcal X$ and ${\rm SM}\,{\rm SM}\to \mathcal X\mathcal X$,
with amplitudes suppressed by powers of the reduced Planck mass.  
In a thermalized visible bath, the gravitational source term has the generic scaling
\begin{equation}
	R_{\rm grav}
	\equiv
	\left.\frac{dn_{\mathcal X}}{dt}\right|_{\rm grav}
	\sim
	\frac{T^8}{\bar M_{\rm Pl}^4}.
	\label{eq:Rgrav_scaling_dark_gluon}
\end{equation}
This follows from \(R\sim n_{\rm bath}^2\langle\sigma v\rangle\), with
\(n_{\rm bath}\sim T^3\) and a graviton-mediated cross section
\(\langle\sigma v\rangle\sim T^2/\bar M_{\rm Pl}^4\)
\cite{GarnyPalessandroSandoraSloth2018PIDM,MambriniOlive2021GravDM}.
For \(m_{\mathcal X}\gtrsim T\), this scaling must be multiplied by the appropriate
threshold or nonthermal reheating factor.
This mechanism is therefore unavoidable but very small and in the present work it is not considered
as the primary origin of the relic abundance.  It is better interpreted as an irreducible
floor on the population of the secluded \(G(2)\)-origin sector.  The benchmark cosmology
keeps the \(\chi\)-parent mechanism as the dominant source, while graviton exchange provides
only a subleading contribution unless the sequestered scalar and phase-transition channels
are absent \cite{ChungKolbRiotto1998NonthermalSupermassive,
	ChungKolbRiotto1999ReheatingProduction}.

\paragraph{Hidden glueball evolution after dark sector population.}

Once a small energy fraction has been deposited into the \(G(2)\)-origin sector, the
subsequent evolution is governed almost entirely by the dark non-Abelian dynamics already
developed in the original \(G(2)\) dark-gluon scenario \cite{Masi2021,Masi2024}.  The
microscopic production mechanism does not need to create the final glueballs directly.  It
only has to populate the secluded sector with heavy \(G(2)/SU(3)_C\) vectors, dark gluonic
degrees of freedom or excitations of the \(G(2)\)-breaking scalar. These states then redistribute their energy through strong self-interactions and, when the dark temperature
falls below the appropriate confinement scale, hadronize into hidden glueball bound states.
The lightest scalar glueball $J=0$
is the natural late-time relic, while heavier glueball excitations cascade to the lightest
state if the allowed dark self-interactions and portal-suppressed decays are sufficiently
efficient.  This is the same physical picture underlying hidden Yang--Mills glueball dark
matter and non-standard hidden-sector cosmologies
\cite{Forestell2017Glueball,Soni:2016gzf,SONI2017379,
	Acharya2017GlueballNonStandard,
	CarenzaPasechnikSalinasWang2022GlueballRevisited,CarenzaPasechnikSalinasWang2023GlueballDM}.
A useful way to describe the hidden sector after population is in terms of the dark temperature ratio $\xi$ or, more generally, in terms of the entropy ratio $\zeta\equiv\frac{s_{\rm DM}}{s_{\rm SM}}$.
Because the visible and dark sectors are sequestered, \(\zeta\) is approximately conserved
after the production epoch, while the dark sector can remain internally thermalized.  At
confinement, the energy stored in the dark plasma is converted into glueball states.  If
\(m_{DM}\) denotes the mass of the lightest dark glueball, the final abundance may be
written schematically as
\begin{equation}
	\Omega_{DM}h^2
	=
	\frac{m_{DM}\,Y_{DM}\,s_0}{\rho_c/h^2},
	\qquad
	Y_{DM}\equiv\frac{n_{DM}}{s_{\rm SM}} .
	\label{eq:glueball_relic_abundance_general}
\end{equation}
The yield \(Y_{DM}\) is not simply the primordial number of particles produced by the aforementioned mechanisms, but it is determined after dark confinement by internal number-changing processes, especially $3\to2$ cannibal reactions.  These reactions heat the dark bath relative to pure radiation
redshifting and reduce the comoving number density until they freeze out.  In the usual
cannibal regime the Boltzmann equation has the schematic form
\begin{equation}
	\dot n_{DM}+3Hn_{DM}
	=
	-\langle\sigma_{3\to2}v^2\rangle
	\left(n_{DM}^3-n_{DM}^2n_{DM,{\rm eq}}\right),
	\label{eq:cannibal_boltzmann_schematic}
\end{equation}
where $\langle\sigma_{3\to2}v^2\rangle$ is the effective three-body number-changing reaction coefficient, with freeze-out determined by
\begin{equation}
	n_{DM}^2
	\langle\sigma_{3\to2}v^2\rangle
	\sim
	H .
	\label{eq:cannibal_freezeout_condition}
\end{equation}
The detailed value of \(\langle\sigma_{3\to2}v^2\rangle\) is nonperturbative and depends on
the hidden confinement dynamics; it must be therefore treated phenomenologically, as in standard
SIMP/cannibal dark-matter analyses
\cite{Bernal_2016,Bernal_2017,Bernal_2019,Heikinheimo_2017}.
Therefore the final relic abundance is controlled
by the combination of this initial fraction, the dark temperature ratio, the confinement
scale and the efficiency of cannibalization. The abundance condition is more robustly imposed at the level of the dark energy density \citep{Aghanim:2018eyx,PDG2024},
\begin{equation}
		\frac{\rho_{\rm DM}}{s_0}
		=
		\Omega_{\rm DM}h^2
		\frac{\rho_c/h^2}{s_0}
		\simeq
		0.12\,
		\frac{1.05\times10^{-5}\,\mathrm{GeV\,cm^{-3}}}
		{2.89\times10^3\,\mathrm{cm^{-3}}}
		\simeq
		4.4\times10^{-10}\,\mathrm{GeV},
		\label{eq:rhoDM_over_s0}
\end{equation}
rather than at the level of the initially produced particle number, so that the physical dark matter population is identified with the late \(J=0\) glueball relic
formed after dark confinement.  \\

The viable production mechanisms are summarized in Table~\ref{tab:dark_production_channels}. The \(G(2)\)-breaking field \(\chi\) is the
immediate parent of the dark ensemble, while higher-dimensional inflaton transfer, \(\chi\)-misalignment and phase-transition production are controlled alternatives.

\begin{table}[h]
	\centering
	\small
	\renewcommand{\arraystretch}{1.25}
	\setlength{\tabcolsep}{4pt}
	\begin{tabularx}{\textwidth}{>{\raggedright\arraybackslash}p{0.23\textwidth}
			>{\raggedright\arraybackslash}p{0.29\textwidth}
			>{\raggedright\arraybackslash}p{0.19\textwidth}
			>{\raggedright\arraybackslash}X}
		\hline
		Mechanism & Main operator/source & Main parameter & Status \\
		\hline
		Direct inflaton to dark gauge bosons
		&
		\(\Phi\, G_{\mu\nu} G^{\mu\nu}/\Lambda\)
		&
		\(c_5/\Lambda\)
		&
		Excluded as dominant channel unless \(c_5\lesssim10^{-11}\).
		\\[1mm]
		
		Higher-dimensional inflaton transfer
		&
		Leading \(d=8\) sequestered operator
		&
		\(c_8/\Lambda^4\)
		&
		Viable if lower-dimensional operators are forbidden.
		\\[1mm]
		
		Inflaton to \(G(2)\)-breaking scalar
		&
		\(g_{\Phi\chi\chi}^{\rm eff}\Phi\chi^2\)
		&
		\(g_{\Phi\chi\chi}^{\rm eff}\)
		&
		Viable controlled branching into the \(\chi\) sector.
		\\[1mm]
		
		\(\chi\)-misalignment
		&
		\(\rho_\chi^{\rm osc}\simeq m_{h_\chi}^2\chi_i^2/2\)
		&
		\(\chi_i\)
		&
		Viable spectator-source mechanism.
		\\[1mm]
		
		\(G(2)\)-breaking phase transition
		&
		\(\epsilon_{\rm DM}\Delta U_\chi\)
		&
		\(\epsilon_{\rm DM},\Delta U_\chi\)
		&
		Viable but model-dependent.
		\\[1mm]
		
		Gravitational production
		&
		\(\Phi\Phi,\,{\rm SM\,SM}\to\mathcal X\mathcal X\) via graviton exchange
		&
		\(\bar M_{\rm Pl}^{-1}\), \(T_{\rm RH}\), \(H_{\rm end}\)
		&
		Irreducible but subleading floor; not the preferred benchmark.
		\\[1mm]
		
		Hidden glueball evolution
		&
		Dark confinement and cannibalization
		&
		\(T_{\rm DM}/T_{\rm SM}\), entropy ratio, confinement scale, canibalization
		&
		Determines final glueball abundance.
		\\[1mm]
		\hline
	\end{tabularx}
	\caption{Cosmological production channels for the secluded \(G(2)\)-origin dark sector.
		The revised benchmark removes the unsuppressed dimension-five inflaton--dark-gauge portal
		and treats \(\chi\), the \(G(2)\)-breaking scalar, as the immediate parent of the dark
		population. $H_{\rm end}$ is Hubble expansion rate at the end of inflation.}
	\label{tab:dark_production_channels}
\end{table}

\subsection{CMB and gravitational-wave signatures of exceptional phase transitions}
\label{sec:CMB_GW}

The staged symmetry breaking of the present exceptional framework may leave imprints in cosmological observables, most notably in the stochastic
gravitational-wave background and, in limited cases, in the CMB \cite{KolbTurner1990}.
The nature and observability of such signals depend on the order of the phase transitions, the associated energy scales and the timing of inflation, because any gravitational-wave signal generated before inflation is exponentially diluted.

The combined transition $SU(3)_A\times U(1)_X\to SU(2)_L\times U(1)_Y$, which is the last one in the chain,
can produce \(SU(3)_A\) monopoles and, after the \(\Theta\) breaking, abelian strings. In the viable cosmological history these defects must be inflated away
and not regenerated after reheating.

Nevertheless, if the transition is sufficiently first order, it can also source a stochastic gravitational-wave background.
A first-order transition completes over a characteristic time scale $\beta^{-1}$, where $\beta$ is the inverse duration parameter. The gravitational waves (GW) spectrum typically peaks at a production-frame frequency $f_* \;\sim\; \frac{\beta}{2\pi}\times \mathcal{O}(1)$,
up to channel-dependent order-one factors (bubble collisions, sound waves, turbulence). The observed frequency today $f_0$ is redshifted by the ratio of Universe expansion scale factors $a$, so that $f_0 \;=\; f_*\,\frac{a_*}{a_0}$. Assuming radiation domination at production temperature $T_*$ and entropy conservation thereafter \cite{KolbTurner1990,DodelsonModernCosmology},
\begin{equation}
	\frac{a_*}{a_0} \;=\; \frac{T_0}{T_*}\left(\frac{g_{*S,0}}{g_{*S,*}}\right)^{1/3},
\end{equation}
with $T_0\simeq 2.35\times 10^{-13}\,$GeV and $g_{*S,0}\simeq 3.91$. Using the standard radiation-era Hubble scale $H_* \;=\; 1.66\,\sqrt{g_*}\,\frac{T_*^2}{M_{\rm Pl}}$,
one obtains the widely used and robust scaling for the peak frequency in terms of $\beta/H_*$ \cite{Caprini2016Review}
\begin{equation}
		f_0 \;\simeq\; 1.65\times 10^{-5}\,\mathrm{Hz}\,
		\left(\frac{\beta}{H_*}\right)\,
		\left(\frac{T_*}{100\,\mathrm{GeV}}\right)\,
		\left(\frac{g_*}{100}\right)^{1/6}
	\label{eq:f0_master}
\end{equation}
 Equation~\eqref{eq:f0_master} shows that \emph{high-scale} phase transitions necessarily produce \emph{ultra-high-frequency} GW backgrounds today \cite{Schmitz2021GW}. For our characteristic scales, 
\begin{equation}
	T_* \sim v_ \Omega \sim 10^{13}\,\mathrm{GeV}
	\qquad \Rightarrow\qquad
	f_0 \simeq 1.65\times 10^8\,\mathrm{Hz}\left(\frac{\beta/H_*}{100}\right),
\end{equation}
\textit{i.e.}\ in the tens-to-hundreds of MHz regime for $\beta/H_*\sim10\div100$, far above the expected sensitivity bands of LISA ($10^{-4}$--$10^{-1}$ Hz), Einstein Telescope / Cosmic Explorer ($\sim 1$--$10^3$ Hz) and PTA experiments ($10^{-9}$--$10^{-7}$ Hz) \cite{CapriniEtAl2020ScienceCaseLISA,Maggiore2000GWReview,caprini20}. Therefore, even in the optimistic case of a strong first-order transition surviving after inflation, the frequency location of the signal places it outside the reach of current and near-future planned interferometers. Consequently the absence of a
detectable gravitational-wave signal will be consistent with the present framework \cite{Schmitz2021GW}.

Concluding, the exceptional symmetry-breaking chain of the present model provides a coherent cosmological description: by arranging inflation to occur after the last monopole-producing transition,
all dangerous defects are diluted, while later symmetry breaking steps do not reintroduce them.
Reheating and baryogenesis \cite{FukugitaYanagida1986} can be consistently implemented,
preserving the dark $G(2)$ sector \cite{Masi2024} and maintaining overall agreement with observational bounds.

\section{Why an Exceptional $E_6$ Embedding: an \textit{a posteriori} Perspective}
\label{sec:whyE6}

In this conclusive section we summarize the structural advantages of the present framework and clarify why the special-embedding $E_6$ construction offers a distinct and predictive unification path \cite{LiebeckSeitz2004Survey,Seitz1991LocalMaxSubgroups,Craven2023MaxSubgroups}, while the \(E_7\) uplift should be understood as the minimal abelian completion of
the special \(E_6\) backbone.
We contrast it with conventional regular-embedding GUTs and emphasize the simultaneous control of particle physics, dark-sector dynamics and cosmology.

\paragraph{Special versus regular embeddings.}

Most unified models based on $E_6$ exploit its \emph{regular} maximal subgroups, such as $E_6 \supset SO(10)\times U(1), E_6 \supset SU(3)\times SU(3)\times SU(3)$.
In these cases the subgroup is generated by deleting nodes from the $E_6$ \cite{GurseyRamondSikivie1976,AchimanStech1978,HewettRizzo1989,Slansky1981}
Dynkin diagram and the resulting gauge bosons typically include leptoquark
vectors that couple directly to quark--lepton currents.
Such couplings generically induce rapid proton decay \cite{Weinberg1979Bviol,WilczekZee1979,SuperK2017,PDG2024} unless the unification
scale is pushed close to the Planck scale.

By contrast, the embedding $E_6 \supset G(2) \times SU(3)_A$ is \emph{special} (non-regular): it is not associated with a Dynkin-node
deletion and instead arises from an exceptional subalgebra structure \cite{Dynkin1957,Slansky1981,Yamatsu2015}.
In classical GUTs (e.g. $SU(5)$, $SO(10)$), the coset vectors typically carry mixed color--electroweak charges, generating tree-level leptoquark interactions and dangerous $d{=}6$ proton-decay operators: along the \emph{special} $E_6\supset G(2) \times SU(3)_A$ route, the broken cosets factorize and, with no bifundamental Higgs, there are \emph{no} renormalizable leptoquark vertices.
This structural feature is the origin of the parametric suppression of baryon-number violating processes and establishes orthogonality of $SU(3)_A$ and $G(2)$ generators, granting the group-theoretic separation between the \(G(2)\)-origin dark sector and
the electroweak ancestor sector.

\paragraph{Minimality of the GUT.}

Among exceptional groups, \(E_6\) is the smallest group that contains the special product $G(2)\times SU(3)_A$. In this precise sense, \(E_6\) is the
minimal exceptional backbone capable of incorporating a \(G(2)\) dark
strong sector and an electroweak progenitor group within a single simple framework. The use of the \(\mathbf{650}_{0,H}\) Higgs sector further provides
an economical way to realize the special breaking and to organize the lower thresholds.

However, the strict \(E_6\)-only realization has a limitation: if hypercharge is required to lie entirely inside the \(SU(3)_A\) weight lattice, the SM charge assignments and the canonical high-scale value of the weak angle cannot be obtained simultaneously with the desired integer normalization. This is not a failure of the special \(E_6\) embedding, rather it identifies the minimal
place where the model needs an abelian completion. The regular exceptional uplift $E_7\to E_6\times U(1)_X$
provides precisely this completion, where the hypercharge is embedded as $Y=\frac13(A+X)$, being $A=\sqrt3\,t_A^8$.

\paragraph{Fermion representations and projected matter.}


In the minimal \(E_6\) picture, the \(\mathbf{27}\) representation provides a
useful reference point for organizing chiral matter and vectorlike partners.
However, the final construction is not a minimal \(3\times\mathbf{27}\)
model.  The low-energy fermions are more accurately described as projected
zero modes selected from \(E_7\)-derived \(E_6\times U(1)_X\) branches.
The reps scan identifies two preferred completions.  In Class A, the low-lepton-representation solution is
\begin{equation}
	\overline{\mathbf{27}}_{+1}
	\oplus
	\mathbf{351}_{-1}
	\oplus
	\mathbf{2430}_{-3}
	\oplus
	\mathbf{2430}_{+3},
\end{equation}
In Class B, the branch-economical solution is
\begin{equation}
	\overline{\mathbf{351}}_{+1}
	\oplus
	\mathbf{351}_{-1}
	\oplus
	\mathbf{2925}_{-3}
	\oplus
	\mathbf{2925}_{+3}.
\end{equation}
This class is especially economical at the branch level and allows the leptonic, sterile-neutrino and \(\Theta\) sectors to be chosen from
\(G(2)\)-singlet branches.
The role of the \(E_7\) macro-parent representation is to certify that these
branches can arise from a common exceptional origin.  Class A is naturally
embedded in the \(\mathbf{86184}_{E_7}\) macro-parent, while Class B can be
embedded already in the smaller \(\mathbf{27664}_{E_7}\) and also in
\(\mathbf{86184}_{E_7}\).  The projection rule keeps the Standard Model
chiral fields and the sterile-neutrino sector, while unwanted exotic
components are assumed to have zero net chirality and to be lifted through
vectorlike pairings at the appropriate high thresholds.
The special \(E_6\) subgroup remains
the structural reason for the separation between the Standard Model sector
and the \(G(2)\) dark sector, while the \(E_7\) uplift supplies the charged
branch structure needed for hypercharge, Yukawa closure, and vectorlike
completion.

\paragraph{Proton decay and flavor protection.}

In regular GUT embeddings, proton decay \cite{Weinberg1979Bviol,WilczekZee1979,SuperK2017,PDG2024} arises from unsuppressed tree-level
exchange of leptoquark gauge bosons.
Here, because the path is \textit{special}, some leptoquark-type gauge couplings that plague regular embeddings can be absent at tree level.
Baryon-number violation is induced only through mixing between light fermions and heavy $G(2)$-charged states and is therefore significantly suppressed compared
with minimal regular embeddings such as $SU(5)$.  Improvements in experimental proton-lifetime limits therefore provide an important indirect probe of the theory \cite{NathFileviezPerez2007}. 
The same mechanism suppresses flavor-changing neutral currents, allowing a common explanation of proton stability and flavor hierarchy.

\paragraph{Running and unification.}

The staged breaking $E_6 \to G(2) \times SU(3)_A \to SU(3)_C\times SU(3)_A \to
SU(3)_C\times SU(2)_L\times U(1)_A$
leads to a calculable piecewise running of the gauge couplings. In the
completed theory this chain is supplemented by the abelian matching $U(1)_A\times U(1)_X\to U(1)_Y$.
The projected spectrum is chosen so that the full \(E_7\) parent multiplets do not
run as light degrees of freedom over large energy intervals; otherwise
perturbativity would be lost. Instead, exotic states are lifted at controlled
thresholds, while the effective running below \(M_{E_6}\) remains governed by
the projected low-energy field content.
The matching conditions at each threshold are fixed by group theory and a benchmark scenario was developed to yield consistent unification at $\sim 10^{15\div16}$ GeV, several orders of magnitude below the Planck scale. 

\paragraph{Visible--dark sector separation.}

The exceptional factor $G(2)$ confines into heavy glueballs \cite{Masi2024} that are neutral under the SM.
Because $G(2)$ is completely separated from the electroweak progenitor group at the first breaking step, the dark sector is protected from large
visible-sector interactions. The \(SU(3)_A\)-coset vectors are \(G(2)\)-singlets and do not hadronize into the dark glue sector. The \(E_7\) uplift adds the \(U(1)_X\) gauge boson and the heavy abelian \(Z'\), but these are also \(G(2)\)-singlets. The \(Z'\) is therefore not a
dark gauge boson. In addition, the abelian-breaking scalar can also be chosen to be a true \(G(2)\)-singlet. 
This naturally realizes a hidden strong sector without introducing ad hoc stabilizing symmetries. The mass of the dark particles ($\sim 10^{13}$--$10^{14}$ GeV) is far above the scope of collider searches, leaving the SM untouched.

\paragraph{Cosmological advantages.}

The exceptional symmetry-breaking chain admits a consistent cosmological history.
Topological defects produced at the highest scales are diluted by inflation \cite{Guth1981,Linde1983}, which may be associated with the $E_6$ breaking Higgs $h_\Phi$ itself,
the dark sector survives reheating, and baryogenesis \cite{FukugitaYanagida1986} can be realized through
heavy right--handed neutrino $E_6$ states via leptogenesis \cite{PhysRevD.104.055007,FongNardiRiotto2012,Morrissey_2012}, while their Majorana masses may arise either from a high-scale singlet direction or from \(\Theta\)-dependent effective operators.
The presence of a confining dark sector further allows macroscopic bosonic compact objects that can produce primordial black holes \cite{Masi2024, Masi2021} or constitute a class of black hole mimickers, linking unification to astrophysics. The implications of these dark boson stars will be addressed in details in a separate work.

\section{Conclusions and Outlook}
\label{sec:conclusions}

A coherent GUT realization has been developed to satisfy fundamental current constraints from unification, proton decay and cosmology, while preserving a confining dark sector. The core of this unified framework is based on a special embedding of the exceptional group $E_6$ in which the first symmetry-breaking step realizes 
\begin{equation}
E_6\to G(2) \times SU(3)_A 
\end{equation}
The construction naturally couples an exceptional enlarged strong theory $G(2)$, which generates standard QCD together with a dark $S^6$--type sector, to an $SU(3)_A$ gauge sector that acts as an ancestor of the electroweak SM group. In the minimal \(E_6\) realization, the electroweak abelian direction is first attempted inside \(SU(3)_A\), with the SM Higgs doublet identified within a higher-dimensional representation. This already
captures the central structural idea of the model: QCD, a secluded dark \(G(2)\)-origin sector and the electroweak ancestor group arise as complementary substructures of a single exceptional embedding.

The staged breaking
$G(2)\to SU(3)_C$ and $SU(3)_A\to SU(2)_L\times U(1)_A$ driven by $\chi\in\mathbf{7}_{G(2)}$ and $ \Omega\in\mathbf{8}_{SU(3)_A}$ provides a coherent path to the SM phenomenology, yielding to a consistent ensemble of heavy gauge bosons and a predictive pattern of gauge coupling running.
The special exceptional embedding suppresses dangerous tree-level leptoquark couplings and exhibits controlled one-loop running with a clean $\alpha_{G(2)}{=}\alpha_A=\alpha_{E_6}$ unification below $M_\mathrm{Pl}$. Flavor-changing and baryon-number-violating processes are simultaneously suppressed by the special embedding and by controlled mixing effects. 

A central lesson of the analysis is that the minimal \(E_6\) construction is
structurally powerful but incomplete in its strict hypercharge realization. If
hypercharge is required to live entirely inside the \(SU(3)_A\) weight lattice,
one encounters a normalization obstruction: the integer \(SU(3)_A\) charges
needed to assign the SM hypercharges do not simultaneously reproduce the
canonical high-scale value $\sin^2\theta_W=\frac38$. This motivates the exceptional completion $E_7\to E_6\times U(1)_X$, which leaves the special \(E_6\to G(2)\times SU(3)_A\) mechanism untouched but
adds the abelian freedom required to complete the electroweak sector. In the
uplifted theory, hypercharge is embedded as a mixed generator,
\begin{equation}
Y=\beta A+\gamma X,
\end{equation}
where $A=\sqrt3\,t_A^8$ is the diagonal \(SU(3)_A\) generator. In the preferred benchmark one may take $Y=\frac13(A+X)$, with the \(U(1)_X\) normalization chosen so that, for
\(g_X(M_\Omega)\simeq g_A(M_\Omega)\), one obtains $\sin^2\theta_W(M_\Omega)=\frac38$.
The final fermion embedding is therefore best understood as a projected \(E_7\)-derived spectrum. 

A coherent $E_6$ reps configuration uses the branch structure contained in $\mathbf{86184}_{E_7}$, with the Standard Model chiral zero modes selected by the projection rule and
all unwanted mirror components assigned zero net chirality. The analysis demonstrated that the hypercharge assignments obstruction can be solved with a correct pattern of representations for quarks and leptons, keeping all scalars $G(2)$-singlets and preserving the special embedding realization.
In addition, in this matter content sterile-neutrino sector is naturally present and supports a seesaw origin of neutrino masses and high-scale leptogenesis, linking the generation of the baryon asymmetry directly to the symmetry-breaking dynamics of the theory.

Exotic states are lifted vectorlike and removed from the low energy SM dynamics. 
This multi-threshold lifting should be compatible with the running analysis and with the projected chiral spectrum.

 \textit{Natural} portal bounds ensure the complete decoupling of non-$G(2)$ sectors from the dark glueballs ensemble. In fact, the other \textit{broken} bosons ($\mathcal{E}$-bosons from $E_6$, $Y$-bosons from $SU(3)_A$ steps ans $Z'$ from $U(1)_X$) are not dark relics: they are neutral under $G(2)$, quarantined from the dark sector by charge orthogonality and absence or suppression of renormalizable mixings, decaying promptly to visible states. Also all the scalar Higgs particle responsible for the symmetries breakings (except for $\chi$) can likewise be taken \(G(2)\)-neutral.
The dark $G(2)$ sector proposed in \cite{Masi2024} emerges here as an intrinsic component of a larger unified theory rather than as an addition dictated only by algebraic reasons, as claimed in \cite{Masi2021}. The present construction embeds that sector inside a special exceptional GUT, explains why it can remain secluded, and identifies the threshold conditions under which the dark glueball ensemble is not destabilized by visible, abelian, or electroweak interactions.

From a cosmological perspective, the model admits a fully consistent history: a \textit{for free} inflaton, topological defects produced at the high energy $SU(3)_A$--$U(1)$ breaking scale are diluted by inflation, the dark $G(2)$ sector survives reheating and baryogenesis
may proceed through heavy-neutrino leptogenesis, with a coherent chronological evolution.  

On the observational side, the interplay between dark glueball phenomenology,
gravitational-wave astronomy and black-hole astrophysics \cite{Masi2024,Cardoso_2019} could offer a rich arena for applying and testing this exceptional unification in the coming years. 
In this sense, the exceptional
group \(E_6\) provides the special non-regular backbone of the construction,
while the \(E_7\) uplift supplies the minimal abelian completion needed for
realistic hypercharge, the canonical weak-angle normalization, and a consistent
projected matter spectrum. The resulting framework gives a concrete bridge
between particle physics, cosmology and dark matter astrophysics, with a
secluded \(G(2)\)-origin dark sector embedded in a controlled exceptional
unification scheme.


\appendix
\section{Explicit \(SU(3)_C\) one-loop coefficient in the intermediate regime}
\label{app:colorcount}


\subsection{Projected color spectrum in the final \(E_7\)-completed theory}
In the intermediate regime the effective gauge group is $
SU(3)_C\times SU(3)_A$, after the breaking $G(2)\to SU(3)_C$ at the scale \(M_\chi\), but before $SU(3)_A\to SU(2)_L\times U(1)_A$ at \(M_\Omega\). 
For \(SU(3)_C\), with the standard normalization $\mathrm{Tr}(T^aT^b)=\frac12\delta^{ab}$
one has \begin{equation}C_2(SU(3)_C)=3,
	\qquad
	T(\mathbf3)=T(\overline{\mathbf3})=\frac12.
\end{equation}
In the original unprojected \(E_6\)-core discussion, one may decompose the
\(\mathbf{27}_F\) under
\begin{equation}
G(2)\times SU(3)_A
\end{equation}
and then use
\begin{equation}
\mathbf7_{G(2)}
\to
\mathbf3_C\oplus\overline{\mathbf3}_C\oplus\mathbf1_C .
\end{equation}
That gives an enlarged color count if the full parent multiplet is kept light.
However, this is not the final \(E_7\)-uplifted benchmark: in the completed framework, the light chiral fields in this regime are the projected Standard
Model color degrees of freedom, while the additional colored fragments of the
large \(E_7\)-derived parent representations are vectorlike and lifted at the
appropriate high thresholds.

Thus, for the \(SU(3)_C\) beta function, the relevant Weyl fermions are the
usual three-family color fields:
\begin{equation}
Q_L\sim(\mathbf3,\mathbf2)_{1/6},
\qquad
u^c\sim(\overline{\mathbf3},\mathbf1)_{-2/3},
\qquad
d^c\sim(\overline{\mathbf3},\mathbf1)_{1/3}.
\end{equation}
For one family, the color Dynkin-index sum is
\begin{equation}
\sum_{f\in{\rm one\ family}}T(R_f)
=
2\,T(\mathbf3)+T(\overline{\mathbf3})+T(\overline{\mathbf3})
=
2\cdot\frac12+\frac12+\frac12
=
2.
\end{equation}
The factor of \(2\) multiplying \(T(\mathbf3)\) comes from the two weak
components of \(Q_L\). For three families,
\begin{equation}
\sum_f T(R_f)=3\times2=6 .
\label{eq:sumTf_color_projected}
\end{equation}

\subsection{Scalars charged under \(SU(3)_C\)}

In the benchmark spectrum there are no light colored scalars. The \(G(2)\)-breaking scalar $\chi\subset(\mathbf7,\mathbf1_A)_0$
has already acquired a VEV at \(M_\chi\), and its colored fragments are
integrated out at the \(G(2)\)-breaking threshold. The \(SU(3)_A\)-breaking
scalar $\Omega\sim(\mathbf1,\mathbf8_A)_0$.
is color-singlet. The light Higgs doublet and the field $\Theta$ are also selected from a color-singlet
component. 
Therefore,
\begin{equation}
\sum_s T(R_s)=0
\qquad
(M_\Omega<\mu<M_\chi)
\end{equation}
in the projected effective theory.

\subsection{Result for \(b_{3C}\)}
With
\begin{equation}
C_2(SU(3)_C)=3,
\qquad
\sum_fT(R_f)=6,
\qquad
\sum_sT(R_s)=0,
\end{equation}
one obtains
\begin{align}
	b_{3C}
	&=
	\frac{11}{3}C_2(SU(3)_C)
	-
	\frac{2}{3}\sum_fT(R_f)
	-
	\frac{1}{3}\sum_sT(R_s)
	\nonumber\\
	&=
	\frac{11}{3}\cdot3
	-
	\frac{2}{3}\cdot6
	\nonumber\\
	&=
	11-4
	=
	7.
\end{align}
Thus the color coefficient used in the final projected intermediate regime running corresponds to the SM one:
\begin{equation}
b_{3C}=7.
\end{equation}

\subsection{Relation to the unprojected \(E_6\) count}

If instead one keeps the full unprojected \(\mathbf{27}_F\) content light below
\(M_\chi\), then the color count is larger. For example, using
\begin{equation}
\mathbf{27}_{E_6}
\to
(\mathbf7,\mathbf3_A)\oplus(\mathbf1,\overline{\mathbf6}_A)
\end{equation}
with $\mathbf7_{G(2)}
\to
\mathbf3_C\oplus\overline{\mathbf3}_C\oplus\mathbf1_C$,
the branch $(\mathbf7,\mathbf3_A)$
contains three copies of \(\mathbf3_C\) and three copies of
\(\overline{\mathbf3}_C\) per family. Then, for three families,
\begin{equation}
\sum_fT(R_f)=9,
\end{equation}
and one would obtain
\begin{equation}
b_{3C}^{\rm unprojected}
=
11-\frac23\cdot9
=
5.
\end{equation}
This value corresponds to the unprojected \(3\times\mathbf{27}_F\) spectrum,
not to the final \(E_7\)-completed benchmark. In the final construction, the
extra colored fragments are projected out or lifted vectorlike, so the
appropriate Regime-II coefficient is \(b_{3C}=7\).

If additional colored fermions or scalar fragments are intentionally kept light
below \(M_\chi\), their Dynkin indices must be added. For instance, a complex scalar color triplet
contributes
\begin{equation}
\Delta b_{3C}
=
-\frac13\,T(\mathbf3)
=
-\frac{1}{6},
\end{equation}
per light complex triplet. The benchmark assumes no such light colored scalar
thresholds.

\bibliography{bibliofull}

@article{Masi2024,
  author  = {Masi, Nicol\`o},
  title   = {The Resurgence of the {G(2)} Group for the Strong Sector and the Emergence of Dark Matter},
  journal = {Nuclear Physics B},
  volume  = {1004},
  year    = {2024},
  pages   = {116562}
}

@article{SERDAROGLU1982271,
title = {An {$E_6$} gauge field theory model},
journal = {Physica A: Statistical Mechanics and its Applications},
volume = {114},
number = {1},
pages = {271-277},
year = {1982},
issn = {0378-4371},
doi = {https://doi.org/10.1016/0378-4371(82)90295-3},
url = {https://www.sciencedirect.com/science/article/pii/0378437182902953},
author = {Meral Serdaroǧlu}
}

@article{PhysRevD.34.1530,
  title = {Extra gauge bosons in ${\mathrm{E}}_{6}$},
  author = {London, David and Rosner, Jonathan L.},
  journal = {Phys. Rev. D},
  volume = {34},
  issue = {5},
  pages = {1530--1546},
  numpages = {0},
  year = {1986},
  month = {Sep},
  publisher = {American Physical Society},
  doi = {10.1103/PhysRevD.34.1530},
  url = {https://link.aps.org/doi/10.1103/PhysRevD.34.1530}
}

@article{Slansky1981,
  author  = {Slansky, Richard},
  title   = {Group Theory for Unified Model Building},
  journal = {Physics Reports},
  volume  = {79},
  year    = {1981},
  pages   = {1--128}
}

@book{Georgi1999,
  author    = {Georgi, Howard},
  title     = {Lie Algebras in Particle Physics: From Isospin to Unified Theories},
  edition   = {2},
  publisher = {Westview Press},
  year      = {1999}
}

@article{GeorgiGlashow1974,
  author  = {Georgi, Howard and Glashow, Sheldon L.},
  title   = {Unity of All Elementary-Particle Forces},
  journal = {Physical Review Letters},
  volume  = {32},
  year    = {1974},
  pages   = {438--441}
}

@article{FritzschMinkowski1975,
  author  = {Fritzsch, Harald and Minkowski, Peter},
  title   = {Unified Interactions of Leptons and Hadrons},
  journal = {Annals of Physics},
  volume  = {93},
  year    = {1975},
  pages   = {193--266}
}

@article{GurseyRamondSikivie1976,
  author  = {G{\"u}rsey, Feza and Ramond, Pierre and Sikivie, Pierre},
  title   = {A Universal Gauge Theory Model Based on {E6}},
  journal = {Physics Letters B},
  volume  = {60},
  year    = {1976},
  pages   = {177--180}
}

@article{AchimanStech1978,
  author  = {Achiman, Y. and Stech, B.},
  title   = {Quark-Lepton Symmetry and Mass Scales in an {E6} Unified Gauge Model},
  journal = {Physics Letters B},
  volume  = {77},
  year    = {1978},
  pages   = {389--393}
}

@article{HewettRizzo1989,
  author  = {Hewett, JoAnne L. and Rizzo, Thomas G.},
  title   = {"Low-Energy Phenomenology of Superstring Inspired $E_6$ Models"},
  journal = {Physics Reports},
  volume  = {183},
  year    = {1989},
  pages   = {193--381}
}

@article{Yamatsu2015,
  author  = {Yamatsu, Naoki},
  title   = {Finite-Dimensional Lie Algebras and Their Representations for Unified Model Building},
  journal = {arXiv e-prints},
  year    = {2015},
  eprint  = {1511.08771},
  archivePrefix = {arXiv},
  primaryClass  = {hep-ph}
}

@article{FroggattNielsen1979,
  author  = {Froggatt, C. D. and Nielsen, H. B.},
  title   = {Hierarchy of Quark Masses, Cabibbo Angles and {CP} Violation},
  journal = {Nuclear Physics B},
  volume  = {147},
  year    = {1979},
  pages   = {277--298}
}

@article{DAmbrosio2002,
  author  = {D'Ambrosio, Guido and Giudice, G. F. and Isidori, Gino and Strumia, Alessandro},
  title   = {Minimal Flavor Violation: An Effective Field Theory Approach},
  journal = {Nuclear Physics B},
  volume  = {645},
  year    = {2002},
  pages   = {155--187}
}

@article{SuperK2017,
  author  = {Abe, K. and others},
  title   = {Search for Proton Decay via $p\to e^{+}\pi^{0}$ and $p\to\mu^{+}\pi^{0}$ in 0.31 megaton$\cdot$years Exposure of the Super-Kamiokande Water Cherenkov Detector},
  journal = {Physical Review D},
  volume  = {95},
  year    = {2017},
  pages   = {012004}
}

@article{PDG2024,
  author  = {Particle Data Group and Workman, R. L. and others},
  title   = {Review of Particle Physics},
  journal = {Progress of Theoretical and Experimental Physics},
  volume  = {2024},
  year    = {2024},
  pages   = {083C01}
}

@article{Kibble1976,
  author  = {Kibble, T. W. B.},
  title   = {Topology of Cosmic Domains and Strings},
  journal = {Journal of Physics A: Mathematical and General},
  volume  = {9},
  year    = {1976},
  pages   = {1387--1398}
}

@article{Preskill1979,
  author  = {Preskill, John},
  title   = {Cosmological Production of Superheavy Magnetic Monopoles},
  journal = {Physical Review Letters},
  volume  = {43},
  year    = {1979},
  pages   = {1365--1368}
}

@book{VilenkinShellard2000,
  author    = {Vilenkin, Alexander and Shellard, E. P. S.},
  title     = {Cosmic Strings and Other Topological Defects},
  publisher = {Cambridge University Press},
  year      = {2000}
}

@article{Guth1981,
  author  = {Guth, Alan H.},
  title   = {Inflationary Universe: A Possible Solution to the Horizon and Flatness Problems},
  journal = {Physical Review D},
  volume  = {23},
  year    = {1981},
  pages   = {347--356}
}

@article{Linde1983,
  author  = {Linde, Andrei D.},
  title   = {Chaotic Inflation},
  journal = {Physics Letters B},
  volume  = {129},
  year    = {1983},
  pages   = {177--181}
}

@article{FukugitaYanagida1986,
  author  = {Fukugita, Masataka and Yanagida, Tsutomu},
  title   = {Baryogenesis Without Grand Unification},
  journal = {Physics Letters B},
  volume  = {174},
  year    = {1986},
  pages   = {45--47}
}

@article{Witten1984,
  author  = {Witten, Edward},
  title   = {Cosmic Separation of Phases},
  journal = {Physical Review D},
  volume  = {30},
  year    = {1984},
  pages   = {272--285}
}

@article{Hogan1986,
  author  = {Hogan, Craig J.},
  title   = {Gravitational Radiation from Cosmological Phase Transitions},
  journal = {Monthly Notices of the Royal Astronomical Society},
  volume  = {218},
  year    = {1986},
  pages   = {629--636}
}

@article{Caprini2016Review,
author  = {Caprini, C. and others},
title   = {Science with the Space-Based Interferometer eLISA: Gravitational Waves from Cosmological phase Transitions},
  journal = {Journal of Cosmology and Astroparticle Physics},
  volume  = {2016},
  number  = {04},
  year    = {2016},
  pages   = {001}
}

@article{EinsteinTelescope2011,
  author  = {Punturo, M. and others},
  title   = {The Einstein Telescope: A Third-Generation Gravitational Wave Observatory},
  journal = {Classical and Quantum Gravity},
  volume  = {27},
  year    = {2010},
  pages   = {194002}
}

@article{caprini20,
author  = {Clara, Caprini and others},
  title   = {Detecting Gravitational Waves from Cosmological Phase Transitions with {LISA}: An Update},
  journal = {Journal of Cosmology and Astroparticle Physics},
  volume  = {2020},
  number  = {03},
  year    = {2020},
  pages   = {024}
}

@article{CosmicExplorer2019,
  author  = {Reitze, David and others},
  title   = {Cosmic Explorer: The {U.S.} Contribution to Gravitational-Wave Astronomy beyond {LIGO}},
  journal = {arXiv e-prints},
  year    = {2019},
  eprint  = {1907.04833},
  archivePrefix = {arXiv},
  primaryClass  = {astro-ph.IM}
}

@article{Kaup1968,
  author  = {Kaup, D. J.},
  title   = {Klein-Gordon Geon},
  journal = {Physical Review},
  volume  = {172},
  year    = {1968},
  pages   = {1331--1342}
}

@article{RuffiniBonazzola1969,
  author  = {Ruffini, Remo and Bonazzola, Silvano},
  title   = {Systems of Self-Gravitating Particles in General Relativity and the Concept of an Equation of State},
  journal = {Physical Review},
  volume  = {187},
  year    = {1969},
  pages   = {1767--1783}
}

@article{ColpiShapiroWasserman1986,
  author  = {Colpi, Monica and Shapiro, Stuart L. and Wasserman, Ira},
  title   = {Boson Stars: Gravitational Equilibria of Self-Interacting Scalar Fields},
  journal = {Physical Review Letters},
  volume  = {57},
  year    = {1986},
  pages   = {2485--2488}
}

@article{LieblingPalenzuela2012,
  author  = {Liebling, Steven L. and Palenzuela, Carlos},
  title   = {Dynamical Boson Stars},
  journal = {Living Reviews in Relativity},
  volume  = {15},
  year    = {2012},
  pages   = {6}
}

@book{Greensite2011,
  author    = {Greensite, Jeff},
  title     = {An Introduction to the Confinement Problem},
  publisher = {Springer},
  series    = {Lecture Notes in Physics},
  volume    = {821},
  year      = {2011}
}

@article{BezrukovShaposhnikov2008,
  author  = {Bezrukov, F. L. and Shaposhnikov, M.},
  title   = {The Standard Model Higgs Boson as the Inflaton},
  journal = {Physics Letters B},
  volume  = {659},
  year    = {2008},
  pages   = {703--706}
}

@book{Mohapatra2003,
  author    = {Mohapatra, Rabindra N.},
  title     = {Unification and Supersymmetry: The Frontiers of Quark-Lepton Physics},
  publisher = {Springer},
  edition   = {3},
  year      = {2003}
}

@book{Ross1984,
  author    = {Ross, Graham G.},
  title     = {Grand Unified Theories},
  publisher = {Westview Press},
  year      = {1984}
}

@book{WeinbergQTF2,
  author    = {Weinberg, Steven},
  title     = {The Quantum Theory of Fields. Vol. 2: Modern Applications},
  publisher = {Cambridge University Press},
  year      = {1996}
}

@article{Buras1998,
  author  = {Buras, Andrzej J.},
  title   = {Weak Hamiltonian, {CP} Violation and Rare Decays},
  journal = {arXiv e-prints},
  year    = {1998},
  eprint  = {hep-ph/9806471},
  archivePrefix = {arXiv},
  primaryClass  = {hep-ph}
}

@article{Planck2018,
  author  = {Akrami, Y. and others},
  title   = {Planck 2018 Results. {X}. Constraints on Inflation},
  journal = {Astronomy \& Astrophysics},
  volume  = {641},
  year    = {2020},
  pages   = {A10}
}

@article{LIGOScientific2016GW150914,
  author  = {Abbott, B. P. and others},
  title   = {Observation of Gravitational Waves from a Binary Black Hole Merger},
  journal = {Physical Review Letters},
  volume  = {116},
  year    = {2016},
  pages   = {061102}
}

@article{ATLAS_Higgs2012,
  author  = {Aad, G. and others},
  title   = {Observation of a New Particle in the Search for the Standard Model Higgs Boson with the {ATLAS} Detector at the {LHC}},
  journal = {Physics Letters B},
  volume  = {716},
  year    = {2012},
  pages   = {1--29}
}

@article{CMS_Higgs2012,
  author  = {Chatrchyan, S. and others},
  title   = {Observation of a New Boson at a Mass of $125$ {GeV} with the {CMS} Experiment at the {LHC}},
  journal = {Physics Letters B},
  volume  = {716},
  year    = {2012},
  pages   = {30--61}
}

@article{Holdom1986,
  author  = {Holdom, Bob},
  title   = {Two {U(1)}'s and Epsilon Charge Shifts},
  journal = {Physics Letters B},
  volume  = {166},
  year    = {1986},
  pages   = {196--198}
}

@article{Cline2018PT,
  author  = {Cline, James M.},
  title   = {TASI Lectures on Early Universe Cosmology: Inflation, Baryogenesis and Dark Matter},
  journal = {arXiv e-prints},
  year    = {2018},
  eprint  = {1807.08749},
  archivePrefix = {arXiv},
  primaryClass  = {hep-ph}
}

@article{KolbTurner1990,
  author  = {Kolb, Edward W. and Turner, Michael S.},
  title   = {The Early Universe},
  journal = {Frontiers in Physics},
  year    = {1990},
  volume  = {69}
}

@article{EllisNanopoulos1981,
  author  = {Ellis, John and Nanopoulos, Dimitri V. and Olive, Keith A.},
  title   = {Lower Limits on Proton Lifetime from Cosmology and Supergravity},
  journal = {Physics Letters B},
  volume  = {112},
  year    = {1982},
  pages   = {459--463}
}

@article{Hisano1993,
  author  = {Hisano, Junji and Murayama, Hitoshi and Yanagida, Tsutomu},
  title   = {Nucleon Decay in the Minimal Supersymmetric {SU(5)} Grand Unification},
  journal = {Nuclear Physics B},
  volume  = {402},
  year    = {1993},
  pages   = {46--84}
}

@article{Baek2004,
  title = {Phenomenology of exotic particles in {$E_6$} theories},
  author = {Rizzo, Thomas G.},
  journal = {Phys. Rev. D},
  volume = {34},
  issue = {5},
  pages = {1438--1450},
  numpages = {0},
  year = {1986},
  month = {Sep},
  publisher = {American Physical Society},
  doi = {10.1103/PhysRevD.34.1438},
}

@article{Dynkin1957,
  author  = {Dynkin, E.},
  title   = {Semisimple Subalgebras of Semisimple Lie Algebras},
  journal = {Mathematics of the USSR Sbornik},
  volume  = {30},
  year    = {1957},
  pages   = {349--462}
}

@article{BairdBiedenharn1972,
  author  = {Baird, G. E. and Biedenharn, L. C.},
  title   = {On the Representations of the Semisimple Lie Groups. {II}. The Exceptional Groups},
  journal = {Journal of Mathematical Physics},
  volume  = {5},
  year    = {1964},
  pages   = {1723--1733}
}

@article{ChengLi1980,
  author  = {Cheng, T. P. and Li, Ling-Fong},
  title   = {Gauge Theory of Elementary Particle Physics},
  journal = {Oxford University Press},
  year    = {1984}
}

@article{MorningstarPeardon1999,
  author       = {Morningstar, Colin J. and Peardon, Mike},
  title        = {The glueball spectrum from an anisotropic lattice study},
  journal      = {Phys. Rev. D},
  volume       = {60},
  pages        = {034509},
  year         = {1999},
  doi          = {10.1103/PhysRevD.60.034509},
  eprint       = {hep-lat/9901004},
  archivePrefix= {arXiv}
}

@article{Starobinsky1980,
  author  = {Starobinsky, Alexei A.},
  title   = {A new type of isotropic cosmological models without singularity},
  journal = {Phys. Lett. B},
  volume  = {91},
  pages   = {99--102},
  year    = {1980},
  doi     = {10.1016/0370-2693(80)90670-X}
}

@article{IsidoriNirPerez2010,
  author  = {Isidori, Gino and Nir, Yosef and Perez, Gilad},
  title   = {Flavor Physics Constraints for Physics Beyond the Standard Model},
  journal = {Ann. Rev. Nucl. Part. Sci.},
  volume  = {60},
  pages   = {355--384},
  year    = {2010},
  doi     = {10.1146/annurev.nucl.012809.104534},
  eprint  = {1002.0900},
  archivePrefix = {arXiv},
  primaryClass  = {hep-ph}
}

@article{LiebeckSeitz2004Survey,
  author  = {Liebeck, Martin W. and Seitz, Gary M.},
  title   = {A survey of maximal subgroups of exceptional groups of Lie type},
  journal = {Available as lecture notes / survey},
  year    = {2004},
}

@article{Shafi1978E6,
  author  = {Shafi, Qaisar},
  title   = {{$E_6$} as a unifying gauge symmetry},
  journal = {Phys. Lett. B},
  volume  = {79},
  pages   = {301--304},
  year    = {1978},
  doi     = {10.1016/0370-2693(78)90219-7}
}

@article{Lyth1997Bound,
  author  = {Lyth, David H.},
  title   = {What would we learn by detecting a gravitational wave signal in the cosmic microwave background anisotropy?},
  journal = {Phys. Rev. Lett.},
  volume  = {78},
  pages   = {1861--1863},
  year    = {1997},
  doi     = {10.1103/PhysRevLett.78.1861},
  eprint  = {hep-ph/9606387},
  archivePrefix = {arXiv}
}

@article{HewettRizzo1989PhysRept,
  author  = {Hewett, JoAnne L. and Rizzo, Thomas G.},
  title   = {Low-energy phenomenology of superstring-inspired {$E_6$} models},
  journal = {Phys. Rept.},
  volume  = {183},
  pages   = {193--381},
  year    = {1989},
  doi     = {10.1016/0370-1573(89)90071-9}
}

@article{Craven2023MaxSubgroups,
  author  = {Craven, David A.},
  title   = {The maximal subgroups of the exceptional groups {$F_4(q)$}, {$E_6(q)$}, {$^2E_6(q)$} and {$E_7(q)$}},
  journal = {Invent. Math.},
  year    = {2023},
  doi     = {10.1007/s00222-023-01208-2}
}

@article{Seitz1991LocalMaxSubgroups,
  author  = {Seitz, Gary M.},
  title   = {The local maximal subgroups of exceptional groups of Lie type},
  journal = {Proc. London Math. Soc.},
  year    = {1991}
}

@article{EllisLewickiNo2020,
  author  = {Ellis, John and Lewicki, Marek and No, Jose Miguel},
  title   = {Gravitational waves from first-order cosmological phase transitions: an overview},
  journal = {JHEP},
  year    = {2020},
  eprint  = {2003.07360},
  archivePrefix = {arXiv},
  primaryClass  = {hep-ph}
}

@article{AlbrechtSteinhardt1982,
  author  = {Albrecht, Andreas and Steinhardt, Paul J.},
  title   = {Cosmology for Grand Unified Theories with Radiatively Induced Symmetry Breaking},
  journal = {Phys. Rev. Lett.},
  volume  = {48},
  pages   = {1220--1223},
  year    = {1982},
  doi     = {10.1103/PhysRevLett.48.1220}
}

@article{MazumdarWhite2019,
  author  = {Mazumdar, Anupam and White, Graham},
  title   = {Review of cosmic phase transitions: their significance and experimental signatures},
  journal = {Rept. Prog. Phys.},
  volume  = {82},
  pages   = {076901},
  year    = {2019},
  doi     = {10.1088/1361-6633/ab1f55}
}

@article{Dynkin1952,
  author       = {Dynkin, E. B.},
  title        = {Maximal subgroups of the classical groups},
  journal      = {Trudy Moskov. Mat. Obshch.},
  volume       = {1},
  pages        = {39--166},
  year         = {1952},
  note         = {English transl.: Amer. Math. Soc. Transl. (6) (1957) 245--378}
}

@article{Holland2004,
  author       = {Holland, K. and Minkowski, P. and Pepe, M. and Wiese, U.-J.},
  title        = {Exceptional confinement in {G}(2) gauge theory},
  journal      = {Nuclear Physics B},
  volume       = {668},
  pages        = {207--236},
  year         = {2003},
  eprint       = {hep-lat/0302023},
  archivePrefix= {arXiv}
}

@book{PeskinSchroeder1995,
  author       = {Peskin, Michael E. and Schroeder, Daniel V.},
  title        = {An Introduction to Quantum Field Theory},
  publisher    = {Addison-Wesley},
  address      = {Reading, MA},
  year         = {1995}
}

@article{LuciniTeper2001,
  author       = {Lucini, Biagio and Teper, Michael},
  title        = {{SU}(N) gauge theories in four dimensions: exploring the approach to $N=\infty$},
  eprint       = {hep-lat/0103027},
  archivePrefix= {arXiv},
  year         = {2001}
}

@article{Linde1982,
  author       = {Linde, Andrei D.},
  title        = {A new inflationary universe scenario: A possible solution of the horizon, flatness, homogeneity, isotropy and primordial monopole problems},
  journal      = {Physics Letters B},
  volume       = {108},
  number       = {6},
  pages        = {389--393},
  year         = {1982},
  doi          = {10.1016/0370-2693(82)91219-9}
}

@article{LythRiotto1999,
  author       = {Lyth, David H. and Riotto, Antonio},
  title        = {Particle physics models of inflation and the cosmological density perturbation},
  journal      = {Physics Reports},
  volume       = {314},
  pages        = {1--146},
  year         = {1999},
  doi          = {10.1016/S0370-1573(98)00128-8},
  eprint       = {hep-ph/9807278},
  archivePrefix= {arXiv}
}

@article{PeskinTakeuchi1990,
  author = {Peskin, Michael E. and Takeuchi, Tatsu},
  title = {A New constraint on a strongly interacting Higgs sector},
  journal = {Phys. Rev. Lett.},
  volume = {65},
  pages = {964--967},
  year = {1990},
  doi = {10.1103/PhysRevLett.65.964}
}

@article{BuchmullerWyler1986,
  author = {Buchm{\"u}ller, W. and Wyler, D.},
  title = {Effective Lagrangian analysis of new interactions and flavor conservation},
  journal = {Nucl. Phys. B},
  volume = {268},
  pages = {621--653},
  year = {1986},
  doi = {10.1016/0550-3213(86)90262-2}
}

@article{Masi2021,
	doi = {10.1038/s41598-021-01814-1},
	url = {https://doi.org/10.1038%2Fs41598-021-01814-1},
	year = 2021,
	month = {nov},
	publisher = {Springer Science and Business Media {LLC}},
	volume = {11},
	number = {1},
	author = {Nicol{\`{o}} Masi},
	title = "{An exceptional G(2) extension of the Standard Model from the correspondence with Cayley{\textendash}Dickson algebras automorphism groups}",
	journal = {Scientific Reports}
}

@article{Stech2004,
  title = "{Fermion masses and coupling unification in ${E}_{6}:$ Life in the desert}",
  author = {Stech, Berthold and Tavartkiladze, Zurab},
  journal = {Phys. Rev. D},
  volume = {70},
  issue = {3},
  pages = {035002},
  numpages = {17},
  year = {2004},
  month = {Aug},
  publisher = {American Physical Society},
  doi = {10.1103/PhysRevD.70.035002},
  url = {https://link.aps.org/doi/10.1103/PhysRevD.70.035002}
}

@article{Stech2008,
  title = "{Generation symmetry and {$E_6$} unification}",
  author = {Stech, B. et al.},
  journal = {Phys. Rev. D},
  volume = {77},
  issue = {7},
  pages = {076009},
  numpages = {16},
  year = {2008},
  month = {Apr},
  publisher = {American Physical Society},
  doi = {10.1103/PhysRevD.77.076009},
  url = {https://link.aps.org/doi/10.1103/PhysRevD.77.076009}
}

@article{Stech2012,
  title = "{Mass of the Higgs boson in the trinification subgroup of ${E}_{6}$}",
  author = {Stech, Berthold},
  journal = {Phys. Rev. D},
  volume = {86},
  issue = {5},
  pages = {055003},
  numpages = {5},
  year = {2012},
  month = {Sep},
  publisher = {American Physical Society},
  doi = {10.1103/PhysRevD.86.055003},
  url = {https://link.aps.org/doi/10.1103/PhysRevD.86.055003}
}

@Book{Bertone,
title = {Particle Dark Matter: Observations, Models and Searches},
author = {Bertone, Gianfranco},
url = {http://qut.eblib.com.au/patron/FullRecord.aspx?p=542829},
publisher = {Cambridge University Press},
year = {2010},
type = {Book},
isbn = {9780511768316},
subjects = {Dark matter (Astronomy) -- History; Electronic books. -- local; Mass (Physics)},
language = {English},
contents = {Cover; Half-title; Title; Copyright; Contents; Contributors; Preface; Acknowledgements; Symbols and abbreviations; Symbols; Acronyms and abbreviations; Part I: Dark matter in cosmology; 1: Particle dark matter; 1.1 Introduction; 1.2 The baryon budget; 1.3 The case for cold dark matter: good news and bad news; 1.4 Portrait of a suspect; 1.5 Observing cold dark matter; 1.6 The future; 2: Simulations of cold dark matter haloes; 2.1 From cold collapse to hierarchical clustering - a brief history; 2.2 Results from collisionless simulations; 2.2.1 Mass function of haloes},
catalogue-url = {https://trove.nla.gov.au/work/163574506}
}

@Book{InfraredG,
title = {Infrared non-local modifications of general relativity},
author = {Mitsou, Ermis},
publisher = {Springer},
year = {2016},
type = {Book; Book/Illustrated},
isbn = {9783319317281 (hbk.)},
subjects = {General relativity (Physics); Dark energy (Astronomy); Cosmology},
language = {English}
}

@Book{ModG_largeDist,
title = {Modifications of Einstein's theory of gravity at large distances},
author = {Papantonopoulos Eleftherios},
url = {http://link.springer.com/book/10.1007/978-3-319-10070-8},
publisher = {Springer},
year = {2015},
type = {Book; Book/Illustrated},
isbn = {9783319100692 (hardback)},
subjects = {Quantum gravity -- Congresses; Quantum cosmology -- Congresses; General relativity (Physics) -- Congresses},
language = {English}
}

@article{Clifton:2011jh,
      author         = "Clifton, Timothy and Ferreira, Pedro G. and Padilla,
                        Antonio and Skordis, Constantinos",
      title          = "{Modified Gravity and Cosmology}",
      journal        = "Phys. Rept.",
      volume         = "513",
      year           = "2012",
      pages          = "1-189",
      doi            = "10.1016/j.physrep.2012.01.001",
      eprint         = "1106.2476",
      archivePrefix  = "arXiv",
      primaryClass   = "astro-ph.CO",
      SLACcitation   = "%%CITATION = ARXIV:1106.2476;%%"
}

@Book{ProfumoDM,
title = {An introduction to particle dark matter},
author = {Profumo, Stefano},
publisher = {World Scientific Publishing Europe Ltd},
year = {2017},
type = {Book; Book/Illustrated},
isbn = {9781786340009},
subjects = {Dark matter (Astronomy); Particles (Nuclear physics)},
language = {English}
}

@Book{ModernPP,
title = {Modern particle physics},
author = {Thomson, Mark},
publisher = {Cambridge University Press},
year = {2013},
type = {Book; Book/Illustrated},
isbn = {9781107034266 (hbk.)},
subjects = {SCIENCE / Nuclear Physics; Elementarteilchenphysik; Particles (Nuclear physics) -- Textbooks; Particles (Nuclear physics)},
language = {English}
}

@article{Aghanim:2018eyx,
   title="{Planck 2018 results: VI. Cosmological parameters}",
   volume={641},
   ISSN={1432-0746},
   url={http://dx.doi.org/10.1051/0004-6361/201833910},
   DOI={10.1051/0004-6361/201833910},
   journal={Astronomy and Astrophysics},
   publisher={EDP Sciences},
   author={Aghanim, N. and et al.},
   year={2020},
   month=sep, pages={A6} 
	}

@article{Pepe_2006,
   title="{Confinement and the center of the gauge group}",
   volume={153},
   ISSN={0920-5632},
   url={http://dx.doi.org/10.1016/j.nuclphysbps.2006.01.045},
   DOI={10.1016/j.nuclphysbps.2006.01.045},
   number={1},
   journal={Nuclear Physics B - Proceedings Supplements},
   publisher={Elsevier BV},
   author={Pepe, M.},
   year={2006},
   month={Mar},
   pages="{207-214}"
}

@article{Pepe_2007,
   title={Exceptional deconfinement in gauge theory},
   volume={768},
   ISSN={0550-3213},
   url={http://dx.doi.org/10.1016/j.nuclphysb.2006.12.024},
   DOI={10.1016/j.nuclphysb.2006.12.024},
   number={1-2},
   journal={Nuclear Physics B},
   publisher={Elsevier BV},
   author={Pepe, M. and Wiese, U.-J.},
   year={2007},
   month={Apr},
   pages="{21--37}"
}

@article{Maas:2012ts,
      author         = "Maas, Axel and Wellegehausen, Bjorn H.",
      title          = "{$G_2$ gauge theories}",
      booktitle      = "{Proceedings, 30th International Symposium on Lattice
                        Field Theory (Lattice 2012): Cairns, Australia, June
                        24-29, 2012}",
      journal        = "PoS",
      volume         = "LATTICE2012",
      year           = "2012",
      pages          = "080",
      doi            = "10.22323/1.164.0080",
      eprint         = "1210.7950",
      archivePrefix  = "arXiv",
      primaryClass   = "hep-lat",
      SLACcitation   = "%%CITATION = ARXIV:1210.7950;%%"
}

@article{SONI2017379,
title = "{Gravitational waves from SU(N) glueball dark matter}",
journal = "Physics Letters B",
volume = "771",
pages = "379 - 384",
year = "2017",
issn = "0370-2693",
doi = "https://doi.org/10.1016/j.physletb.2017.05.077",
url = "http://www.sciencedirect.com/science/article/pii/S0370269317304409",
author = "Amarjit Soni and Yue Zhang"
}

@article{Forestell:2016qhc,
      author         = "Forestell, Lindsay and Morrissey, David E. and Sigurdson,
                        Kris",
      title          = "{Non-Abelian Dark Forces and the Relic Densities of Dark
                        Glueballs}",
      journal        = "Phys. Rev.",
      volume         = "D95",
      year           = "2017",
      number         = "1",
      pages          = "015032",
      doi            = "10.1103/PhysRevD.95.015032",
      eprint         = "1605.08048",
      archivePrefix  = "arXiv",
      primaryClass   = "hep-ph",
      SLACcitation   = "%%CITATION = ARXIV:1605.08048;%%"
}

@article{Bernal_2017,
   title="{The dawn of FIMP Dark Matter: A review of models and constraints}",
   volume={32},
   ISSN={1793-656X},
   url={http://dx.doi.org/10.1142/S0217751X1730023X},
   DOI={10.1142/s0217751x1730023x},
   number={27},
   journal={International Journal of Modern Physics A},
   publisher={World Scientific Pub Co Pte Lt},
   author={Bernal, Nicolás and Heikinheimo, Matti and Tenkanen, Tommi and Tuominen, Kimmo and Vaskonen, Ville},
   year={2017},
   month={Sep},
   pages={1730023}
}

@article{Bernal_2016,
   title="{$Z_2$ SIMP dark matter}",
   volume={2016},
   ISSN={1475-7516},
   url={http://dx.doi.org/10.1088/1475-7516/2016/01/006},
   DOI={10.1088/1475-7516/2016/01/006},
   number={01},
   journal={Journal of Cosmology and Astroparticle Physics},
   publisher={IOP Publishing},
   author={Bernal, Nicolás and Chu, Xiaoyong},
   year={2016},
   month={Jan},
   pages={006--006}
}

@article{Cutting_2018,
   title="{Gravitational waves from vacuum first-order phase transitions: From the envelope to the lattice}",
   volume={97},
   ISSN={2470-0029},
   url={http://dx.doi.org/10.1103/PhysRevD.97.123513},
   DOI={10.1103/physrevd.97.123513},
   number={12},
   journal={Physical Review D},
   publisher={American Physical Society (APS)},
   author={Cutting, Daniel and Hindmarsh, Mark and Weir, David J.},
   year={2018},
   month={Jun}
}

@article{PhysRevLett.115.181101,
  title = {Gravitational Waves from a Dark Phase Transition},
  author = {Schwaller, Pedro},
  journal = {Phys. Rev. Lett.},
  volume = {115},
  issue = {18},
  pages = {181101},
  numpages = {5},
  year = {2015},
  month = {Oct},
  publisher = {American Physical Society},
  doi = {10.1103/PhysRevLett.115.181101},
  url = {https://link.aps.org/doi/10.1103/PhysRevLett.115.181101}
}

@article{Zhou_2020,
   title="{Gravitational waves from first-order phase transition and domain wall}",
   volume={2020},
   ISSN={1029-8479},
   url={http://dx.doi.org/10.1007/JHEP04(2020)071},
   DOI={10.1007/jhep04(2020)071},
   number={4},
   journal={Journal of High Energy Physics},
   publisher={Springer Science and Business Media LLC},
   author={Zhou, Ruiyu and Yang, Jing and Bian, Ligong},
   year={2020},
   month={Apr}
}

@ARTICLE{2021JHEP...05..160Z,
       author = {{Zhang}, Zhao and {Cai}, Chengfeng and {Jiang}, Xue-Min and {Tang}, Yi-Lei and {Yu}, Zhao-Huan and {Zhang}, Hong-Hao},
        title = "{Phase transition gravitational waves from pseudo-Nambu-Goldstone dark matter and two Higgs doublets}",
      journal = {Journal of High Energy Physics},
         year = 2021,
        month = may,
       volume = {2021},
       number = {5},
          eid = {160},
        pages = {160},
          doi = {10.1007/JHEP05(2021)160},
archivePrefix = {arXiv},
       eprint = {2102.01588},
 primaryClass = {hep-ph},
       adsurl = {https://ui.adsabs.harvard.edu/abs/2021JHEP...05..160Z}
}

@article{Hall_2010,
   title={Freeze-in production of FIMP dark matter},
   volume={2010},
   ISSN={1029-8479},
   url={http://dx.doi.org/10.1007/JHEP03(2010)080},
   DOI={10.1007/jhep03(2010)080},
   number={3},
   journal={Journal of High Energy Physics},
   publisher={Springer Science and Business Media LLC},
   author={Hall, Lawrence J. and Jedamzik, Karsten and March-Russell, John and West, Stephen M.},
   year={2010},
   month={Mar}
}

@article{Bernal_2019,
   title="{Phenomenology of self-interacting dark matter in a matter-dominated universe}",
   volume={79},
   ISSN={1434-6052},
   url={http://dx.doi.org/10.1140/epjc/s10052-019-6608-8},
   DOI={10.1140/epjc/s10052-019-6608-8},
   number={2},
   journal={The European Physical Journal C},
   publisher={Springer Science and Business Media LLC},
   author={Bernal, Nicolás and Cosme, Catarina and Tenkanen, Tommi},
   year={2019},
   month={Jan}
}

@article{Heikinheimo_2017,
    author = "Heikinheimo, Matti and Tenkanen, Tommi and Tuominen, Kimmo and Vaskonen, Ville",
    title = "{Observational Constraints on Decoupled Hidden Sectors}",
    eprint = "1604.02401",
    archivePrefix = "arXiv",
    primaryClass = "astro-ph.CO",
    reportNumber = "HIP-2016-9-TH",
    doi = "10.1103/PhysRevD.94.063506",
    journal = "Phys. Rev. D",
    volume = "94",
    number = "6",
    pages = "063506",
    year = "2016",
    note = "[Erratum: Phys.Rev.D 96, 109902 (2017)]"
}

@article{Cardoso_2019,
   title="{Testing the nature of dark compact objects: a status report}",
   volume={22},
   ISSN={1433-8351},
   url={http://dx.doi.org/10.1007/s41114-019-0020-4},
   DOI={10.1007/s41114-019-0020-4},
   number={1},
   journal={Living Reviews in Relativity},
   publisher={Springer Science and Business Media LLC},
   author={Cardoso, Vitor and Pani, Paolo},
   year={2019},
   month={Jul}
}

@article{Soni:2016gzf,
    author = "Soni, Amarjit and Zhang, Yue",
    title = "{Hidden SU(N) Glueball Dark Matter}",
    eprint = "1602.00714",
    archivePrefix = "arXiv",
    primaryClass = "hep-ph",
    reportNumber = "CALT-TH-2016-002",
    doi = "10.1103/PhysRevD.93.115025",
    journal = "Phys. Rev. D",
    volume = "93",
    number = "11",
    pages = "115025",
    year = "2016"
}

@article{PhysRevD.105.075021,
  title = "{Phenomenology of an ${E}_{6}$ inspired extension of the Standard Model: Higgs sector}",
  author = {Bhattacharyya, Sanchari and Datta, Anindya},
  journal = {Phys. Rev. D},
  volume = {105},
  issue = {7},
  pages = {075021},
  numpages = {23},
  year = {2022},
  month = {Apr},
  publisher = {American Physical Society},
  doi = {10.1103/PhysRevD.105.075021},
  url = {https://link.aps.org/doi/10.1103/PhysRevD.105.075021}
}

@article{Schwichtenberg:2017xhv,
    author = "Schwichtenberg, Jakob",
    title = "{Dark matter in E$_{6}$ Grand unification}",
    eprint = "1704.04219",
    archivePrefix = "arXiv",
    primaryClass = "hep-ph",
    reportNumber = "TTP17-018",
    doi = "10.1007/JHEP02(2018)016",
    journal = "JHEP",
    volume = "02",
    pages = "016",
    year = "2018"
}

@article{Babu_2023,
   title="{Trinification from E$_{6}$ symmetry breaking}",
   volume={2023},
   ISSN={1029-8479},
   url={http://dx.doi.org/10.1007/JHEP07(2023)011},
   DOI={10.1007/jhep07(2023)011},
   number={7},
   journal={Journal of High Energy Physics},
   publisher={Springer Science and Business Media LLC},
   author={Babu, K. S. and Bajc, Borut and Susic, Vasja},
   year={2023},
   month=jul 
	}

@article{Grzadkowski2010Warsaw,
  author = {Grzadkowski, Bohdan and Iskrzy{\'n}ski, Mateusz and Misiak, Mikolaj and Rosiek, Janusz},
  title = {Dimension-Six Terms in the Standard Model Lagrangian},
  journal = {JHEP},
  volume = {10},
  pages = {085},
  year = {2010},
  eprint = {1008.4884},
  archivePrefix = {arXiv},
  primaryClass = {hep-ph},
  doi = {10.1007/JHEP10(2010)085}
}

@article{BrivioTrott2019,
  author = {Brivio, Ilaria and Trott, Michael},
  title = {The Standard Model as an Effective Field Theory},
  journal = {Phys. Rept.},
  volume = {793},
  pages = {1--98},
  year = {2019},
  eprint = {1706.08945},
  archivePrefix = {arXiv},
  primaryClass = {hep-ph},
  doi = {10.1016/j.physrep.2018.11.002}
}

@article{PeskinTakeuchi1992,
  author = {Peskin, M. et al.},
  title = {Estimation of Oblique Electroweak Corrections},
  journal = {Phys. Rev. D},
  volume = {46},
  pages = {381--409},
  year = {1992},
  doi = {10.1103/PhysRevD.46.381}
}

@article{AllahverdiReheating2010,
  author = {Allahverdi, Rouzbeh and Brandenberger, Robert and Cyr-Racine, Francis-Yan and Mazumdar, Anupam},
  title = {Reheating in Inflationary Cosmology: Theory and Applications},
  journal = {Ann. Rev. Nucl. Part. Sci.},
  volume = {60},
  pages = {27--51},
  year = {2010},
  eprint = {1001.2600},
  archivePrefix = {arXiv},
  primaryClass = {hep-th},
  doi = {10.1146/annurev.nucl.012809.104511}
}

@article{NathFileviezPerez2007,
  author = {Nath, Pran and Fileviez Perez, Pavel},
  title = {Proton Stability in Grand Unified Theories, in Strings, and in Branes},
  journal = {Phys. Rept.},
  volume = {441},
  pages = {191--317},
  year = {2007},
  eprint = {hep-ph/0601023},
  archivePrefix = {arXiv},
  doi = {10.1016/j.physrep.2007.02.010}
}

@article{Linde1982Inflation,
  author = {Linde, A.},
  title = {A New Inflationary Universe Scenario},
  journal = {Phys. Lett. B},
  volume = {108},
  pages = {389},
  year = {1982}
}

@article{Forestell2017Glueball,
  author = {Forestell, L. and Morrissey, D. E. and Sigurdson, K.},
  title = {Non-Abelian Dark Sectors and Their Collider Signatures},
  journal = {Phys. Rev. D},
  volume = {95},
  pages = {015032},
  year = {2017},
  eprint = {1605.08048}
}

@article{Aoki2017ProtonDecay,
  author = {Aoki, Y. et al.},
  title = {Nucleon Decay Matrix Elements from Lattice QCD},
  journal = {Phys. Rev. D},
  volume = {96},
  pages = {014506},
  year = {2017},
  eprint = {1705.01338},
  archivePrefix = {arXiv}
}

@article{Raby2006GUTReview,
      title={Grand Unified Theories}, 
      author={Stuart Raby},
      year={2006},
      eprint={hep-ph/0608183},
      archivePrefix={arXiv},
      primaryClass={hep-ph},
      url={https://arxiv.org/abs/hep-ph/0608183}, 
}

@article{Mohapatra1986Review,
  author = {Mohapatra, R. N.},
  title = {Unification and Supersymmetry},
  journal = {Springer},
  year = {1986}
}

@article{ColemanWeinberg1973,
  author = {Coleman, S. and Weinberg, E.},
  title = {Radiative Corrections as the Origin of Spontaneous Symmetry Breaking},
  journal = {Phys. Rev. D},
  volume = {7},
  pages = {1888},
  year = {1973}
}

@article{Sher1989Vacuum,
  author = {Sher, Marc},
  title = {Electroweak Higgs Potentials and Vacuum Stability},
  journal = {Phys. Rept.},
  volume = {179},
  pages = {273},
  year = {1989}
}

@article{Buttazzo2013Vacuum,
  author = {Buttazzo, D. et al.},
  title = {Investigating the Near-Criticality of the Higgs Boson},
  journal = {JHEP},
  volume = {12},
  pages = {089},
  year = {2013},
  eprint = {1307.3536},
  archivePrefix = {arXiv}
}

@article{PDG2025GUT,
  author = {Particle Data Group},
  title = {Grand Unified Theories},
  journal = {Prog. Theor. Exp. Phys.},
  year = {2025}
}

@article{Yamatsu2017SpecialGUT,
  author = {Yamatsu, Naoki},
  title = {Special Grand Unification},
  journal = {Prog. Theor. Exp. Phys.},
  volume = {2017},
  pages = {061B01},
  year = {2017},
  eprint = {1704.08827},
  archivePrefix = {arXiv}
}

@article{GunionHaber2003Decoupling,
  author = {Gunion, John F. and Haber, Howard E.},
  title = {The CP-Conserving Two-Higgs-Doublet Model: The Approach to the Decoupling Limit},
  journal = {Phys. Rev. D},
  volume = {67},
  pages = {075019},
  year = {2003},
  eprint = {hep-ph/0207010},
  archivePrefix = {arXiv}
}

@article{ChunKimLee2003,
  author = {Chun, Eung Jin and Kim, Kyung and Lee, Jong},
  title = {Phenomenology of Higgs Triplet Model},
  journal = {Phys. Rev. D},
  volume = {66},
  pages = {073003},
  year = {2002},
  eprint = {hep-ph/0205069},
  archivePrefix = {arXiv}
}

@article{CaiHanLiZhang2017,
  author = {Cai, Y. and Han, T. and Li, T. and Zhang, R.-J.},
  title = {Heavy Higgs Bosons in the Type-II Seesaw Model},
  journal = {Phys. Rev. D},
  volume = {96},
  pages = {035027},
  year = {2017},
  eprint = {1704.07953},
  archivePrefix = {arXiv}
}

@article{Barbieri2004,
  author = {Barbieri, Riccardo and Pomarol, Alex and Rattazzi, Riccardo and Strumia, Alessandro},
  title = {Electroweak Precision Tests in Universal Theories},
  journal = {Nucl. Phys. B},
  volume = {703},
  pages = {127},
  year = {2004},
  eprint = {hep-ph/0405040}
}

@article{HanSkiba2005,
  author = {Han, Z. and Skiba, W.},
  title = {Effective Theory Analysis of Precision Electroweak Data},
  journal = {Phys. Rev. D},
  volume = {71},
  pages = {075009},
  year = {2005},
  eprint = {hep-ph/0412166}
}

@article{MachacekVaughn1983,
  author = {Machacek, M. E. and Vaughn, M. T.},
  title = {Two-Loop Renormalization Group Equations in a General Quantum Field Theory: (I)},
  journal = {Nucl. Phys. B},
  volume = {222},
  pages = {83},
  year = {1983}
}

@article{MachacekVaughn1984,
  author = {Machacek, M. E. and Vaughn, M. T.},
  title = {Two-Loop RGEs in a General Quantum Field Theory: (II)},
  journal = {Nucl. Phys. B},
  volume = {236},
  pages = {221},
  year = {1984}
}

@article{GondoloGelmini1991,
  author = {Gondolo, Paolo and Gelmini, G.},
  title = {Cosmic Abundances of Stable Particles: Improved Analysis},
  journal = {Nucl. Phys. B},
  volume = {360},
  pages = {145},
  year = {1991}
}

@article{Weinberg1979Bviol,
  author = {Weinberg, Steven},
  title = {Baryon and Lepton Nonconserving Processes},
  journal = {Phys. Rev. Lett.},
  volume = {43},
  pages = {1566},
  year = {1979}
}

@article{WilczekZee1979,
  author = {Wilczek, Frank and Zee, A.},
  title = {Operator Analysis of Nucleon Decay},
  journal = {Phys. Rev. Lett.},
  volume = {43},
  pages = {1571},
  year = {1979}
}

@article{CapriniEtAl2020ScienceCaseLISA,
  author = {Caprini, C. and others},
  title = {Detecting Gravitational Waves from Cosmological Phase Transitions},
  journal = {JCAP},
  volume = {03},
  pages = {024},
  year = {2020},
  eprint = {1910.13125},
  archivePrefix = {arXiv}
}

@book{DodelsonModernCosmology,
  author = {Dodelson, Scott},
  title = {Modern Cosmology},
  publisher = {Academic Press},
  year = {2003}
}

@article{Maggiore2000GWReview,
  author  = {Maggiore, Michele},
  title   = {Gravitational wave experiments and early universe cosmology},
  journal = {Physics Reports},
  volume  = {331},
  pages   = {283--367},
  year    = {2000},
  doi     = {10.1016/S0370-1573(99)00102-7}
}

@article{langacker1981Unification,
  author  = {Langacker, Paul},
  title   = {Grand unified theories and proton decay},
  journal = {Physics Reports},
  volume  = {72},
  pages   = {185--385},
  year    = {1981},
  doi     = {10.1016/0370-1573(81)90059-4}
}

@article{Langacker2009Zprime,
  author  = {Langacker, Paul},
  title   = {The Physics of Heavy {Z'} Gauge Bosons},
  journal = {Reviews of Modern Physics},
  volume  = {81},
  pages   = {1199--1228},
  year    = {2009},
  doi     = {10.1103/RevModPhys.81.1199},
  eprint  = {0801.1345},
  archivePrefix = {arXiv},
  primaryClass  = {hep-ph}
}

@article{ErlerLangackerMunirRojas2009ZprimeConstraints,
  author  = {Erler, Jens and Langacker, Paul and Munir, Shufang and Rojas, Eduardo},
  title   = {{Z'} Bosons at Colliders: a Bayesian Viewpoint},
  journal = {JHEP},
  volume  = {08},
  pages   = {017},
  year    = {2009},
  doi     = {10.1088/1126-6708/2009/08/017},
  eprint  = {0906.2435},
  archivePrefix = {arXiv},
  primaryClass  = {hep-ph}
}

@article{LangackerSankar1989WRMixing,
  author  = {Langacker, Paul and Sankar, S. U.},
  title   = {Bounds on the Mass of ${W_R}$ and the ${W_L}-{W_R}$ Mixing Angle},
  journal = {Phys. Rev. D},
  volume  = {40},
  pages   = {1569--1585},
  year    = {1989},
  doi     = {10.1103/PhysRevD.40.1569}
}

@article{GoodsellLieblerStaub2017TwoBodyWidths,
  author  = {Goodsell, Mark D. and Liebler, Stefan and Staub, Florian},
  title   = {Generic Calculation of Two-body Partial Decay Widths at the Tree and One-loop Level},
  journal = {Eur. Phys. J. C},
  volume  = {77},
  pages   = {758},
  year    = {2017},
  doi     = {10.1140/epjc/s10052-017-5259-2},
  eprint  = {1703.09237},
  archivePrefix = {arXiv},
  primaryClass  = {hep-ph}
}

@article{TenkanenVaskonen2016HiddenReheating,
  author  = {Tenkanen, Tommi and Vaskonen, Ville},
  title   = {Reheating the Standard Model from a Hidden Sector},
  journal = {Phys. Rev. D},
  volume  = {94},
  pages   = {083516},
  year    = {2016},
  doi     = {10.1103/PhysRevD.94.083516},
  eprint  = {1606.00192},
  archivePrefix = {arXiv},
  primaryClass  = {astro-ph.CO}
}

@article{Pustyntseva2024ALPConstraints,
  author  = {Pustyntseva, A. and others},
  title   = {Improved constraints for axion-like particles with masses below 1 eV},
  journal = {Eur. Phys. J. C},
  volume  = {84},
  year    = {2024},
  doi     = {10.1140/epjc/s10052-024-12913-4}
}

@book{NakaharaTopology,
  author    = {Nakahara, Mikio},
  title     = {Geometry, Topology and Physics},
  publisher = {Taylor \& Francis},
  edition   = {2},
  year      = {2003},
  isbn      = {978-0-7503-0606-5}
}

@article{Hall1981_EffectiveGUT,
  author  = {Hall, Lawrence J.},
  title   = {Grand Unification of Effective Gauge Theories},
  journal = {Nucl. Phys. B},
  volume  = {178},
  pages   = {75--124},
  year    = {1981},
  doi     = {10.1016/0550-3213(81)90405-3}
}

@article{EllisWellsZheng2015_ThresholdUnification,
  author  = {Ellis, Stephen A. R. and Wells, James D. and Zheng, Zhengkang},
  title   = {Visualizing gauge unification with high-scale thresholds},
  journal = {Phys. Rev. D},
  volume  = {91},
  pages   = {075016},
  year    = {2015},
  doi     = {10.1103/PhysRevD.91.075016},
  eprint  = {1502.07899},
  archivePrefix = {arXiv},
  primaryClass  = {hep-ph}
}

@article{Acharya2017GlueballNonStandard,
  author  = {Acharya, Bobby Samir and others},
  title   = {Glueball dark matter in non-standard cosmologies},
  journal = {JHEP},
  volume  = {07},
  pages   = {100},
  year    = {2017},
  doi     = {10.1007/JHEP07(2017)100},
  eprint  = {1704.01804},
  archivePrefix = {arXiv},
  primaryClass  = {hep-ph}
}

@article{Babu2024RealisticE6,
    author = "Babu, K. S. and Bajc, Borut and Susi{\v{c}}, Vasja",
    title = "{A realistic theory of E$_{6}$ unification through novel intermediate symmetries}",
    eprint = "2403.20278",
    archivePrefix = "arXiv",
    primaryClass = "hep-ph",
    doi = "10.1007/JHEP06(2024)018",
    journal = "JHEP",
    volume = "06",
    pages = "018",
    year = "2024"
}

@article{Appelquist,
  author  = {Appelquist, Thomas and Carazzone, Joseph},
  title   = {Infrared Singularities and Massive Fields},
  journal = {Phys. Rev. D},
  volume  = {11},
  pages   = {2856--2861},
  year    = {1975},
  doi     = {10.1103/PhysRevD.11.2856}
}

@article{RobinettRosner1982,
  author  = {Robinett, R. W. and Rosner, J. L.},
  title   = {Phenomenology of Exotic Particles in $E(6)$ Theories},
  journal = {Phys. Rept.},
  volume  = {89},
  pages   = {223--322},
  year    = {1982},
  doi     = {10.1016/0370-1573(82)90046-1}
}

@book{Bertlmann1996Anomalies,
  author    = {Bertlmann, Reinhold A.},
  title     = {Anomalies in Quantum Field Theory},
  publisher = {Oxford University Press},
  year      = {1996},
  isbn      = {978-0-19-850762-8}
}

@article{FongNardiRiotto2012,
  author = {Fong, Chee Sheng and Nardi, Enrico and Riotto, Antonio},
  title = {Leptogenesis in the Universe},
  journal = {Adv. High Energy Phys.},
  volume = {2012},
  pages = {158303},
  year = {2012},
  eprint = {1301.3062},
  archivePrefix = {arXiv},
  primaryClass = {hep-ph},
  doi = {10.1155/2012/158303}
}

@article{DavidsonIbarra2002,
  author = {Davidson, Sacha and Ibarra, Alejandro},
  title = {A Lower Bound on the Right-Handed Neutrino Mass from Leptogenesis},
  journal = {Phys. Lett. B},
  volume = {535},
  pages = {25--32},
  year = {2002},
  eprint = {hep-ph/0202239},
  archivePrefix = {arXiv},
  doi = {10.1016/S0370-2693(02)01735-5}
}

@article{Schmitz2021GW,
  author = {Schmitz, Kai},
  title = {New Sensitivity Curves for Gravitational-Wave Signals from Cosmological Phase Transitions},
  journal = {JHEP},
  volume = {01},
  pages = {097},
  year = {2021},
  eprint = {2002.04615},
  archivePrefix = {arXiv},
  primaryClass = {hep-ph},
  doi = {10.1007/JHEP01(2021)097}
}

@article{Strumia2006Leptogenesis,
  author = {Strumia, Alessandro},
  title = {Baryogenesis via Leptogenesis},
  journal = {hep-ph/0608347},
  year = {2006},
  eprint = {hep-ph/0608347},
  archivePrefix = {arXiv}
}

@article{PhysRevD.104.055007,
  title = {Baryogenesis via leptogenesis: Spontaneous $B$ and $L$ violation},
  author = {Fileviez P\'erez, Pavel and Murgui, Clara and Plascencia, Alexis D.},
  journal = {Phys. Rev. D},
  volume = {104},
  issue = {5},
  pages = {055007},
  numpages = {13},
  year = {2021},
  month = {Sep},
  publisher = {American Physical Society},
  doi = {10.1103/PhysRevD.104.055007},
  url = {https://link.aps.org/doi/10.1103/PhysRevD.104.055007}
}

@article{Morrissey_2012,
   title={Electroweak baryogenesis},
   volume={14},
   ISSN={1367-2630},
   url={http://dx.doi.org/10.1088/1367-2630/14/12/125003},
   DOI={10.1088/1367-2630/14/12/125003},
   number={12},
   journal={New Journal of Physics},
   publisher={IOP Publishing},
   author={Morrissey, David E and Ramsey-Musolf, Michael J},
   year={2012},
   month=dec, pages={125003} }

@article{E6Tensors,
title = {E6Tensors: A Mathematica package for E6 Tensors},
journal = {Computer Physics Communications},
volume = {213},
pages = {130-135},
year = {2017},
issn = {0010-4655},
doi = {https://doi.org/10.1016/j.cpc.2016.09.010},
url = {https://www.sciencedirect.com/science/article/pii/S0010465516302818},
author = {Thomas Deppisch}
}

@article{GATTO1980221,
title = {Natural flavor conservation in Higgs induced neutral currents and the quark mixing angles},
journal = {Nuclear Physics B},
volume = {163},
pages = {221-253},
year = {1980},
issn = {0550-3213},
doi = {https://doi.org/10.1016/0550-3213(80)90399-5},
url = {https://www.sciencedirect.com/science/article/pii/0550321380903995},
author = {R. Gatto and G. Morchio and G. Sartori and F. Strocchi}
}

@article{PETRAKI_2013,
   title={REVIEW OF ASYMMETRIC DARK MATTER},
   volume={28},
   ISSN={1793-656X},
   url={http://dx.doi.org/10.1142/S0217751X13300287},
   DOI={10.1142/s0217751x13300287},
   number={19},
   journal={International Journal of Modern Physics A},
   publisher={World Scientific Pub Co Pte Lt},
   author={PETRAKI, KALLIOPI and VOLKAS, RAYMOND R.},
   year={2013},
   month=jul, pages={1330028} }

@article{Blennow_2011,
   title={Aidnogenesis via leptogenesis and dark sphalerons},
   volume={2011},
   ISSN={1029-8479},
   url={http://dx.doi.org/10.1007/JHEP03(2011)014},
   DOI={10.1007/jhep03(2011)014},
   number={3},
   journal={Journal of High Energy Physics},
   publisher={Springer Science and Business Media LLC},
   author={Blennow, Mattias and Dasgupta, Basudeb and Fernandez-Martinez, Enrique and Rius, Nuria},
   year={2011},
   month=mar }

@article{PESKIN2025116971,
title = {What is the Hierarchy Problem?},
journal = {Nuclear Physics B},
volume = {1018},
pages = {116971},
year = {2025},
issn = {0550-3213},
doi = {https://doi.org/10.1016/j.nuclphysb.2025.116971},
url = {https://www.sciencedirect.com/science/article/pii/S0550321325001804},
author = {Michael E. Peskin}
}

@article{Enqvist_2013,
doi = {10.1088/1475-7516/2013/10/057},
url = {https://doi.org/10.1088/1475-7516/2013/10/057},
year = {2013},
month = {oct},
publisher = {},
volume = {2013},
number = {10},
pages = {057},
author = {Kari Enqvist and Tuukka Meriniemi and Sami Nurmi},
title = {Generation of the Higgs condensate and its decay after inflation},
journal = {Journal of Cosmology and Astroparticle Physics}
}

@article{LieArt,
title = {LieART 2.0 – A Mathematica application for Lie Algebras and Representation Theory},
journal = {Computer Physics Communications},
volume = {257},
pages = {107490},
year = {2020},
issn = {0010-4655},
doi = {https://doi.org/10.1016/j.cpc.2020.107490},
url = {https://www.sciencedirect.com/science/article/pii/S0010465520302290},
author = {Robert Feger and Thomas W. Kephart and Robert J. Saskowski}
}

@article{Yamatsu2018SpecialFamily,
  author  = {Yamatsu, Naoki},
  title   = {Family unification in special grand unification},
  journal = {Progress of Theoretical and Experimental Physics},
  volume  = {2018},
  number  = {9},
  pages   = {091B01},
  year    = {2018},
  doi     = {10.1093/ptep/pty093}
}

@article{AcharyaWitten2002,
  author  = {Acharya, Bobby S. and Witten, Edward},
  title   = {Chiral fermions from manifolds of \(G_2\) holonomy},
  journal = {Journal of High Energy Physics},
  volume  = {2001},
  number  = {11},
  pages   = {045},
  year    = {2001},
  doi     = {10.1088/1126-6708/2001/11/045}
}

@article{BerglundBrandhuber2002,
  author  = {Berglund, Per and Brandhuber, Andreas},
  title   = {Matter from \(G_2\) manifolds},
  journal = {Nuclear Physics B},
  volume  = {641},
  number  = {1--2},
  pages   = {351--375},
  year    = {2002},
  doi     = {10.1016/S0550-3213(02)00612-0}
}

@article{HabaShimizu2003E7Orbifold,
  author  = {Haba, Naoyuki and Shimizu, Yasuhiro},
  title   = {Gauge--Higgs-boson unification in the five-dimensional \(E_6\), \(E_7\), and \(E_8\) grand unified theories on the orbifold},
  journal = {Physical Review D},
  volume  = {67},
  number  = {9},
  pages   = {095001},
  year    = {2003},
  doi     = {10.1103/PhysRevD.67.095001}
}

@article{HebeckerRatz2003,
  author  = {Hebecker, Arthur and Ratz, Michael},
  title   = {Group-theoretical aspects of orbifold and conifold GUTs},
  journal = {Nuclear Physics B},
  volume  = {670},
  number  = {1--2},
  pages   = {3--26},
  year    = {2003},
  doi     = {10.1016/j.nuclphysb.2003.07.021}
}

@article{LiTaylor2022PRD,
  author  = {Li, Shing Yan and Taylor, Washington},
  title   = {Natural F-theory constructions of standard model structure from \(E_7\) flux breaking},
  journal = {Physical Review D},
  volume  = {106},
  number  = {6},
  pages   = {L061902},
  year    = {2022},
  doi     = {10.1103/PhysRevD.106.L061902}
}

@article{LiTaylor2022JHEP,
  author  = {Li, Shing Yan and Taylor, Washington},
  title   = {Gauge symmetry breaking with fluxes and natural Standard Model structure from exceptional GUTs in F-theory},
  journal = {Journal of High Energy Physics},
  volume  = {2022},
  number  = {11},
  pages   = {089},
  year    = {2022},
  doi     = {10.1007/JHEP11(2022)089}
}

@article{LiTaylor2024JHEP,
  author  = {Li, Shing Yan and Taylor, Washington},
  title   = {Towards natural and realistic \(E_7\) GUTs in F-theory},
  journal = {Journal of High Energy Physics},
  volume  = {2024},
  number  = {5},
  pages   = {334},
  year    = {2024},
  doi     = {10.1007/JHEP05(2024)334}
}

@article{MizoguchiYata2013,
  author  = {Mizoguchi, Shun'ya and Yata, Masaya},
  title   = {Family unification via quasi-Nambu--Goldstone fermions in string theory},
  journal = {Progress of Theoretical and Experimental Physics},
  volume  = {2013},
  number  = {5},
  pages   = {053B01},
  year    = {2013},
  doi     = {10.1093/ptep/ptt019}
}

@article{Mizoguchi2014FTheoryFamily,
  author  = {Mizoguchi, Shun'ya},
  title   = {F-theory family unification},
  journal = {Journal of High Energy Physics},
  volume  = {2014},
  number  = {7},
  pages   = {018},
  year    = {2014},
  doi     = {10.1007/JHEP07(2014)018}
}

@article{Sato2022E7LmuLtau,
  author  = {Sato, Joe},
  title   = {Aiming for unification of \(L_{\mu}-L_{\tau}\) and the standard model gauge symmetries},
  journal = {Journal of High Energy Physics},
  volume  = {2022},
  number  = {7},
  pages   = {011},
  year    = {2022},
  doi     = {10.1007/JHEP07(2022)011}
}

@article{nym6-vpms,
  title = {Superconducting strings in ${E}_{6}$},
  author = {Maji, Rinku and Shafi, Qaisar},
  journal = {Phys. Rev. D},
  volume = {112},
  issue = {10},
  pages = {L101903},
  numpages = {6},
  year = {2025},
  month = {Nov},
  publisher = {American Physical Society},
  doi = {10.1103/nym6-vpms},
  url = {https://link.aps.org/doi/10.1103/nym6-vpms}
}

@article{MILAGRE2024116542,
title = {Unitarity constraints on large multiplets of arbitrary gauge groups},
journal = {Nuclear Physics B},
volume = {1004},
pages = {116542},
year = {2024},
issn = {0550-3213},
doi = {https://doi.org/10.1016/j.nuclphysb.2024.116542},
url = {https://www.sciencedirect.com/science/article/pii/S0550321324001081},
author = {André Milagre and Luís Lavoura}
}

@article{ClaudsonWiseHall1982,
  author = {Claudson, Mark and Wise, Mark B. and Hall, Lawrence J.},
  title = {Chiral Lagrangian for Deep-Mine Physics},
  journal = {Nuclear Physics B},
  volume = {195},
  pages = {297--307},
  year = {1982},
  doi = {10.1016/0550-3213(82)90428-6}
}

@article{Aoki2008NucleonDecayMatrix,
  author = {Aoki, Y. and others},
  title = {Proton decay matrix elements on the lattice},
  journal = {Physical Review D},
  volume = {78},
  pages = {054505},
  year = {2008},
  eprint = {0806.1031},
  archivePrefix = {arXiv},
  doi = {10.1103/PhysRevD.78.054505}
}

@article{Dorsner,
  author        = {Dor{\v{s}}ner, Ilja and others},
  title         = {Gauge and scalar boson mediated proton decay in a predictive grand unified model},
  journal       = {Physical Review D},
  volume        = {109},
  pages         = {075023},
  year          = {2024},
  doi           = {10.1103/PhysRevD.109.075023}
}

@article{ChungKolbRiotto1999ReheatingProduction,
  author        = {Chung, Daniel J. H. and Kolb, Edward W. and Riotto, Antonio},
  title         = {Production of massive particles during reheating},
  journal       = {Physical Review D},
  volume        = {60},
  pages         = {063504},
  year          = {1999},
  doi           = {10.1103/PhysRevD.60.063504},
  eprint        = {hep-ph/9809453},
  archivePrefix = {arXiv}
}

@article{ChungKolbRiotto1998NonthermalSupermassive,
  author        = {Chung, Daniel J. H. and Kolb, Edward W. and Riotto, Antonio},
  title         = {Nonthermal supermassive dark matter},
  journal       = {Physical Review Letters},
  volume        = {81},
  pages         = {4048--4051},
  year          = {1998},
  doi           = {10.1103/PhysRevLett.81.4048},
  eprint        = {hep-ph/9805473},
  archivePrefix = {arXiv}
}

@article{KofmanLindeStarobinsky1994Reheating,
  author        = {Kofman, Lev and Linde, Andrei D. and Starobinsky, Alexei A.},
  title         = {Reheating after inflation},
  journal       = {Physical Review Letters},
  volume        = {73},
  pages         = {3195--3198},
  year          = {1994},
  doi           = {10.1103/PhysRevLett.73.3195},
  eprint        = {hep-th/9405187},
  archivePrefix = {arXiv}
}

@article{KofmanLindeStarobinsky1997ReheatingTheory,
  author        = {Kofman, Lev and Linde, Andrei D. and Starobinsky, Alexei A.},
  title         = {Towards the theory of reheating after inflation},
  journal       = {Physical Review D},
  volume        = {56},
  pages         = {3258--3295},
  year          = {1997},
  doi           = {10.1103/PhysRevD.56.3258},
  eprint        = {hep-ph/9704452},
  archivePrefix = {arXiv}
}

@article{Turner1983CoherentOscillations,
  author  = {Turner, Michael S.},
  title   = {Coherent scalar-field oscillations in an expanding universe},
  journal = {Physical Review D},
  volume  = {28},
  pages   = {1243--1247},
  year    = {1983},
  doi     = {10.1103/PhysRevD.28.1243}
}

@article{CarenzaPasechnikSalinasWang2022GlueballRevisited,
  author        = {Carenza, Pierluca and Pasechnik, Roman and Salinas, Gustavo and Wang, Zhi-Wei},
  title         = {Glueball Dark Matter Revisited},
  journal       = {Physical Review Letters},
  volume        = {129},
  pages         = {261302},
  year          = {2022},
  doi           = {10.1103/PhysRevLett.129.261302},
  eprint        = {2207.13716},
  archivePrefix = {arXiv},
  primaryClass  = {hep-ph}
}

@article{CarenzaPasechnikSalinasWang2023GlueballDM,
  author        = {Carenza, Pierluca and Pasechnik, Roman and Salinas, Gustavo and Wang, Zhi-Wei},
  title         = {Glueball dark matter},
  journal       = {Physical Review D},
  volume        = {108},
  pages         = {123027},
  year          = {2023},
  doi           = {10.1103/PhysRevD.108.123027},
  eprint        = {2306.09510},
  archivePrefix = {arXiv},
  primaryClass  = {hep-ph}
}

@article{BakerKoppLong2020FilteredDM,
  author        = {Baker, Michael J. and Kopp, Joachim and Long, Andrew J.},
  title         = {Filtered Dark Matter at a First Order Phase Transition},
  journal       = {Physical Review Letters},
  volume        = {125},
  pages         = {151102},
  year          = {2020},
  doi           = {10.1103/PhysRevLett.125.151102},
  eprint        = {1912.02830},
  archivePrefix = {arXiv},
  primaryClass  = {hep-ph}
}

@article{GiudiceLeePomarolShakya2024FOPTDM,
  author        = {Giudice, Gian F. and Lee, Hyun Min and Pomarol, Alex and Shakya, Bibhushan},
  title         = {Nonthermal heavy dark matter from a first-order phase transition},
  journal       = {Journal of High Energy Physics},
  volume        = {2024},
  number        = {12},
  pages         = {190},
  year          = {2024},
  doi           = {10.1007/JHEP12(2024)190},
  eprint        = {2403.03252},
  archivePrefix = {arXiv},
  primaryClass  = {hep-ph}
}

@article{GarnyPalessandroSandoraSloth2018PIDM,
  author        = {Garny, Mathias and Palessandro, Andrea and Sandora, McCullen and Sloth, Martin S.},
  title         = {Theory and Phenomenology of Planckian Interacting Massive Particles as Dark Matter},
  journal       = {Journal of Cosmology and Astroparticle Physics},
  volume        = {2018},
  number        = {02},
  pages         = {027},
  year          = {2018},
  doi           = {10.1088/1475-7516/2018/02/027},
  eprint        = {1709.09688},
  archivePrefix = {arXiv},
  primaryClass  = {hep-ph}
}

@article{MambriniOlive2021GravDM,
  author        = {Mambrini, Yann and Olive, Keith A.},
  title         = {Gravitational Production of Dark Matter during Reheating},
  journal       = {Physical Review D},
  volume        = {103},
  pages         = {115009},
  year          = {2021},
  doi           = {10.1103/PhysRevD.103.115009},
  eprint        = {2102.06214},
  archivePrefix = {arXiv},
  primaryClass  = {hep-ph}
}
\bibliographystyle{IEEEtran}

\end{document}